%% file: main.tex
\documentclass[10pt,pre,twocolumn,tightenlines,longbibliography,notitlepage,superscriptaddress,nofootinbib]{revtex4-2}
\pdfoutput=1

\usepackage{xcolor}
\usepackage{amsmath}
\usepackage{verbatim}
\usepackage{graphicx}
\usepackage{natbib}
\usepackage{siunitx}
\usepackage{float}
\usepackage{xr}
\usepackage{amsfonts}

\usepackage[textsize=tiny]{todonotes}

\begin{document}

\title{From toroids to helical tubules: Kirigami-inspired programmable assembly of two-periodic curved crystals}


\author{Mason Price}
\affiliation{Martin A. Fisher School of Physics, Brandeis University, Waltham, Massachusetts 02453, USA}
\author{Daichi Hayakawa}
\affiliation{Martin A. Fisher School of Physics, Brandeis University, Waltham, Massachusetts 02453, USA}
\author{Thomas E. Videb\ae k}
\affiliation{Martin A. Fisher School of Physics, Brandeis University, Waltham, Massachusetts 02453, USA}
\author{Rupam Saha}
\affiliation{Martin A. Fisher School of Physics, Brandeis University, Waltham, Massachusetts 02453, USA}
\author{Botond Tyukodi} 
\affiliation{Department of Physics, Babe\c{s}-Bolyai University, 400084 Cluj-Napoca, Romania}
\affiliation{Centre International de Formation et de Recherche Avanc\'{e}es en Physique, 077125 Bucharest-M\={a}gurele, Romania}
\affiliation{Martin A. Fisher School of Physics, Brandeis University, Waltham, Massachusetts 02453, USA}
\author{Michael F. Hagan}
\affiliation{Martin A. Fisher School of Physics, Brandeis University, Waltham, Massachusetts 02453, USA}
\author{Seth Fraden}
\affiliation{Martin A. Fisher School of Physics, Brandeis University, Waltham, Massachusetts 02453, USA}
\author{Gregory M. Grason}
\affiliation{Department of Polymer Science and Engineering, University of Massachusetts, Amherst, Massachusetts 01003, USA}
\author{W. Benjamin Rogers}
\email{wrogers@brandeis.edu}
\affiliation{Martin A. Fisher School of Physics, Brandeis University, Waltham, Massachusetts 02453, USA}

\begin{abstract}
Biology is full of intricate molecular structures whose geometries are inextricably linked to their function. Many of these structures exhibit varying curvature, such as the helical structure of the bacterial flagellum, which is critical for their motility. Because synthetic analogues of these shapes could be valuable platforms for nanotechnologies, including drug delivery and plasmonics, controllable synthesis of variable-curvature structures of various material systems, from fullerenes to supramolecular assemblies, has been a long-standing goal. Like two-dimensional crystals, these structures have a two-periodic symmetry, but unlike standard two-dimensional crystals, they are embedded in three dimensions with complex, spatially-varying curvatures that cause the structures to close upon themselves in one or more dimensions.  Here, we develop and implement a design strategy to program the self-assembly of a complex spectrum of {\it two-periodic curved crystals} with variable periodicity, spatial dimension, and topology, spanning from toroids to achiral serpentine tubules to both left- and right-handed helical tubules. Notably, our design strategy exploits a kirigami-based mapping of a modular class of 2D planar tilings to 3D curved crystals that preserves the periodicity, two-fold rotational symmetries, and subunit dimensions by modulating the arrangement of disclination defects. We survey the modular geometry of these curved crystals and infer the addressable subunit interactions required to assemble them from triangular subunits. To demonstrate this design strategy in practice, we program the self-assembly of toroids, helical- and serpentine-tubules from DNA origami subunits, deriving the distinct kirigami foldings of a single two-periodic planar tiling. A simulation model of the assembly pathways reveals physical considerations for programming the geometric specificity of angular folds in the curved crystal required to avoid defect-mediated misassembly.

\end{abstract}

\maketitle

Programmable self-assembly aims to direct the structure and composition of materials by encoding information about the target organization into the local attributes of nanoscale building blocks~\cite{whitesides2002self}.  The most sought-after---and arguably the most functional---targets for programmable assembly are periodic crystalline structures, which are characterized by well-defined, repeating intersubunit spacings and orientations~\cite{Zheng2009Sep,liu_self-organized_2016, Macfarlane2011Oct,wang2015crystallization}.  Over the past few decades, there has been tremendous progress in targeting crystals with ever-increasing complexity, such as open three-dimensional diamond crystals~\cite{liu2016diamond, he2020colloidal, Posnjak2024May} or two-dimensional Kagome lattices~\cite{chen2011directed}, by designing colloidal particles with directional, specific attractions. The ability to further combine orientation-specific with type-specific interactions has opened the door to even more complex, ``supracrystals'', with multi-component unit cells that far exceed the size of the subunits themselves~\cite{Russo_2022, kahn2025arbitrary, Hayakawa2024, Lee2025Jan, jacobs2025assembly}.

While these examples of two- and three-dimensional crystals are exceptional achievements of synthetic crystal engineering in their own right, they leave out a remarkable class of periodic structures that we call ``curved crystals''. Curved crystals can largely be conceptualized by considering locally rolling or bending 2D planar crystals into curved 3D geometries.  Arguably, the simplest example is a 2D crystal that has been rolled into a cylinder, which is the template of various nanomaterials from carbon nanotubes (CNT) to filamentous protein assemblies, such as microtubules~\cite{iijima1991helical, chretien1995structure}. Programmable assembly of synthetic DNA- and de novo protein-based building blocks has also succeeded in targeting cylindrical curved crystals~\cite{Rothemund2004tube, mohammed2013directing, shen_novo_2018, Hayakawa2022Oct, Dong2022Dec}, as well as another canonical uniform-curvature geometry---spherical shells---creating structures akin to fullerenes and virus capsids~\cite{bale_accurate_2016, tikhomirov_triangular_2018, sigl_programmable_2021, dowling2024hierarchical}.  
However, these examples of curved crystals belong to a limited class corresponding to uniform Gaussian-curvature shapes---zero for cylinders and positive for spheres---in which the Gaussian curvature is the same everywhere on the surface.
Notably, while biological assemblies have evolved means to assemble curved crystals with spatially variable Gaussian curvature, such as the helical bacterial flagellum~\cite{nakamura2019flagella}, developing design strategies to assemble related structures with precisely controlled geometries from synthetic components remains a long-standing challenge~\cite{ dunlap1992connecting, tamura2005positive, giomi2008defective, chuang2009generalized, beuerle2011optical, sarapat2019review, brisson2024nanoscale}.

\begin{figure*}[htb]
 \centering
 \includegraphics[width=0.99\linewidth]{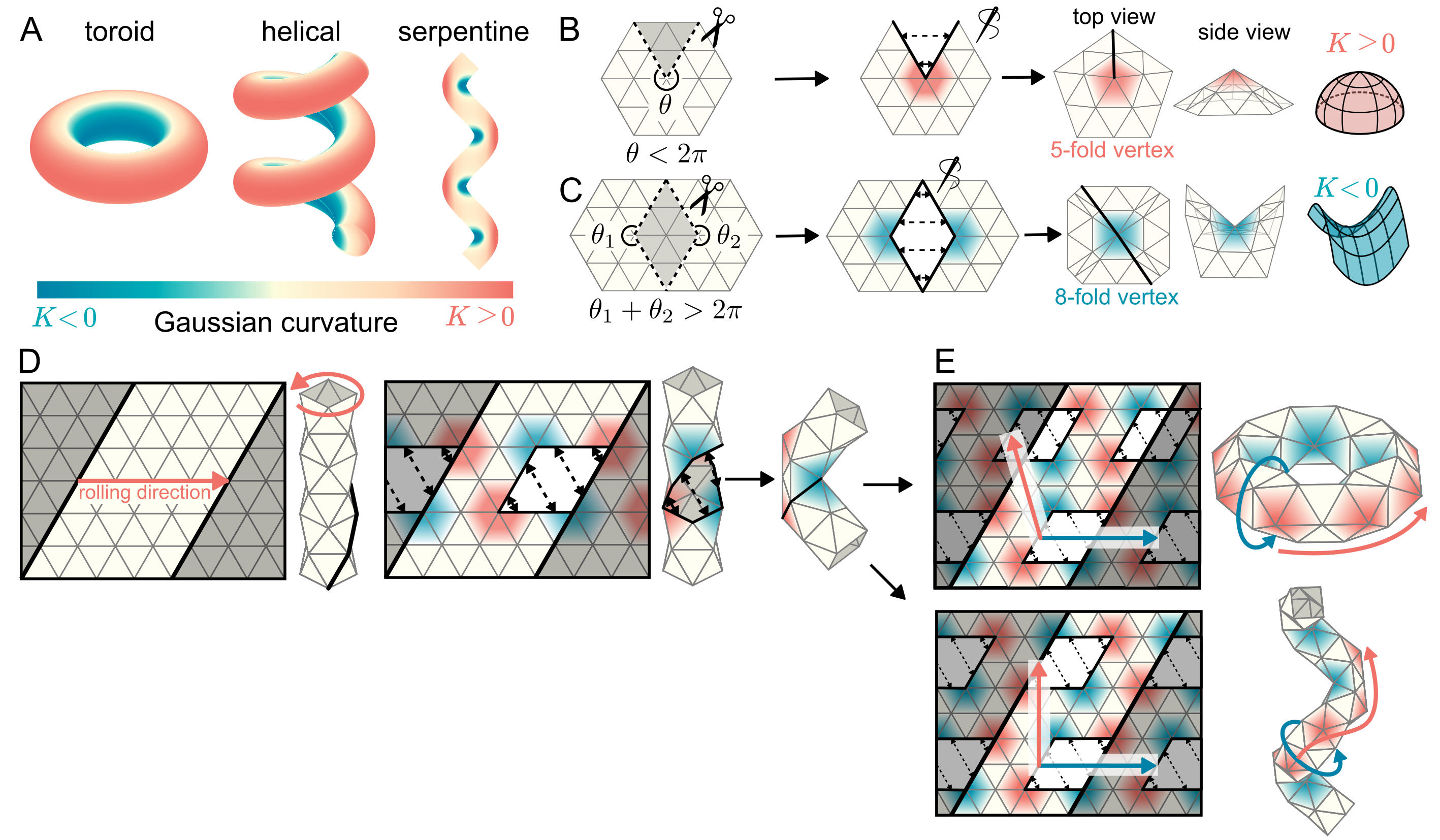}
 \caption{\textbf{Kirigami construction of two-periodic curved crystals.}
(A) The spatially varying Gaussian curvature ($K$) of a toroid, helical tubule, and serpentine tubule. (B and C) Illustration of removing patches from a lattice and stitching them closed to create local (B) positive or (C) negative Gaussian curvature. (B) Removing a patch and stitching it closed with an angle in deficit of $2\pi$ creates a defect of positive Gaussian curvature. (C) Conversely, closing an angle in excess of $2\pi$ creates a defect with negative Gaussian curvature~\cite{perotti2019kirigami, koning2016crystals}. 
(D) We apply kirigami to a cylindrical tubule to form a ``bent'' tubule segment. (E) The holes in the 2D tiling repeat with two fixed translation generators that control the final geometry. The generators map to curvilinear paths that follow along and around the 3D folded structure. We show a toroid and a helical tubule as examples, which have positive Gaussian curvature along the outer equator and negative Gaussian curvature along the inner equator. 
}
 \label{fig:1}
\end{figure*}

In this article, we develop a general approach to programming the assembly of a family of two-periodic curved crystals with spatially patterned Gaussian curvature, targeting modular 3D shapes spanning from toroids to serpentine tubules to helical tubules of controlled helicity (Fig.~\ref{fig:1}A). 
Exploiting the fact that this family of target shapes has net-zero Gaussian curvature, we develop a kirigami-based design strategy applied to planar 2D tilings. 
Our design scheme is based on the removal of specific domains from programmable 2D tilings to create controlled arrangements of disclination defects that act as local sources of positive and negative curvature (Fig.~\ref{fig:1}B, C)~\cite{Castle2016Sep} that `bend' and `twist' an otherwise Euclidean assembly (i.e. a cylinder) (Fig.~\ref{fig:1}D,E). We demonstrate this design approach by designing and self-assembling toroids, serpentine, and helical tubules composed of DNA-origami subunits whose edge-edge interactions and local curvature can be specified through the design of the DNA base sequences~\cite{saha2025modular}. We further use kinetic Monte Carlo (KMC) simulations to explore the physical regimes of on-target assembly for toroids and helical tubules, which reveal that type specificity alone is insufficient to guarantee high-fidelity assembly of the target structure and that geometric specificity is required to suppress off-target polymorphs.

\section*{Results}

\begin{figure*}[]
 \centering
 \includegraphics[width=0.99\linewidth]{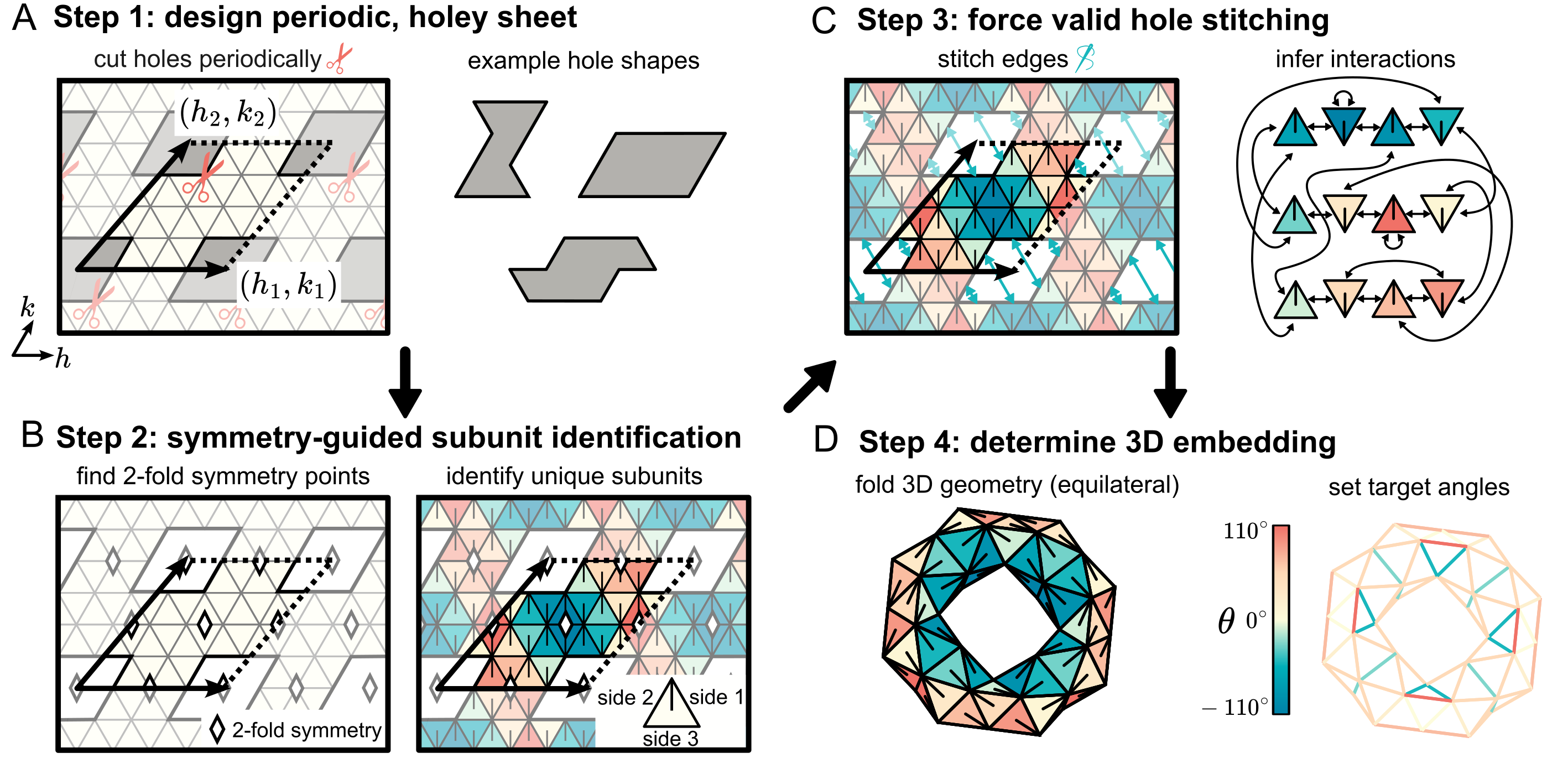}
 \caption{\textbf{Kirigami design for deriving interactions and binding angles of assembling subunits.}
(A) Remove patches with user-specified size, shape, and periodicity. (B) Identify the symmetries of the resultant tiling, color the triangles, and assign their orientations according to the symmetry, so that all triangles that can be mapped to each other through symmetry operations have the same color. (C) Stitch vertices across the holes and identify edge-to-edge interactions between neighboring subunits. (D) Map the colors and orientations onto a stitched-and-folded 3D configuration and measure the binding angles between neighboring subunits. 
}
 \label{fig:2}
\end{figure*}


\subsection{Kirigami design of curved crystals}

In kirigami, one makes a sequence of cuts, folds, and (possibly) stitches to transform a flat sheet into a three-dimensional shape. As shown previously for ``lattice-based kirigami'', the cutting and resewing of 2D tilings results in controlled arrangements of disclination defects that act as local sources of positive and negative curvature (Fig.~\ref{fig:1}B, C)~\cite{Castle2016Sep}. 
While this coupling between disclinations and Gaussian curvature is well-appreciated for 3D shapes of 2D crystals~\cite{chuang2009generalized, giomi2008defective,bausch2003grain, irvine2010pleats, sun2025colloidal}, the ability to design and program the specific locations of such ``defects'' in self-assembled crystals, to thereby modulate their 3D global shapes, has so far not been demonstrated. 
Below, we articulate a design strategy for filling this conceptual gap by using interaction specificity to guide the placement of defects in an assembly, focusing on the design of a family of curved cylindrical crystals that includes both toroids and helical tubules.

Our approach is to apply kirigami to a 2D triangular lattice and `roll' the resultant sheet into a closed tubule along a specified rolling direction.  By using kirigami to introduce local regions of positive and negative Gaussian curvature, we can `bend' a closed tubule into a toroid or helical tubule with the appropriate curvature, controlled by the relative locations of the holes. This process is detailed in four distinct steps (Fig.~\ref{fig:2}A--D).

In step one, we cut holes in a flat sheet of triangles (Fig.~\ref{fig:2}A). We choose two translation vectors that correspond to the two periodic directions of the curved tubule. The first vector $(h_1,k_1)$ is equal to the rolling vector of a corresponding tubule, and the second vector $(h_2,k_2)$ repeats with some component along the length of the tubule. These two vectors define a unit cell on the tiling. At the corners of this unit cell, we cut out four identical holes. In this design, we require the shape of these holes to be two-fold symmetric, allowing for many choices of their shape. We also focus on the case where the rolling vector $(h_1, k_1)$ is along the lattice direction (that is, $k_1=0$), but we note that this is not necessary and explore variations in SI section V D).

In step two, we find the minimal number of subunit types and their arrangement in the holey tiling (Fig.~\ref{fig:2}B). In general, the most economical design exploits tilings of the highest order symmetry~\cite{Hayakawa2024}. Because tubules can be assembled from tilings with either o or 2222 symmetry, we choose designs with unit cells having four two-fold rotational symmetry points (see SI section II B). One constraint in this design is that we enforce a two-fold point to align with the holes. Then, leveraging a recently developed algorithm~\cite{Hayakawa2024}, we find the associated coloring of the triangles and assign their orientations (see SI section II B). Crucially, this step tells us the minimum number of subunit species that are required for a given assembly. 

In step three, we determine the interaction matrix that specifies the graph of specific interparticle attractions (Fig.~\ref{fig:2}C). We first stitch closed the holes. A stitching is a set of binding rules, in which pairs of edges are matched to each other. For a stitching to be valid, all edges in the hole must be part of a pair, and, if one were to draw lines between paired edges, no lines would cross. Once the stitching is determined, we infer the graph of all inter-subunit edge-edge interactions from the stitched tiling (Fig.~\ref{fig:2}C). 

Finally, in step four, we find the binding angles between all pairs of bound subunits in the final 3D geometry. To identify the associated binding angles, we fold the holey tiling into the corresponding 3D geometry by simulating the surface as an elastic triangular mesh: After merging the edges according to the stitching rules, we relax the triangular mesh into its ground-state configuration, so that all edge lengths are equal (see SI section II A). Then, we measure the binding angles between bound triangles in the 3D geometry (Fig.~\ref{fig:2}D). 

These four steps yield the set of specific interactions and the binding angles that fully define the local rules that need to be encoded in the building blocks to program the self-assembly of the associated curved tubule. Beyond the toroid shown in Figure~\ref{fig:2}, this systematic approach enables the design of a whole zoo of curved cylindrical crystals. We describe the design principles and some example classes in SI sections II and V.

\subsection{Systematic modulation of 3D geometries from 2D holey tilings}

\begin{figure}[]
 \centering
 \includegraphics[width=0.99\linewidth]{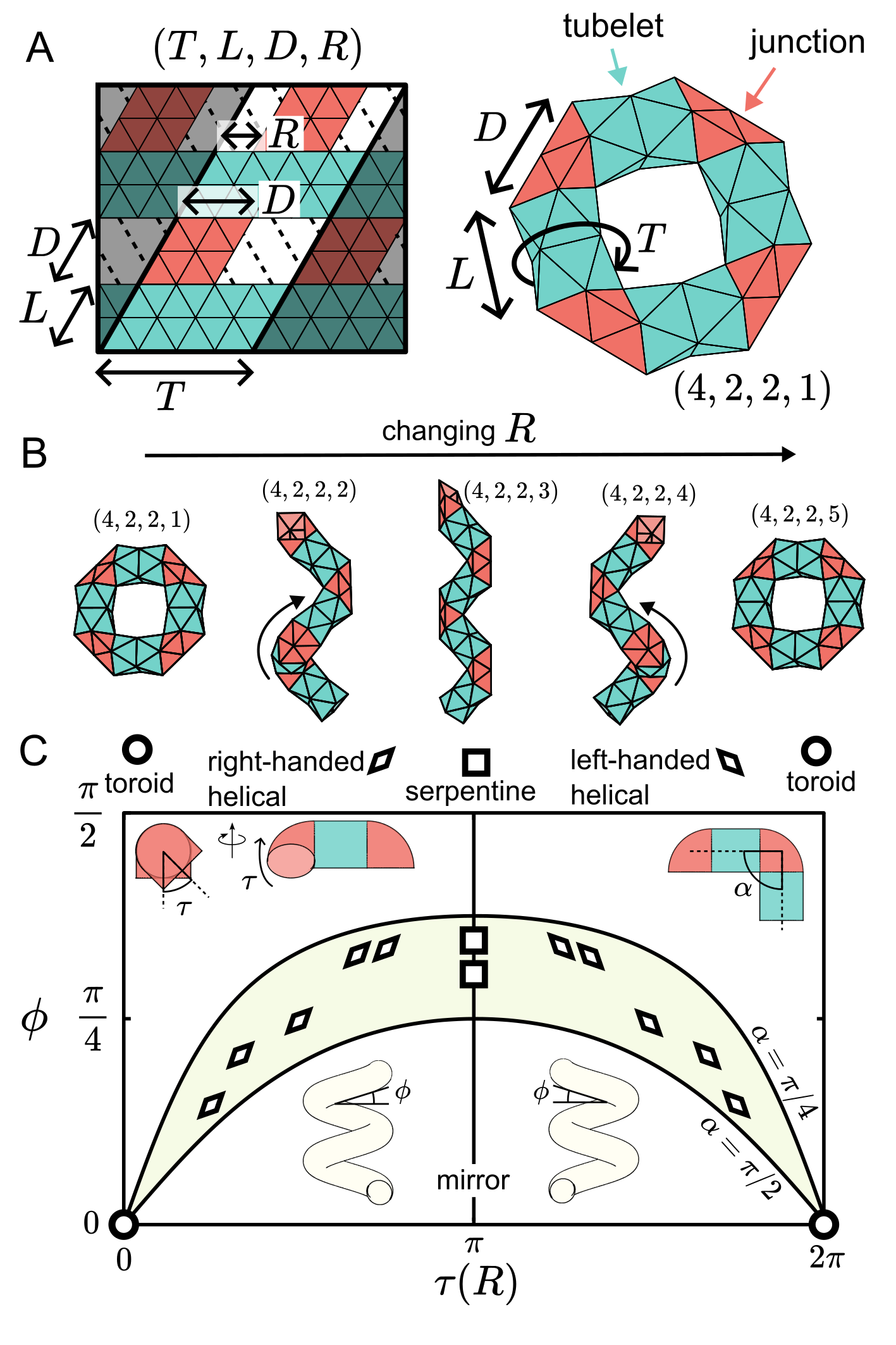}
 \caption{\textbf{Geometry of a subclass of curved tubules.}
(A) Parameters $(T,L,D,R)$ define the size and locations of excised patches. Toroids consist of tubelet and junction regions. (B) A family of helical tubules that result from varying $R$. (C) The helix angle $\phi$ transitions smoothly from toroid to helical tubule to serpentine tubule and back as the inter-tubelet twist $\tau$ varies from 0 to $2\pi$. The data points show examples for $T=3,4$, $L = 1,2,3$, and $D$ equal to the integer part $T/2$. We measure the actual pitch angle $\phi$ for a given geometry by fitting a parametrized surface to the set of vertices (see SI section III).
}
 \label{fig:geometry}
\end{figure}

To better understand the connection between the 2D design and the 3D curved geometry, we introduce a subclass of 2D holey tilings, which we term the \textit{TLDR} tilings. We specifically consider curved tubules that consist of a small tubule region, which we call the tubulet, and a diamond-shaped junction region that connects two tubelets  (Fig.~\ref{fig:geometry}A). This class of curved tubules can be fully specified by four parameters, $T, L, D,$ and $R$, which denote the tubulet circumference, tubulet length, junction size, and the spacing between cuts in the rolling direction, respectively. For simplicity, we impose a set of stitching rules that connect triangles across the shorter diagonal of the holes (the possible stitching rules are further described in SI section V). 

We find that the design space defined by this construction includes toroids, both right- and left-handed helical tubules, and achiral serpentine tubules, as determined by the inter-cut spacing $R$ (Fig.~\ref{fig:geometry}B). In the folded 3D geometry, we find that changing $R$ corresponds to rotating the junctions with respect to the tubelet axes, leading to the variety of curved tubules.  As an example, the toroid described in Fig.~\ref{fig:2} is classified as (4,2,2,1).

By examining the $TLDR$ design space, we identify the design choices that lead to the formation of toroids. When $D\leq T/2$, the specific value of $R$ that corresponds to a toroid can be found using the positioning of disclination points with respect to the 2-fold rotational symmetry points (see SI section II C), and is given by 
\begin{equation}\label{eq:s_toroid}
R_{\text{toroid}} \equiv -D-L/2\pmod{T}.
\end{equation}
Note that the surface is equivalent if $R$ is offset by a factor of $T$ because the tubelet is periodic around its circumference. Furthermore, since $R$ is an integer, we conclude from Eq.~(\ref{eq:s_toroid}) that tubelet sections are coplanar if and only if $L$ is even. This condition is necessary for a toroid state, but does not guarantee it, since closing all the bonds may require stretching of edges.

Shifting the inter-cut spacing $R$ away from the toroid configuration ``twists'' the arrangement of tubelets out-of-plane and into various helical configurations. This transformation leads to helices because adjacent tubelets have a fixed pair-wise relative orientation, and yet any triplet of tubelets is not coplanar~\cite{lord2002helical, read2022calculating}. In our design scheme, offsetting the inter-cut spacing $R$ rotates each junction about the tubelet axes by an angle $\tau$ (Fig.~\ref{fig:geometry}C). Assuming then that the tubelets are rigid segments, the twist angle $\tau$ is given by
\begin{equation}
\tau (R) = 2 \pi \cdot \frac{R-R_{\text{toroid}}}{T},
\end{equation}
so that $\tau(R)=0$ corresponds to a planar toroid configuration. 

This abstraction of the input parameter $R$ into a rotation angle $\tau(R)$ leads to a simple geometric interpretation. The twist between tubelets determines the handedness of the corresponding helical tubule. For $0 < \tau(R) < \pi$, the helical tubules are right-handed while for $\pi< \tau(R) < 2\pi$, they are left-handed (Fig.~\ref{fig:geometry}C). For $\tau(R)=\pi$, we find an interesting achiral serpentine geometry, which exists if $T-L$ is even, and is the state with the maximum pitch angle $\phi$. Finally, at $\tau(R) = 2\pi $, the surface returns to a toroid. Figure~\ref{fig:geometry}C shows how the pitch of helical tubules changes over the domain of $\tau$ for examples with $T=3,4$ and $L=1,2,3$, where $D$ is taken to be the integer part of $T/2$ for simplicity (see SI section II D for a gallery of the corresponding structures). The highlighted region represents the design space of possible states assuming a rigid segmented helix model, where the upper- and lower-bounds are defined by the approximation $\phi = \arctan \left[ \sin(\tau/2) \sin \alpha / (1-\cos \alpha) \right]$ for a co-planar angle $\alpha$ of adjacent tubelets between $\pi/4$ and $\pi/2$, corresponding to 8- and 4-fold symmetric toroids, respectively (see SI section II E).  Beyond reflecting our broad observations above, this relationship illustrates a smooth transition of helical states as the twist angle $\tau$ goes from 0 to $2\pi$, highlighting that these surfaces indeed belong to a family of related geometries.


\subsection{Programmable self-assembly of curved tubules using DNA origami triangles}

\begin{figure*}
 \centering
 \includegraphics[width=0.98\linewidth]{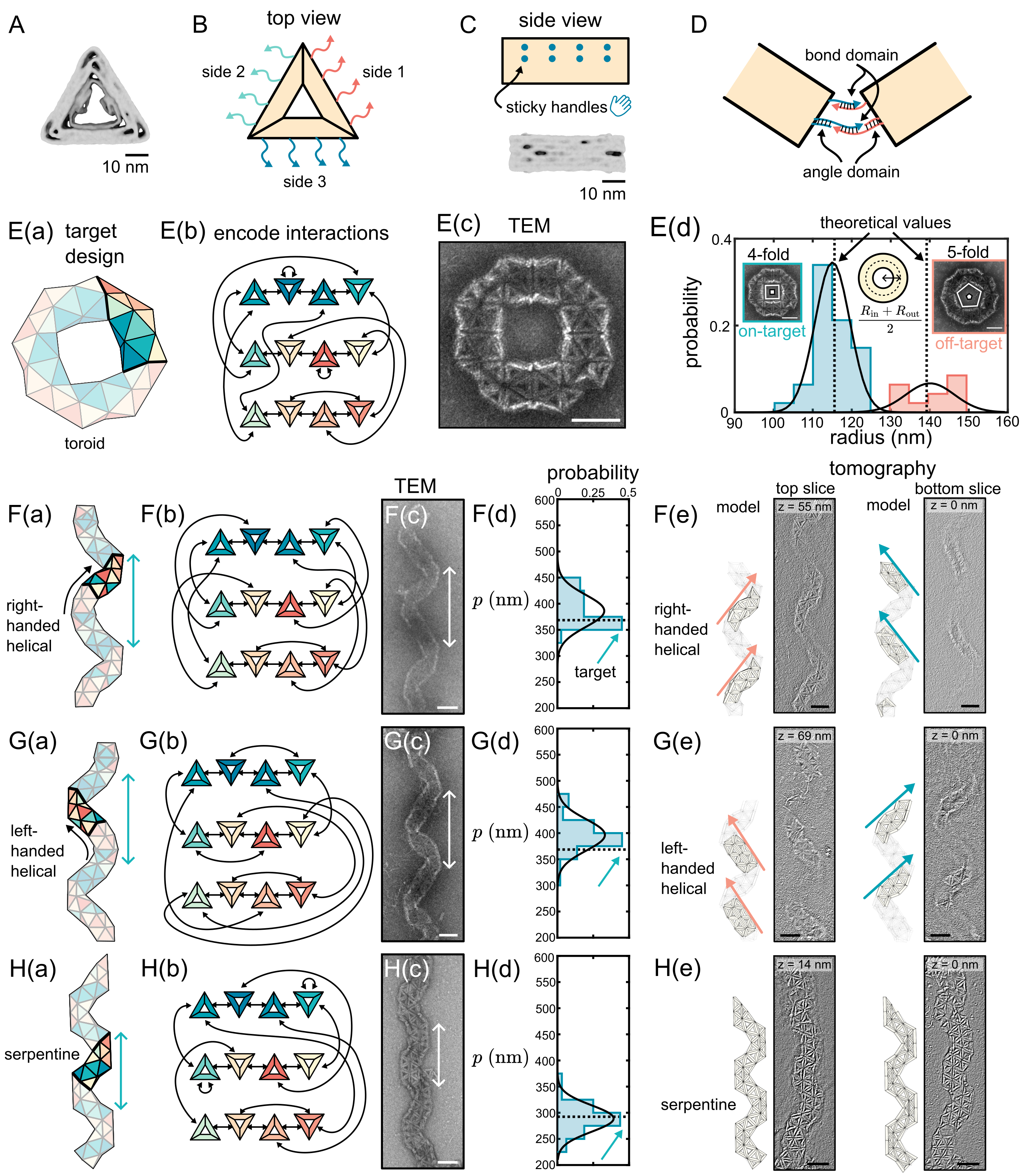}
 \caption{\textbf{Design implementation using modular DNA-origami subunits.} 
(A) Shows a cryo-EM reconstruction of the subunit. (B) Side-specific ssDNA sequences protrude from each edge of the DNA origami subunit. (C) The handles protrude in two rows on each side. (D) Interactions are controlled by the bond domain, while binding angles are controlled by changing the length of the angle domain on the middle row. (E--H)(a) Target designs from which we infer interactions and binding angles. (b) Graph of interactions between species. (c) Negative-stain TEM micrographs of assemblies. (E)(d) Probability distribution of toroid radii shows both 4-fold and 5-fold toroids. The x-axis denotes the average of the inner- and outer-radii. There were 47 measurements used for this distribution. (F-H)(d) Measured values of pitch for helical tubules compared to the target value. The number of measurements taken for these distributions are (F) 38, (G) 51, (H) 32. (F-H)(e) Tomographic reconstruction of assembled curved tubules. Scale bars are 100~nm.}
 \label{fig:experiment}
\end{figure*}

We next switch our attention to implementing our design scheme in practice. Inspired by the recent examples of self-assembly of tubules and shells ~\cite{sigl_programmable_2021, Hayakawa2022Oct, videbaek2024economical, saha2025modular}, we demonstrate the utility of our design approach by assembling all four curved tubules illustrated in Figure~\ref{fig:geometry}B from 50-nm-sized equilateral triangular subunits synthesized using DNA origami (Fig.~\ref{fig:experiment}A). All of these assemblies can be constructed from twelve distinct particle species, but with varying pairwise interactions and binding angles, which we program using single-stranded DNA (ssDNA) sequences that protrude from the triangle edges along two rows (Fig.~\ref{fig:experiment}B,C). The ``bond domains'' consist of 5 base-pair sequences and encode specific edge-edge interactions via Watson–Crick base pairing (Fig.~\ref{fig:experiment}D). The relative lengths of the ``angle domains'' encode the binding angles between neighboring subunits (Fig.~\ref{fig:experiment}D). Importantly, this method of controlling interparticle interactions and binding angles from a modular subunit allows us to encode a large number of different species without redesigning the entire subunit~\cite{saha2025modular}, which would otherwise be intractable given the number of distinct components required to assemble structures of this complexity.

To assemble each design, we encode for the interactions and binding angles based on our design (Fig.~\ref{fig:experiment}E-H(a,b)), and then mix the different species at the same molar concentration and tune the temperature to control the intersubunit attractions. For toroids, we perform assembly by slowly cooling the mixture from 40--25~$^\circ$C over four days to remove secondary structures. For helical and serpentine tubules, we perform assembly isothermally at 25~$^\circ$C to limit the nucleation rate and assemble longer curved tubules. After assembly, we image the resultant structures using negative-stain transmission electron microscopy (TEM). See Methods for further experimental details.

In toroid assembly, we observe the desired on-target geometry, but also an unexpected, closed, off-target structure. Among the fully-closed structures, a majority of the assemblies (about 80\%) are the intended 4-fold symmetric toroid containing 96 subunits (Fig.~\ref{fig:experiment}E(c)). Surprisingly, in addition to the on-target geometry, we also observe the assembly of an off-target, 5-fold symmetric toroid with 120 subunits (Fig.~\ref{fig:experiment}E(d)). We provide two hypotheses for the formation of this off-target structure. First, sufficiently large thermal bending fluctuations in the assembling curved tubule could permit the formation of off-target self-closing sizes, as observed previously for straight tubules and icosohedral shells~\cite{videbaek2024economical,tyukodi2024magic}. Second, in addition to prescribing the angles, the bond and angle domains in our subunit design add extra distance between particles, especially at regions with negative curvatures, which would tend to open up the toroids away from their target radius~\cite{saha2025modular}.  Beyond these two closed geometries, we also observe a range of incomplete oligomers, likely due to a kinetic deadlock encountered at the end of our cooling ramp~\cite{Deeds2012Feb}. 

In addition to the toroid, all three helical-tubule designs successfully assemble, as evidenced by the unique wave-like structures viewed under negative-stain TEM (Fig.~\ref{fig:experiment}F-H(c)). At first glance, the right- and left-handed helical tubules and the serpentine tubule look qualitatively similar. However, differences in measurements of their pitch and subtle variations in their staining reveal key structural differences. First, the average pitch of both chiral tubules is roughly 400 nm, whereas the average pitch of the serpentine tubule is only about 300 nm, with all three values in close agreement with our predictions (Fig.~\ref{fig:experiment}F-H(d)). Second, we observe periodic patterns of dark and light stain on the right- and left-handed helical tubules, whereas the stain around the serpentine tubules is uniform. We attribute these regions of more or less stain to the coiled structure of the right- and left-handed helical tubules, in which some regions of the tubules lie flat on the grid and others are suspended above the grid and can therefore accumulate different amounts of stain. 

To verify the helical nature of the chiral tubules and determine their handedness, we perform tomographic reconstruction  (Fig.~\ref{fig:experiment}F--H(e)). We find that the top-most slices of the right- and left-handed helical tubules are mirror images of one another, indicating that they are the opposite handedness (Fig.~\ref{fig:experiment}F,G(e)). Furthermore, we find that the top-most and bottom-most slices of the two structures individually point in opposite directions, as one would expect for a coil (Fig.~\ref{fig:experiment}F,G(e)). In contrast, the serpentine tubule is achiral and lies flat on a TEM grid (Fig.~\ref{fig:experiment}H(e)). We also verify that the helical tubules have the correct handedness. While determining the chirality from tomographic reconstruction typically requires a detailed understanding of the microscope setup and the analysis~\cite{briegel_challenge_2013}, we infer the correct handedness of our helical tubules by referencing a chiral straight tubule with known handedness from our previous studies~\cite{videbaek2024economical}. These results clearly show that both the pitch and handedness of the helical tubules are programmable via our design approach.

Interestingly, while toroids exhibit off-target closed assemblies, helical tubules do not exhibit analogous off-target states. This key difference may be because helical tubules have no nearby off-target closed states that are accessible by thermal fluctuations. However, such off-target states may still exist. So, although the agreement between design and experiment is promising, the existence of off-target toroid states begs a few questions: What types of off-target structures exist for a given set of interactions and binding angles? When do such off-target structures become prevalent? And what are the kinetic pathways for such off-target structures? We explore these questions in the next section.

\subsection{Exploring the physical design space with KMC simulations}

\begin{figure*}[t]
 \centering
 \includegraphics[width=0.99\linewidth]{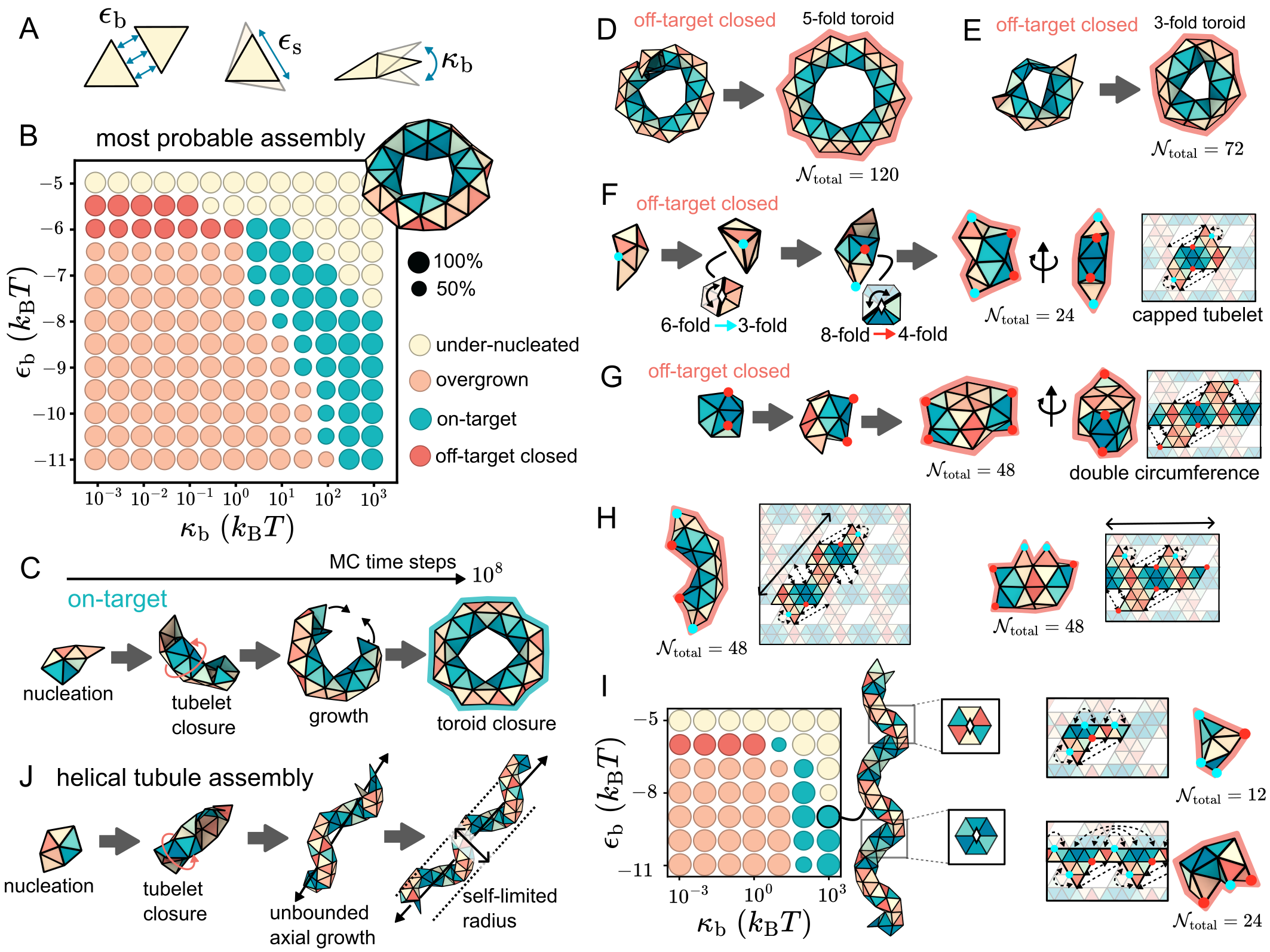}
 \caption{\textbf{Simulations of assembly of toroids and helical tubules.} 
 (A) Illustration of the three energy terms: binding energy $\epsilon_{\rm b}$, stretching modulus $\epsilon_{\rm s}$, and bending modulus $\kappa_{\rm b}$. (B) Simulating the assembly of the $(4,2,2,1)$ toroid reveals multiple assembly outcomes. Each point represents the dominant assembly outcome, and the size of the point represents the relative probability of that outcome. (C) Illustration of the on-target assembly pathway of a toroid, which starts with nucleation and ends with the assembly closing into a toroid. (D and E) Examples of off-target toroids that close with the incorrect symmetry about the toroid's axis, forming (D) 5-fold and (E) 3-fold symmetric toroids. (F) Other off-target assembly pathways incorporate vertex defects and close with the wrong geometry and topology. These assemblies correspond to a subset of the holey tiling that is closed in ways that are compatible with the local interactions of the holey tiling. (G) Another example of an off-target closed assembly forms when a tubelet closes with twice the intended circumference, further mediated by vertex defects. (H) Off-target closures can also be repeated periodically. Here are examples of off-target states that arise from repeating the tiling subset of (F) along either of the translation vectors of the holey tiling. (I) Assembly of helical tubules shows the same trends as toroids and has similar mechanisms for off-target closure. (J) The on-target assembly pathway of helical tubules begins with tubelet closure and continues with unbounded growth along its axis. All simulations were performed using a chemical potential $\mu = -4.5~k_{\rm B}T$ for all species. To approximate rigid particles, we fix the stretching modulus as $\epsilon_\text{s}=100~k_{\rm B}T/l_0^2$, where $l_0$ is the edge length of the triangles. We run 50 simulations for each condition with a maximum of $10^8$ Monte Carlo sweeps per simulation, terminating early if the assembly has no free edges.
 }
 \label{fig:5}
\end{figure*}

To explore the potential off-target structures accessible within our design space, we adapt a model from previous works on constant curvature \cite{Rotskoff2018, Panahandeh2018, Wagner2015, duque2024limits, Tyukodi2022Jun, videbaek2022tiling, Fang2022, Mohajerani2022a, tyukodi2024magic}, and perform KMC simulations of toroid and helical-tubule assembly, varying the bond strength $\epsilon_\text{b}$ and the bending modulus $\kappa_\text{b}$.  In these simulations, we model the subunits as coarse-grained triangles with flexible edges with stretching modulus $\epsilon_{\rm s}$ that can bind to one another along complementary edges with a binding energy $\epsilon_{\rm b}$. Bound triangles have a preferred binding angle $\theta_0$, which sets the local curvature, and a corresponding bending modulus $\kappa_{\rm b}$ (Fig. \ref{fig:5}A). The total energy of an assembly has three components: (1) the sum of stretching energy over all edges; (2) the sum of binding energy over bound edge pairs; and (3) the sum of bending energy over bound edge pairs: 

\begin{equation}
\label{eq: energy}
    E= \frac{1}{2} \sum_i \epsilon_{\rm s} \left(l_i - l_i^{(0)}\right)^2 + \frac{1}{2}\sum_{\langle ij \rangle} \epsilon_{\rm b} + \frac{1}{2} \sum_{\langle ij \rangle} \kappa_{\rm b} \left(\theta_{ij}-\theta_{ij}^{(0)}\right)^2
\end{equation}
We perform the simulations at constant subunit chemical potential, meaning that we follow the growth of a single assembly in contact with a bath of free monomers at fixed chemical potential of each species. Subunits can exchange with the bath at rates consistent with binding rates for orientation-specific subunit interactions. For further details on the simulation methods, including the allowed Monte Carlo moves, see SI section VI and our previous work~\cite{Tyukodi2022Jun,tyukodi2024magic, duque2024limits, videbaek2022tiling, Fang2022}.

The toroid simulation exhibits four general assembly outcomes depending on the binding energy and bending modulus: 1) under-nucleated, 2) overgrown, 3) on-target, and 4) off-target closed assemblies (Fig.~\ref{fig:5}B). First, if the binding energy $\epsilon_\text{b}$ is too weak, the assembly fails to nucleate at all. Second, if the binding energy is too strong, the assembly grows too fast, incorporating defects at a rate that renders it impossible to find the equilibrium configuration, and therefore overgrows. These two regimes are characteristic of nearly all examples of self-assembly~\cite{Hagan2021, Whitelam2015, Perlmutter2015, Zandi2020}. The third and fourth general outcomes are more revealing.

For a high enough bending modulus ($\kappa_{\rm b} \gtrsim 1  \; k_{\rm B}T$), in between the under-nucleated and overgrown states, there lies a working range of binding energies over which the assembly closes into the on-target toroid state with high yield. As shown in Fig.~\ref{fig:5}C, a typical on-target assembly pathway starts with nucleation, then tubelet closure, and finally toroid ring closure. Assuming a rough estimate of 20 $k_{\rm B}T$ for the bending modulus in experiment~\cite{saha2025modular, videbaek2025measuring}, we expect that our experimental system reaches this regime at low enough temperatures, which is supported by the presence of assembled toroids. 

Finally, and perhaps most interestingly, for a low bending modulus ($\kappa_{\rm b} \lesssim 1 \; k_{\rm B}T$), in-between the under-nucleated and overgrown states, there is a fourth type of assembly that leads to off-target, closed structures. Notably, these off-target states fully satisfy the interactions between subunits, yet exhibit the incorrect geometry because the cost for binding at off-target angles is lower for small $\kappa_{\rm b}$. These off-target `metastable' states compete with the target design in the low bending-modulus regime, and therefore limit the yield of the on-target assemblies. This observation highlights the importance of having control over the geometric specificity (i.e., having the correct binding angles and sufficiently high bending modulus) for the programmable self-assembly of a specific structure that is specified by its binding angles in addition to the interaction matrix.

Upon closer investigation, we find that there are a variety of off-target structures, and that all involve the change of a local symmetry axis. For toroids, the main symmetry axis is the one that goes through its center and sets its diameter. Instead of just forming the designed 4-fold toroid, we find that 3-fold and 5-fold toroids assemble as well (Fig.~\ref{fig:5}D,E). Since each segment of the toroid is identical, the ends of any segments are allowed to bind together, leading to defects that involve integer insertions or deletions of tubelet-junction regions, leading to a change in symmetry of the primary 4-fold axis. 

We also find defects about the 2-fold symmetric vertices. If a vertex has 2-fold symmetry, the interactions within the vertex allow a vertex with half of the number of triangles to form. For the toroid, 2-fold vertices appear in the junction region, and reduction of their symmetry can lead to the premature capping of a tubelet. This pathway is shown in Fig.~\ref{fig:5}F, where both the 6- and 8-fold vertices of the junction close with reduced symmetry, resulting in uniquely specified 3- and 4-fold vertices, respectively. Though not as obvious, there is also a local symmetry axis that goes through the axis of the tubelet region. Figure~\ref{fig:5}G shows an assembly where the tubelet region closes with twice the circumference that is intended, and is capped by the junction region. The compatibility of these, and other, off-target closed structures with the given set of interactions can be rationalized by taking a subset of the corresponding holey tiling, and imposing different stitching rules that are consistent with the interactions (Fig.~\ref{fig:5}F,G,H). We further characterize the frequency of these different failure modes in SI section IV, and we find that vertex disclinations mediate off-target assembly for most cases. In summary, we find that low geometrical specificity (i.e. bending modulus) leads to failure in closure at different scales: vertex closure, tubule closure, and toroid closure. 

We also simulate helical tubule assembly, and similar to toroidal assemblies, we observe off-target closed assemblies at low bending modulus that compete with the desired structure. For the simulation results shown in Fig.~\ref{fig:5}I, we change the interaction matrix and binding angles to target the assembly of the left-handed helical tubule from above. The phase diagram that we generate looks qualitatively similar to the toroid phase diagram: between the under-nucleated and overgrown states, on-target assembly dominates for high bending modulus, while off-target closed assemblies dominate for low bending modulus. Furthermore, the kinetic pathway for correct helical assemblies is similar to toroids, proceeding through nucleation, tubelet closure, and finally, unbounded growth (Fig.~\ref{fig:5}J). The off-target states for the helical tubule also follow the same principles as we found for toroids. Interestingly, since the left-handed helical tubule has additional vertices with 2-fold symmetry, a smaller subset of the corresponding holey tiling can self-close, leading to a minimal structure with only $\mathcal{N}_{\text{total}} = 12$ triangles (Fig.\ref{fig:5}I). As before, this subset can also repeat in the two periodic directions of the holey tiling, with different stitching rules that satisfy the interaction matrix, giving rise to a plethora of off-target structures. 

Taken together, the two examples explored here give us insight into the universal physical principles for potential failure modes in highly specific assemblies and how to avoid them: A failed closure leads to off-target assembly. The failure mechanism of all of the off-target closed structures we have seen can be classified into failed toroid closure, vertex closure, or tubule closure. Around each of these closures, we can identify the rotational symmetries, which inform possible self-closing subsets of triangles. Similar failure mechanisms specifically for vertex and tubule closure have been found elsewhere~\cite{videbaek2022tiling, Fang2022, videbaek2024economical, tyukodi2024magic}

\section*{Discussion and Conclusions}

In conclusion, we developed a design approach for creating programmable, 2-periodic self-closing assemblies by combining ideas from kirigami, symmetry, and programmable tilings. We demonstrated that these designs could be implemented in practice using DNA origami. By combining our experimental observations with KMC simulations, we further discovered that failure in correctly closing any unbound edge leads to misassemblies, inviting further research into economic ways to predict and eliminate these types of off-target defects. 

We note that similar theoretical constructions have been developed for toroids and helical tubules in the context of carbon nanotubes~\cite{dunlap1992connecting, chuang2009generalized, tamura2005positive, beuerle2011optical}. In particular, toroidal carbon nanotubes (TCNT) and coiled carbon nanotubes are molecules of trivalent carbon atoms whose ground states have the structures of a toroid or helical tubule, respectively. The key difference between TCNTs and the structures considered here is that while a specific ground-state configuration of a TCNT may exist in principle, it is impossible to specifically program the self-assembly of carbon atoms into a user-specified target TCNT geometry. In our implementation, we overcome this fundamental limitation by using supramolecular building blocks instead of atoms, for which we can precisely specify the interactions and binding geometries between DNA-origami subunits~\cite{saha2025modular}. 

Furthermore, the helical tubules developed here differ in important ways from other helical DNA-based assemblies ~\cite{maier2017self, zhang2017placing, Grome2018May, iwaki2016programmable}. First, the pitch, chirality, and size of the cavity enclosed by our helical tubules can be tuned precisely with our design scheme by specifying the interactions and binding angles between the subunits. Second, the helical assemblies reported herein are self-assembled from collections of discrete subunits and can therefore grow unbounded, in contrast to helical DNA origami, which is folded from a single scaffold and therefore does not repeat. Further, the helical tubules that we assemble are rigid, as evidenced by their near-targeted geometry even after drying on the EM grid. This observation is in contrast to thinner helical assemblies that are more flexible and whose geometries therefore cannot be measured reliably using EM~\cite{iwaki2016programmable, maier2017self}. Hence, the helical tubules assembled here are both rigid and can grow indefinitely with a consistent, programmed helicity. 

Looking forward, we speculate that kirigami-inspired approaches could be useful for the inverse design of other curved surfaces~\cite{Chuang2011Jan}. For example, while this approach yields a wide variety of structures, future designs could consider how other shaped holes---not necessarily 2-fold symmetric holes---could be stitched and folded to create new geometries. We explore some variations in SI section V, where other hole shapes give rise to different toroids with unique symmetries. Additionally, we speculate that by applying kirigami to Frieze patterns, instead of 2D tilings, we can design helicoids---a related class of surfaces which are 1-periodic. While the approach we consider here---cutting-and-stitching 2D tilings---limits the design space to geometries with net-zero Gaussian curvature, going forward, this limitation can be circumvented by considering embeddings of hyperbolic tilings ~\cite{conway2016symmetries,duque2024limits}, or by taking only a subset of the tiling~\cite{sigl_programmable_2021}, in which the curvature is embedded along the boundary of the net. These extensions of our approach could open new vistas in the construction of functional, curved-crystalline nanomaterials by self-assembly.

\begin{acknowledgments}
TEM images were prepared and imaged at the Brandeis Electron Microscopy facility. This work is supported by the Brandeis University Materials Research Science and Engineering Center, which is funded by the National Science Foundation (NSF) under award number DMR-2011846. MP acknowledges the Jerome A. Schiff Undergraduate Fellowship. BT acknowledges support from the Romanian Ministry of Research, Innovation, and Digitization under PNRR-I8/C9-CF105 - contract no. 760099 and CNCS - UEFISCDI, project number PN-IV-P2-2.1-TE-2023-0558 within PNCDI IV. WBR acknowledges support from the NSF under DMR-2214590. MFH acknowledges support from DMR-2309635. Computing resources were provided by the NSF ACCESS allocation TG-MCB090163 and the Brandeis HPCC which is partially supported by the NSF through DMR-MRSEC 2011846.
\end{acknowledgments}

\bibliography{main}

\end{document}


\title{Supplemental Information for ``From toroids to helical tubules: Kirigami-inspired programmable assembly of two-periodic curved crystals"}
\author{Mason Price, Daichi Hayakawa, Thomas E. Videb\ae k, Rupam Saha, Botond Tyukodi, Michael F. Hagan, Seth Fraden, Gregory M. Grason, and W. Benjamin Rogers}

\maketitle

\section{Experimental methods}\label{sec:methods}

\subsection{Folding DNA origami}\label{subsec:folding}
Each DNA origami particle is folded by mixing 50~nM of p8064 scaffold DNA (Tilibit) and 200~nM each of staple strands with folding buffer and annealed through a temperature ramp starting at 65~$^{\circ}$C for 15 minutes, then 58 to 50~$^{\circ}$C, $-1~^{\circ}$C per hour. Our folding buffer, contains 5~mM Tris Base, 1~mM EDTA, 5~mM NaCl, and 15~mM MgCl$_2$. We use a Tetrad (Bio-Rad) thermocycler for annealing the solutions. 

\subsection{Agarose gel electrophoresis}\label{subsec:electrophoresis}
To assess the outcome of folding, we separate the folding mixture using agarose gel electrophoresis. Gel electrophoresis requires the preparation of the gel and the buffer. The gel is prepared by heating a solution of 1.5\% w/w agarose, 0.5x TBE to boiling in a microwave. The solution is cooled to 60~$^{\circ}$C. At this point, we add MgCl$_2$ solution and SYBR-safe (Invitrogen) to adjust the concentration of the gel to 5.5~mM MgCl$_2$ and 0.5x SYBR-safe. The solution is then quickly cast into an Owl B2 gel cast, and further cooled to room temperature. The buffer solution contains 0.5x TBE and 5.5~mM MgCl$_2$, and is chilled to 4~$^{\circ}$C before use. Agarose gel electrophoresis is performed at 110 V for 1.5 to 2 hours in a cold room kept at 4~$^{\circ}$C. The gel is then scanned with a Typhoon FLA 9500 laser scanner (GE Healthcare).

\subsection{Gel purification and resuspension}\label{subsec:purification}
After folding, DNA origami particles are purified to remove all excess staples and misfolded aggregates using gel purification. The folded particles are run through an agarose gel (now at a 1xSYBR-safe concentration for visualization) using a custom gel comb, which can hold around 4~mL of solution per gel. We use a blue fluorescent table to identify the gel band containing the monomers. The monomer band is then extracted using a razor blade, which is further crushed into smaller pieces by passing through a syringe. We place the gel pieces into a Freeze `N' Squeeze spin column (Bio-Rad), freeze it in a -80~$^\circ$C freezer for 30 minutes, thaw at room temperature, and then spin the solution down for 5 minutes at 13 krcf. 

Since the concentration of particles obtained after gel purification is typically not high enough for assembly, we concentrate the solution through ultrafiltration~\cite{wagenbauer_how_2017}. First, a 0.5~mL Amicon 100kDA ultrafiltration spin column is equilibrated by centrifuging down 0.5~mL of the folding buffer at 5~krcf for 7 minutes. Then, the DNA origami solution is added up to 0.5~mL and centrifuged at 14~krcf for 15 minutes. Finally, we flip the filter upside down into a new Amicon tube and spin down the solution at 1~krcf for 2 minutes. The concentration of the DNA origami particles is measured using a Nanodrop (Thermofisher), assuming that the solution consists only of monomers, where each monomer has 8064 base pairs.

\subsection{Assembly}\label{subsec:assembly}
All assembly experiments are conducted at a total DNA origami particle concentration of 30~nM. All triangular species are mixed together in an equal ratio, with 2.5~nM per species, because they have equal stoichiometry in the target assemblies and 12 unique species for each structure considered here. By mixing the concentrated DNA origami solution after purification with buffer solution, we make 24~$\upmu$L of 30~nM DNA origami at 20~mM MgCl$_2$. The solution is carefully pipetted into 0.2~mL strip tubes (Thermo Scientific) and annealed through different temperature protocols using a Tetrad (Bio-Rad) thermocycler.

\subsection{Negative stain TEM}\label{subsec:TEM}
We first prepare a solution of uranyl formate (UFo). ddH$_2$O is boiled to deoxygenate it and then mixed with uranyl formate powder to create a 2\% w/w UFo solution. The solution is covered with aluminum foil to avoid light exposure, then vortexed vigorously for 20 minutes. The solution is filtered using a 0.2~$\upmu$m filter. The solution is divided into 0.2~mL aliquots, which are stored in a --80~$^\circ$C freezer until further use.

Prior to each negative-stain TEM experiment, a 0.2~mL aliquot is taken out from the freezer to thaw at room temperature. We add 4~$\upmu$L of 1~M NaOH and vortex the solution vigorously for 15 seconds. The solution is centrifuged at 4~$^\circ$C and 16~krcf for 8 minutes. We extract 170~$\upmu$L of the supernatant for staining and discard the rest. 

The EM samples are prepared using FCF400-Cu grids (Electron Microscopy Sciences). We glow discharge the grid prior to use at --20~mA for 30 seconds at 0.1~mbar, using a Quorum Emitech K100X glow discharger. We place 4~$\upmu$L of the sample on the grid for 1 minute to allow adsorption of the sample to the grid. During this time 5~$\upmu$L and 18~$\upmu$L droplets of UFo solution are placed on a piece of parafilm. After the adsorption period, the remaining sample solution is blotted on a Whatman filter paper. We then touch the carbon side of the grid to the 5~$\upmu$L drop and blot it away immediately to wash away any buffer solution from the grid. This step is followed by picking up the 18~$\upmu$L UFo drop onto the carbon side of the grid and letting it rest for 30 seconds to deposit the stain. The UFo solution is then blotted to remove excess fluid. Grids are allowed to dry for a minimum of 15 minutes before insertion into the TEM.

We image the grids using an FEI Morgagni TEM operated at~80 kV with a Nanosprint5 CMOS camera (AMT). Images are acquired between x8,000 to x28,000 magnification. The images are high-pass filtered and the contrast is adjusted using Fiji~\cite{schindelin2012fiji}. 

\subsection{TEM tomography}

To obtain a tilt-series, we use an FEI F20 equipped with a Gatan Ultrascane 4kx4k CCD camera, operated at 200 kV. The grid is observed at x18000 magnification from -50 degrees to 50 degrees in 2-degree increments. The data is analyzed and the z-stack is reconstructed using IMOD~\cite{kremer_computer_1996}.

\clearpage

\section{Designing toroids and helical tubules with geometric parameters $(T,L,D,R)$} \label{sec:tldr}

\subsection{Generating 3D geometries from holey tilings}\label{subsec:stitching}

\begin{figure*}[htb]
 \centering
 \includegraphics[width=0.99\linewidth]{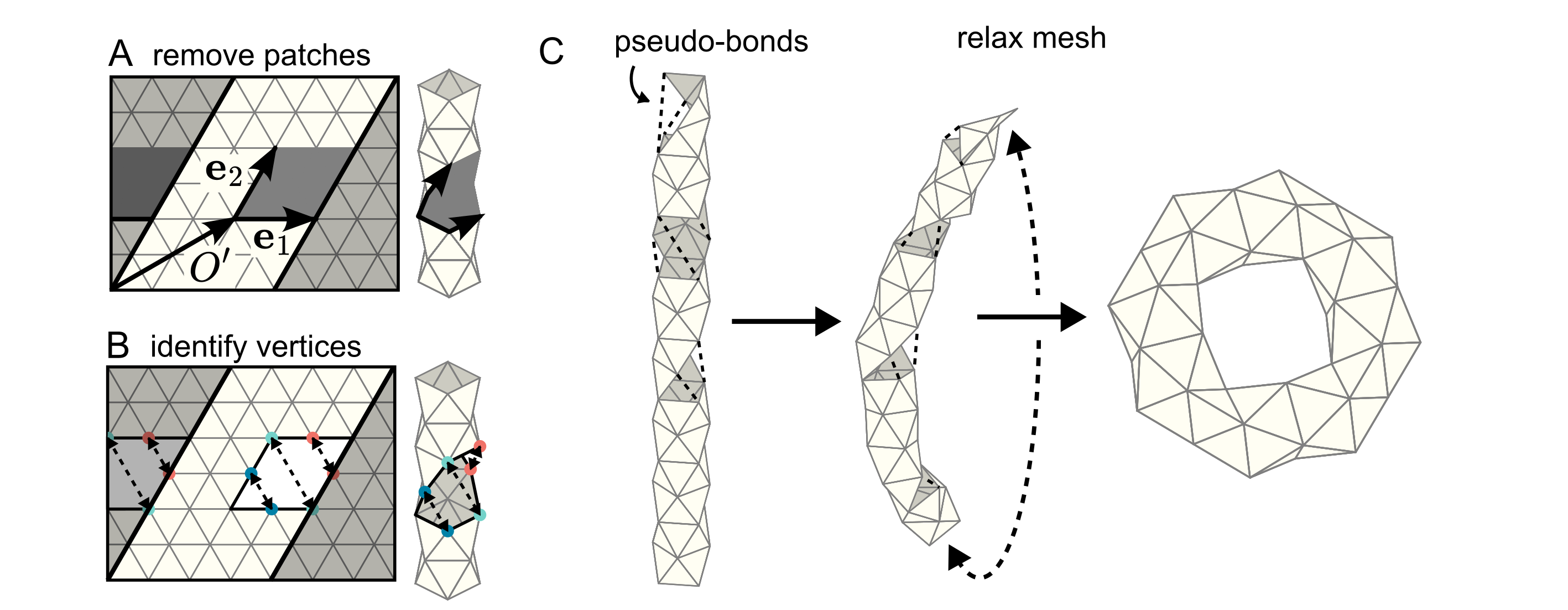}
 \caption{\textbf{Energy minimization scheme.} (A) First, we remove parameter-specified patches in the lattice of a tubule. (B) Second, we identify vertices across the excised patches to stitch together. (C) We implement stitching rules numerically by applying pseudo-bonds between identified vertices with a rest length of zero. Next, we merge identified vertex pairs into a single vertex to close the holes. For toroids we also merge the top and bottom boundary components. Finally, we allow the mesh to relax.  
}
 \label{SIfig:construction}
\end{figure*}
Here, we define our approach to generate the geometry for $(T,L,D,R)$ toroids and helical tubules (Fig.~\ref{SIfig:construction}). We begin this construction by generating a triangular lattice for a $(T,0)$ achiral tubule. Second, we remove the parallelogram patch whose corner is at $ O' := (D,L)$ and has edge-vectors given by $\textbf{e}_1 = (T-D,0)$ and $\textbf{e}_2 = (0, D)$. Then, we remove all equivalent patches up to the translation vectors $(h_1,k_1)=(T,0)$ and $(h_2, k_2)=(R,L+D)$, where the $h$-coordinates are taken $\text{mod } T$. Next, we map the remaining triangles onto a cylinder to obtain a tubule with excised patches (reference~\cite{Hayakawa2022Oct} for tubule constructions). The radius of the cylinder is 
$$ r = \frac{1}{2 \sin \left( \frac{\pi}{T} \right) },$$
and the number of triangles along its height is 
$$ H = N(L+D)$$
where $N$ is the number of tubelets. 

We map the vertices we want to ``stitch'' together using the following matching rules. For $i \in \{1, \dots, d\}$, let
$$ O' + (0,i) \mapsto O' + (i,0),$$
and 
$$  O' + \textbf{e}_1 + \textbf{e}_2 - (0, i) \mapsto O' + \textbf{e}_1 + \textbf{e}_2 - (i,0). $$
If $T > 2D+1$, then we need to stich vertices between tubelets as well, so for $i \in \{1, \dots, T-2D\}$, let 
$$ O' + (D+i, 0) \mapsto O' + (i, D).$$
This is how we identify associated vertices. 

Now, to transform the tubule into a toroid or helical tubule, we treat the surface like an elastic mesh and minimize the energy. The coordinates of the vertices are updated so that each edge is treated like an overdamped spring with a rest length of 1 (so that the subunit edge-length $l_0$ is normalized), and associated vertices share a pseudo-bond with a rest length of zero. The energy minimization script then proceeds in three stages. First, we apply the stitching pseudo-bonds to bring the tubule geometry near that of a toroid or helical tubule, while forcing the tubelet regions to retain a circular circumference and pushing outward on the vertices in the junction regions to prevent buckling (using additional pseudo-bonds). Next, we merge associated pairs of vertices, and for the toroid we merge the top- and bottom-boundary components (based on the periodic boundary conditions along the length of the tubule). Finally, we stop enforcing all additional pseudo-bonds and allow the mesh to relax into its final 3-dimensional configuration.

Once the mesh is relaxed, we measure the binding angle $\beta_{i,j}$ between bound triangles $i$ and $j$ as
$$ \beta_{i,j} = \text{atan2}\left(\| \hat{n}_i \times \hat{n}_j\|, \langle \hat{n}_i,\hat{n}_j \rangle \right), $$
where $\hat{n}_i$ and $\hat{n}_j$ are the oriented unit normal vectors of triangles $i$ and $j$, respectively. 

\subsection{Inverse design of toroids and helical tubules} \label{subsec:holeyTilings}

\begin{figure}[ht]
\includegraphics[width=0.9\linewidth]{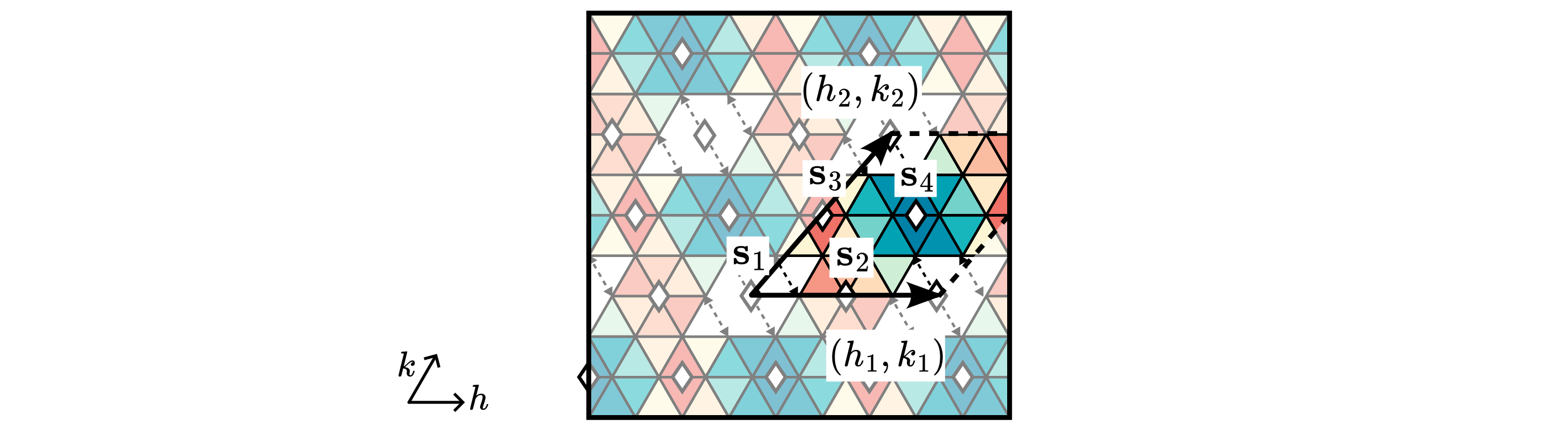}
\caption{\textbf{Example holey tiling}}
\label{SIfig:PU}
\end{figure}

To tile the surface of toroids and helical tubules, we exploit the symmetry of 2D tilings. Once cut open, both toroids and helical tubules can be converted into periodic 2D tilings, with the vectors of the primitive unit (PU) cell  given by 
$$(h_1, k_1) = (T,0)$$
and
$$(h_2, k_2)=(R, L+D).$$
Previously, we developed a symmetry-guided inverse design strategy for programming self-assembly systems~\cite{Hayakawa2024}. In short, the species and the orientations of particles in a symmetric assembly structure can be assigned algorithmically by exploiting its symmetry. The symmetry of any 2D periodic tiling can be classified into one of the 17 Wallpaper groups. Due to the restriction of particles from flipping out of plane, and the lack of right angles in a triangular tilings, the periodic triangular tilings can only exhibit o, 2222, 333, or 632 symmetries (in Orbifold notation~\cite{conway2016symmetries}). Here, we are additionally constrained by the folded configuration; through cutting and stitching, our triangular tilings should be convertible to (holey) tubules. The 3D geometry of the tubule allows for 2-fold rotational symmetry, which corresponds to flipping of the tubules. However, 3- and 6-fold rotational symmetries are not allowed. The geometrical constraint of the tubules manifests in the 2D tilings as well; since the only symmetry operation allowed is the 2-fold rotational symmetry, the only allowed symmetries in the tilings are o and 2222. In the following, we consider tilings with 2222 symmetry, due to their higher economy. For details, also refer~\cite{videbaek2024economical}.

Then, the 2-fold rotational symmetry points have locations given by
\begin{align*}
& \textbf{s}_1= \left( 0,0 \right), \\
& \textbf{s}_2=  \left( \frac{T}{2}, 0 \right), \\
& \textbf{s}_3=\left( \frac{R}{2}, \frac{L+D}{2} \right), \\
& \textbf{s}_4=\left(\frac{R+T}{2}, \frac{L+D}{2} \right).
\end{align*}
See the example in Fig.~\ref{SIfig:PU}. Here, we require that one of the 2-fold rotational symmetry points reside in the hole, and specifically, in the middle of the hole. Further, this constrains the hole to be 2-fold rotationally symmetric. Although this specific location of the 2-fold rotational symmetry points is required for our construction of the toroids, this is not necessarily the requirement of the holey 2D tiling itself.

\begin{figure}[htb]
\includegraphics[width=0.9\linewidth]{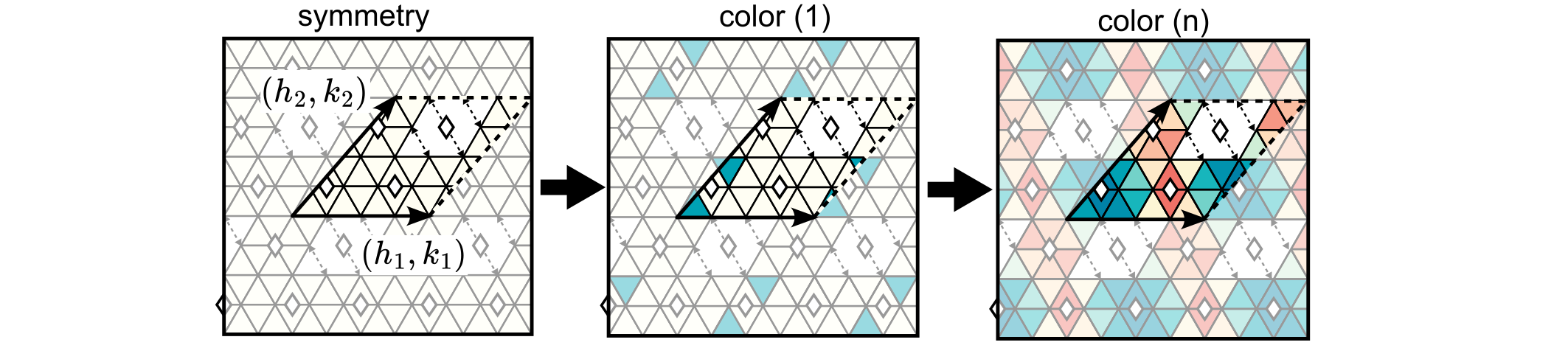}
\caption{\textbf{Coloring triangles by symmetry}}
\label{SIfig:coloring}
\end{figure}

Once we know the symmetry points, we can identify the different species. We start by coloring a single subunit. Then we trace out all equivalent triangles in the PU cell by applying a symmetry transformation and coloring the corresponding image of the subunit (Fig.~\ref{SIfig:coloring}). We repeat this until we have exhausted all subunits (Fig.~\ref{SIfig:coloring}). The minimal number of unique species for a given $(T,L,D,R)$ is then
\begin{equation} n = TL + D^2.
\end{equation}
For more in depth procedure, refer ref.~\cite{Hayakawa2024}.

\subsection{Mirror symmetry for identifying the geometry of toroids} \label{subsec:mirror}

\begin{figure}[ht]
\includegraphics[width=0.9\linewidth]{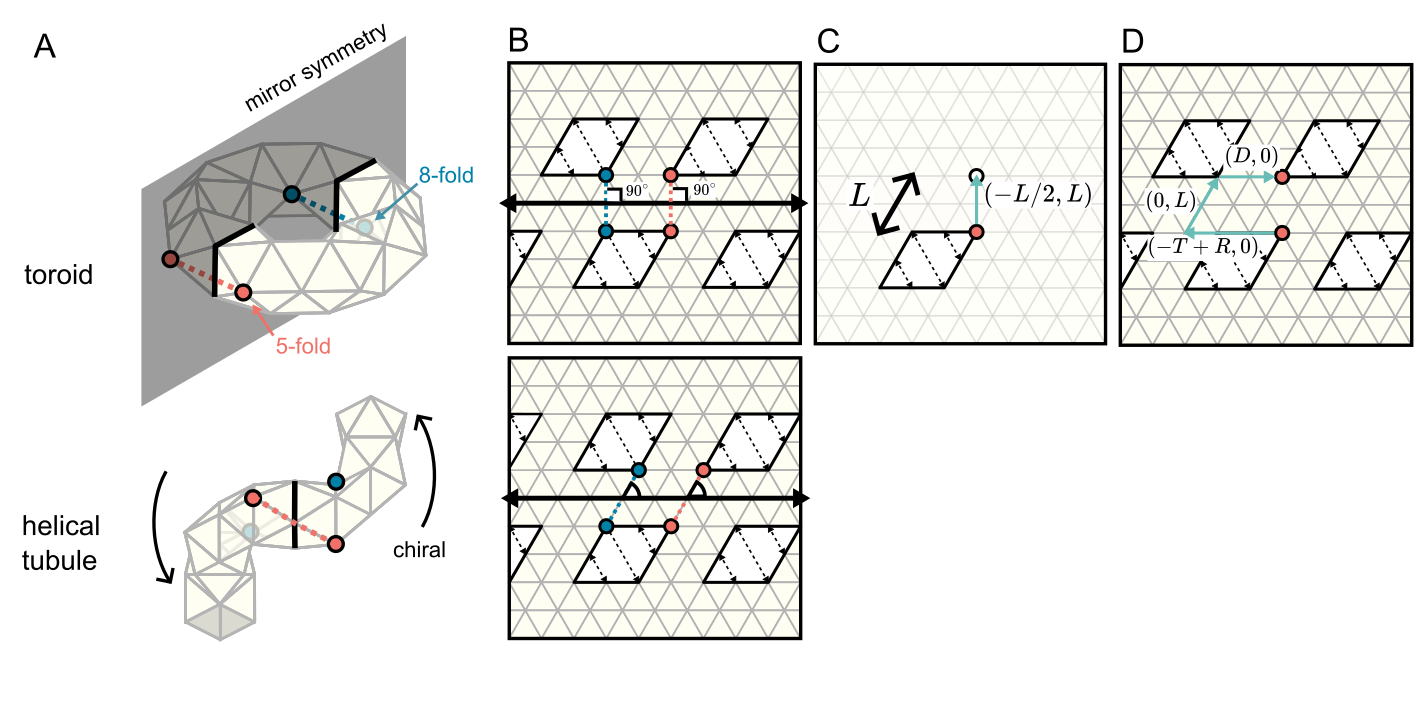}
\caption{\textbf{Mirror symmetry of a toroid.} (A) The $(T,L,D,R)$ toroids are mirror-symmetric, while helical tubules are chiral. (B) For toroids, the lines connecting similar defects ($5\to5$ or $8\to8$) cross the horizontal $(1,0)$ vector at a $90^{\circ}$ angle, while for helical tubules they cross at non-right angles. (C) The point directly above a given 5-fold vertex is given by an offset vector $(-L/2, L)$. (D) Consecutive 5-folds are found by walking along the tubelet and junction regions.}
\label{SIfig:mirrorSym}
\end{figure}

To understand when $(T,L,D,R)$ will form a toroid or else a helical tubule, we exploit the mirror symmetry of a toroid (Fig.~\ref{SIfig:mirrorSym}). For a toroid, we note that 5-fold and 8/7-fold vertices lie on opposite sides of a mirror plane slicing through the circumference of the tubelet segments in the 3D folded geometry. In contrast, helical tubules are chiral and hence not mirror symmetric across any plane (Fig.~\ref{SIfig:mirrorSym}A). Then, the corresponding line connecting two 5-folds or 8-folds will cross the circumference (horizontal) line at a $90^\circ$ angle for a toroid in the 2D holey tiling, but not for a helical tubule (Fig.~\ref{SIfig:mirrorSym}B). Now, to derive a formula to express this idea, let $v_5$ denote a 5-fold vertex. Then, if $L$ is even, the point lying vertically above $v_5$ along a right angle to the circumference direction will fall on a vertex of the lattice (Fig.~\ref{SIfig:mirrorSym}C) and is given by 
\begin{equation}\label{v5above}
    v_5 + \left( -\frac{L}{2}, L\right).
\end{equation}
On the other hand, the next 5-fold vertex is determined by the parametrization (Fig.~\ref{SIfig:mirrorSym}D) and is given by 
\begin{equation}\label{v5next}
    v_5+(D+R-T, L)
\end{equation}
So, to find the right value of $R$ for a toroid, we set (\ref{v5above}) and (\ref{v5next}) equal and find 
$$ R = T-D-L/2$$
Finally, since $R$ is periodic by an offset of $T$, we find the value of $R$ that corresponds to a toroid is given by
$$ R_{\text{toroid}} \equiv -D -L/2 \pmod{T}. $$

The appearance of mirror symmetry from a 3D geometry in a 2D holey tiling by considering the location of disclinations raise an interesting question: can we engineer other 3D symmetries using the kirigami method presented here? If so, it would be interesting to learn what parameters specify different symmetries.

\subsection{Design library using the $(T,L,D,R)$ scheme} \label{subsec:library}

To summarize, the $(T,L,D,R)$ scheme represents a class of toroids,  helical and serpentine tubules made from sequences of tubulets and junctions connected together. For this scheme, the two periodic vectors defining the placement of holes are given by $(h_1, k_1)=(T,0)$ and $(h_2, k_2) = (R, L+D)$, meaning the first vector lies along the $(1,0)$ lattice direction, while the second vector can be rastered along the $(1,0)$ lattice direction by varying $R$. Importantly, the 2-fold rotational symmetry points are aligned with the center of the holes in the tiling. For simplicity, we further impose that the junctions are diamond shaped, which allows for any parallelogram shaped holes (not necessarily diamond) such that the sides are aligned with the lattice directions. Then, we take a vertical strip of the holey tiling of width $T$ and wrap it into a tubule. Finally, we impose the stitching rules, and for simplicity we stitch the holes closed across the shorter diagonal of the holes. Further, for the $(T,L,D,R)$ structure to be a toroid, it has to satisfy additional conditions: $L$ must be even, and $R$ must be chosen so that $R \equiv -D-L/2\pmod{T}$.
  
Here we showcase a collection of toroids and helical tubules generated using the $(T,L,D,R)$ scheme. We measure all geometric parameters based on the following subsection for surface fitting. We calculate the corresponding values of pitch angle $\phi$ as 
$$ \phi_H = \arctan\left(\frac{p}{2\pi R}\right)$$
for helical tubules, and as 
$$ \phi_S = \arctan\left(\frac{p}{4 R}\right)$$
for serpentine tubules. 

\begin{figure*}[htb]
 \centering
 \includegraphics[width=0.99\linewidth]{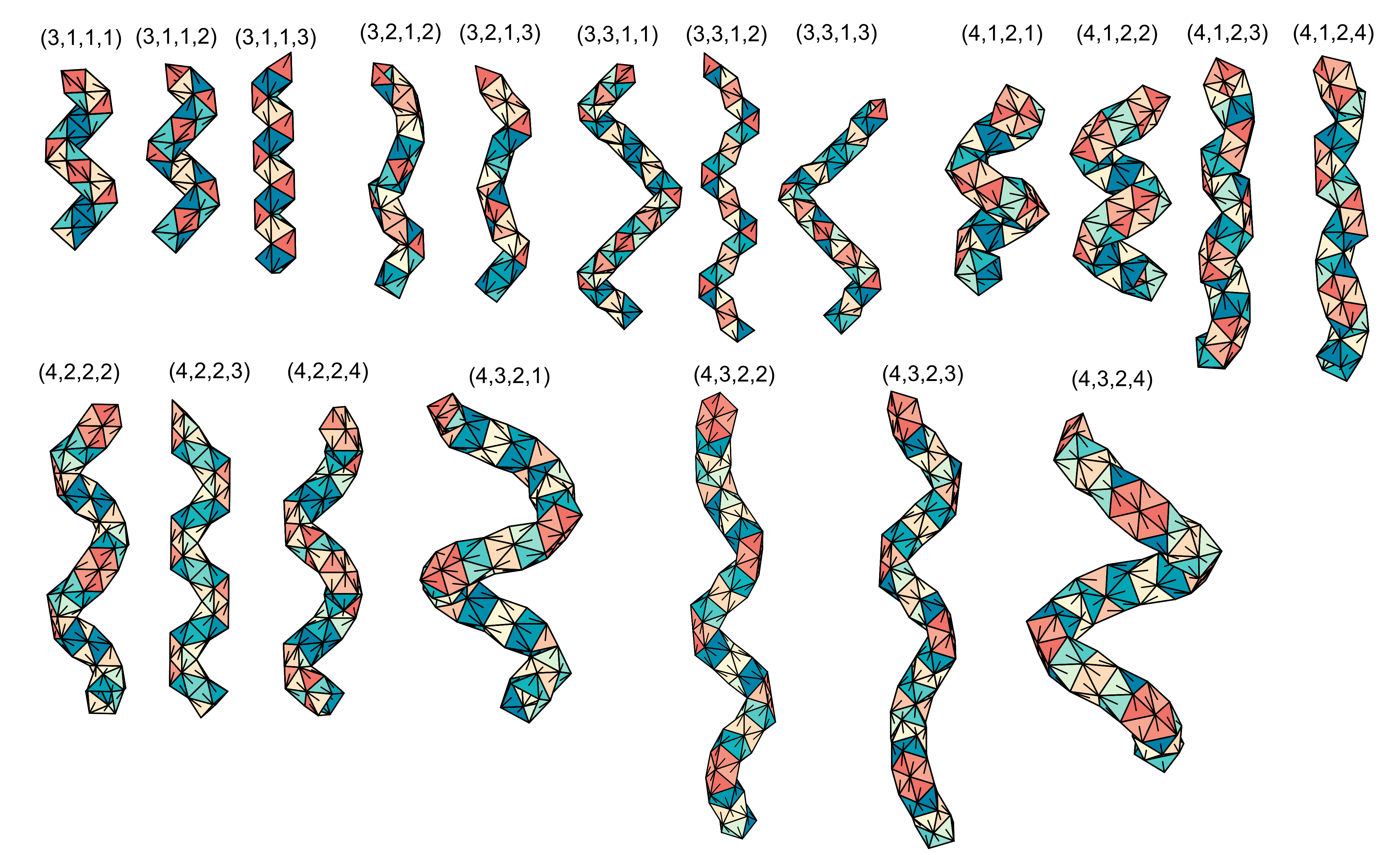}
 \caption{\textbf{Examples of different $(T,L,D,R)$ helical/serpentine tubules. }
}
 \label{}
\end{figure*}

\begin{longtable}{p{0.12\linewidth} p{0.17\linewidth} p{0.12\linewidth} p{0.13\linewidth}p{0.12\linewidth}p{0.13\linewidth}p{0.07\linewidth}}
\caption{\textbf{Helical tubule geometry data.} All distances are normalized with respect to the triangle edge length. } \label{tab:HTGeometry} \\
\input{Tab-geometry_HT}
\end{longtable}

\clearpage

\begin{figure*}[htb]
 \centering
 \includegraphics[width=0.99\linewidth]{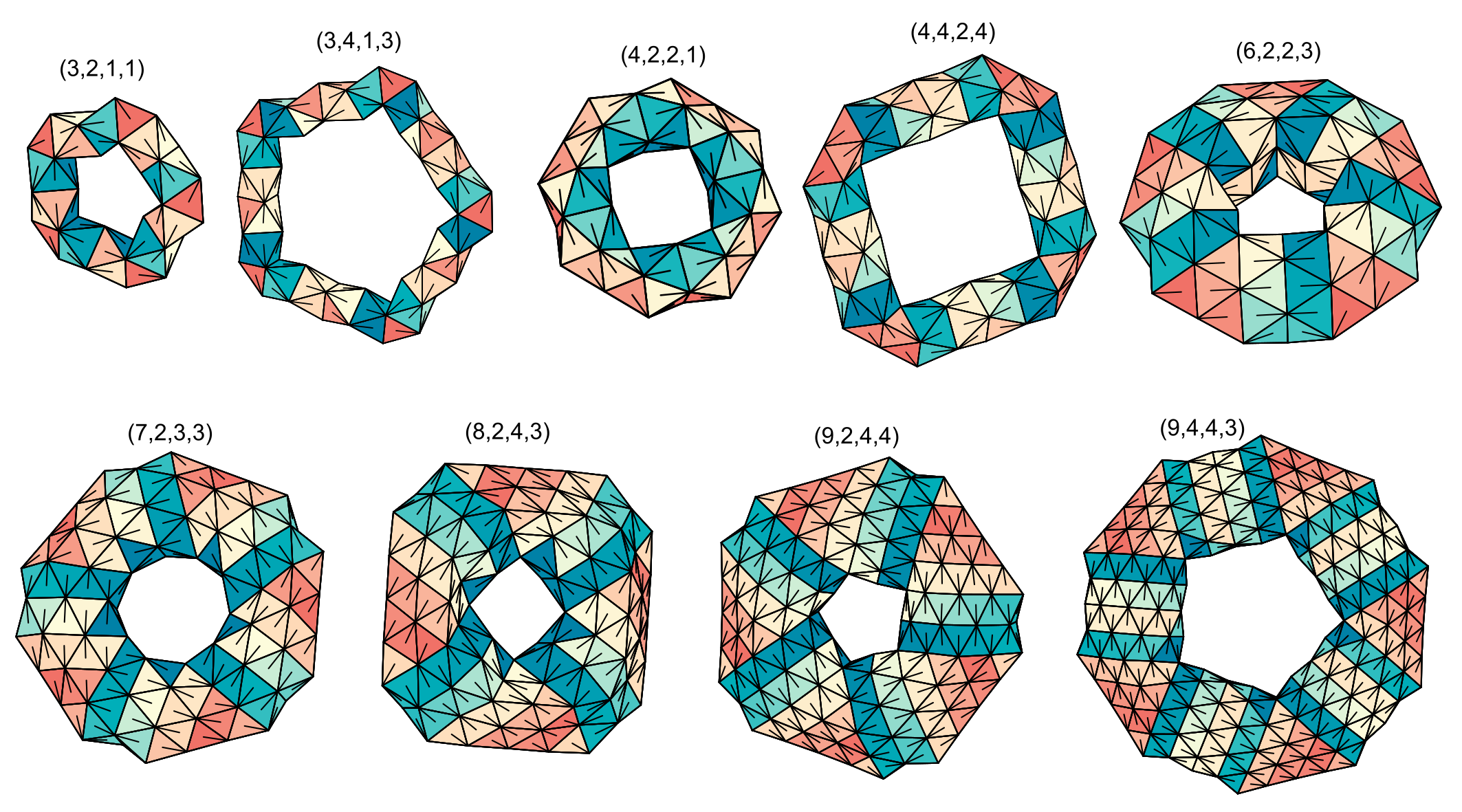}
 \caption{\textbf{Examples of different $(T,L,D,R)$ toroids. }
}
 \label{}
\end{figure*}

\begin{longtable}{p{0.12\linewidth} p{0.17\linewidth} p{0.10\linewidth} p{0.15\linewidth}p{0.15\linewidth}p{0.15\linewidth} }
\caption{\textbf{Toroid geometry data.} All distances are normalized with respect to the triangle edge length. If a toroid is commensurate, then there is no edge stretching in the final closed configuration.} \label{tab:toroidGeometry} \\
\input{Tab-geometry_toroids}
\end{longtable}

\subsection{Segmented helix model} \label{subsec:seghelix}

\begin{figure*}[htb]
 \centering
 \includegraphics[width=0.99\linewidth]{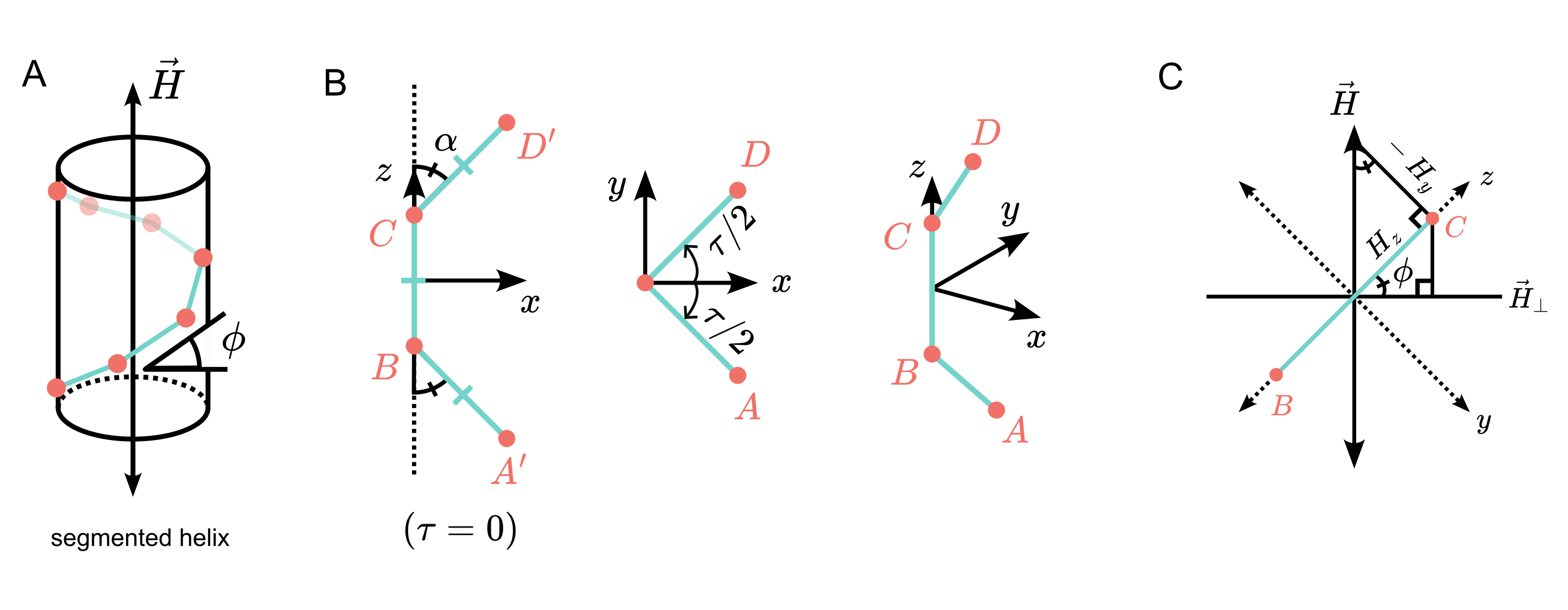}
 \caption{\textbf{The segmented helix approximation.} (A) A segmented helix includes discretely spaced vertices along a helix, connected by line segments of equal length. (B) Projections of the vertices onto the $xz$- and $xy$-planes reveal the coordinates of the vertices based on the angles $\alpha$ and $\tau$. (C) The signed projection of the helical axis onto the $y$- and $z$-axes yields the pitch angle $\phi$ as a function of $\alpha$ and $\tau$. 
}
 \label{SIfig:geometryTwist}
\end{figure*}

To understand the relationship between the input parameters and the resulting geometry of helical tubules, we adapt a model for segmented helices from ref.~\cite{read2022calculating}, focusing on the question of the helix parameter $\phi$ defining the pitch angle. We approximate the tubelets as rigid segments, connected together at two angles: the coplanar angle $\alpha$ and a twist out-of-plane $\tau$ (Fig.~\ref{SIfig:geometryTwist}). Importantly, each segment has the same length $L$ and adjacent segments have the same angle $\alpha$. We start by considering three connected segments arranged symmetrically about the $z$-axis, with end-points $A,B,C,D$. Supposing initially that $\tau=0$, the angle $\alpha$ determines the orientation between segments in the $xz$-plane (Fig.~\ref{SIfig:geometryTwist}B). Then, twisting the segments $AB$ and $CD$ about the $z$-axis symmetrically as in Fig.~\ref{SIfig:geometryTwist}B, the endpoints of the segments are given by
\begin{align*}
    & A = L\begin{pmatrix} \cos(\tau/2)\sin\alpha \\ \sin(\tau/2)\sin\alpha \\ -1/2-\cos\alpha \end{pmatrix}, \\
   & B = L\begin{pmatrix}
       0\\ 0 \\ -1/2
   \end{pmatrix}, \\
   & C = L\begin{pmatrix}
       0\\ 0 \\ 1/2 
   \end{pmatrix},  \\
   & D = L\begin{pmatrix} \cos(\tau/2)\sin\alpha \\ -\sin(\tau/2)\sin\alpha \\ 1/2+\cos\alpha \end{pmatrix}.
\end{align*}

Next, the angle bisectors $\vec{B}_b$ and $\vec{C}_b$ between $\angle ABC$ and $\angle BCD$ are given by 
\begin{align*}
    & \vec{B}_b = \frac{A+C}{2} - B = \frac{L}{2}\begin{pmatrix}
        \cos(\tau/2)\sin\alpha \\ \sin(\tau/2)\sin\alpha \\
        1-\cos\alpha
    \end{pmatrix}, \\
    & \vec{C}_b = \frac{B+D}{2} - C = \frac{L}{2}\begin{pmatrix}
        \cos(\tau/2)\sin\alpha \\ -\sin(\tau/2)\sin\alpha \\
        -1+\cos\alpha
    \end{pmatrix}.
\end{align*}

So, we can find the helical axis $\vec{H}$ by taking the cross-product of the angle bisectors, which are mutually orthogonal with the helical axis (except in the special cases of the toroid and serpentine tubule), yielding
$$ \vec{H} = \vec{B}_b\times\vec{C}_b = \frac{L^2}{2}\begin{pmatrix}
    0 \\ \cos(\tau/2)\sin\alpha \cdot(1-\cos\alpha) \\ \sin(\tau/2)\cos(\tau/2)\sin^2\alpha
\end{pmatrix}.$$
Note that the $x$-component of $\vec{H}$ is always zero. So to find the helical pitch angle $\phi$, we can do a signed projection of the helix axis $\vec{H}$ onto the $z$- and $y$-axes (Fig~\ref{SIfig:geometryTwist}C). This yields the following formula for the pitch angle $\phi$ in terms of $\tau$ and $\alpha$
\begin{equation}
\phi = \arctan \left( \frac{\sin(\tau/2) \sin \alpha}{1-\cos \alpha} \right)
\end{equation}

This completes the derivation for the approximation used in the `Systematic modulation of 3D geometries from 2D holey tilings' subsection of the main text. 

\clearpage

\section{Surface fitting} \label{sec:surffitting}

To measure the geometric parameters of a toroidal or helical polyhedron generated using our approach, such as the radius or pitch, we fit parametrized surfaces to the given structure. Consider a surface parametrized by the function 
$$ f : U \to \mathbb{R}^3 $$
where $U \subseteq \mathbb{R}^2$. For example, see Table~\ref{tab:surfaceParms}. Then, the goal is to fit this parametrized surface to a given set of vertices $V$. For this, we need to optimize the parameters defining the surface. For a torus, define the parameter tuple
$$\textbf{p}_T := (r,R,x_0, y_0,z_0, \theta_0, \alpha_0, \gamma_0), $$
where $r$ and $R$ are the minor- and major-radii of the torus (the geometry-defining parameters), and $x_0, y_0,z_0$ and $\theta_0, \alpha_0, \gamma_0$ represent an offset translation and rotation, respectively (the trivial parameters). Likewise, for a helical tubule define 
$$\textbf{p}_H := (r,R, p, u_{\text{min}}, u_{\text{max}}, x_0, y_0,z_0, \theta_0, \alpha_0, \gamma_0), $$
where $r$ is the tubelet radius, $R$ is the helix radius, $p$ is the pitch, and $u_{\text{min}}$ and $u_{\text{max}}$ are the minimum and maximum heights of the surface.
For a surface with a parameter tuple $\textbf{p}$, we then define the following transformed version of the surface.
\begin{align*}
f_{\textbf{p}} : & \; U \to \mathbb{R}^3 \\
& (u,v) \mapsto \textbf{R}_z(\gamma_0)\textbf{R}_y(\alpha_0)\textbf{R}_x(\theta_0)f(u,v) + (x_0, y_0, z_0)
\end{align*}
where $U \subseteq \mathbb{R}^2$ is the domain of the surface, $f$ is the standard parametrization, and $\textbf{R}_x$, $\textbf{R}_y$, $\textbf{R}_z$ are rotation matrices about the $x$-, $y$- and $z$-axes, respectively.

\begin{table}[ht]
    \centering
    \begin{tabular}{c|c|c}
         & Torus & Helical tube \\
        \hline

        Parametrization $f(u,v)$ & $\begin{pmatrix} ( R + r \cos u) \cos v \\
(R+r\cos u) \sin v \\
r \sin u \end{pmatrix}$ & $\begin{pmatrix} (R-r\cos(v))\sin(u) - pr \frac{\cos(u)\sin(v)}{\sqrt{R^2 + p^2}} \\
(R - r\cos(v))\cos(u) + pr \frac{\sin(u)\sin(v)}{\sqrt{R^2+p^2}} \\
pu + Rr\frac{\sin(v)}{\sqrt{R^2+p^2}} \end{pmatrix}$ \\
         Domain $U$ & $[0, 2\pi] \times [0, 2\pi]$ & $[u_{\text{min}}, u_{\text{max}}] \times [0, 2\pi]$     
    \end{tabular}
    \caption{\textbf{Surface parametrizations}}
    \label{tab:surfaceParms}
\end{table}

To characterize how closely the parametrized surface approximates the set of vertices $V$ for given parameters $\textbf{p}$, we define the objective function by 
\begin{equation}\label{eq:lossFunc}
\mathcal{L}(f_{\textbf{p}}) := \sum_{\textbf{v}\in V} \min\left(\| \textbf{v} - f_{\textbf{p}}(u,v) \|^2 \; : \; (u,v) \in U \right),
\end{equation}
to represent the sum of distances of vertices from the surface. We minimize this function over the parameter tuple $\textbf{p}$ using scipy.optimize~\cite{2020SciPy-NMeth}.

\begin{figure*}[htb]
 \centering
 \includegraphics[width=0.99\linewidth]{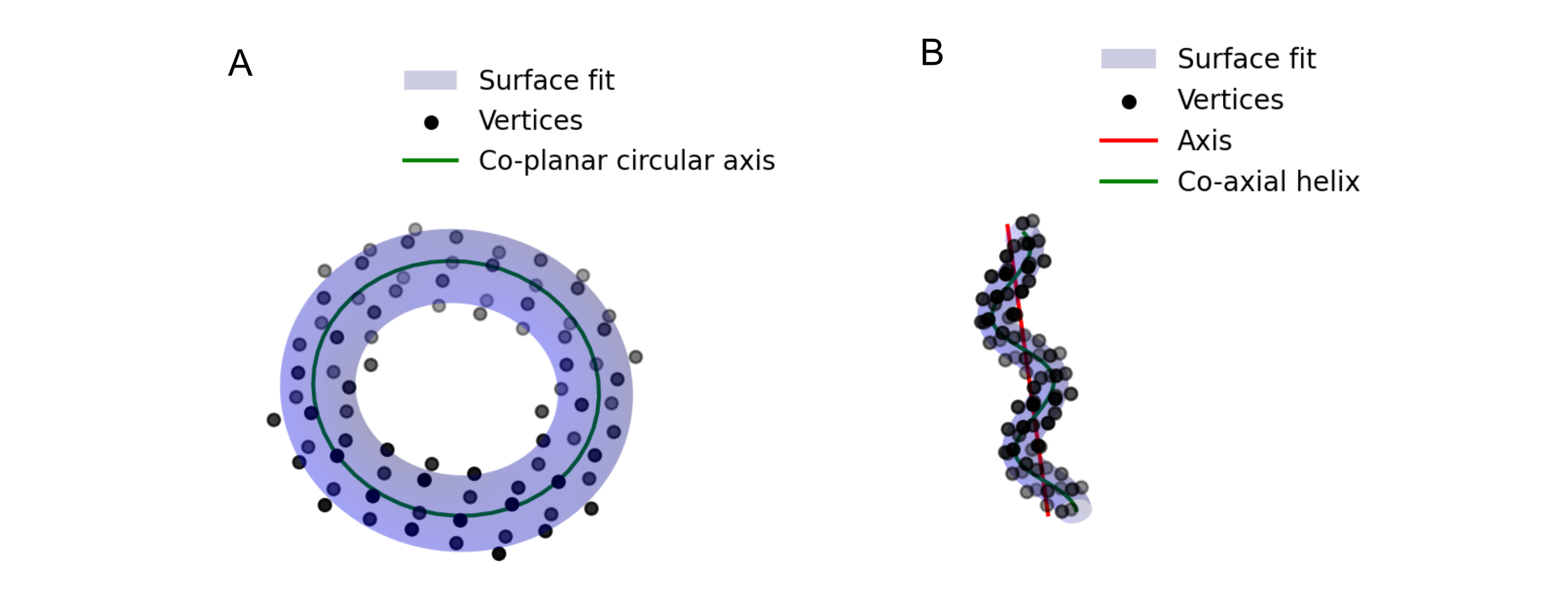}
 \caption{\textbf{Surface fits.} (A) Toroid $(4,4,2,0)$ with $r/l_0 = 0.71$ and $R/l_0 = 2.64$. (B) Helical tubule $(4,2,2,2)$ with $r/l_0 = 0.69$, $R/l_0 = 0.87$ and $p/l_0 = 5.42$.
}
 \label{SIfig:torusFit}
\end{figure*}

Minimizing this function over possible parameters is, in general, a challenging optimization problem and in practice there are many nearby local minima. To simplify this process, we break the problem into two easier steps. First, we fit suitable curve(s) such as the co-planar circular axis for a toroid, or the co-axial helix for a helical tubule. Then, essentially all that's left to find is the tube radius $r$, so we extract ideal initial parameters from the curve, and fit the surface to the set of vertices  (Fig.~\ref{SIfig:torusFit}).

\clearpage

\section{Further analysis of simulation results} \label{subsec:disclinationAnalysis}

\begin{figure*}[htb]
 \centering
 \includegraphics[width=0.99\linewidth]{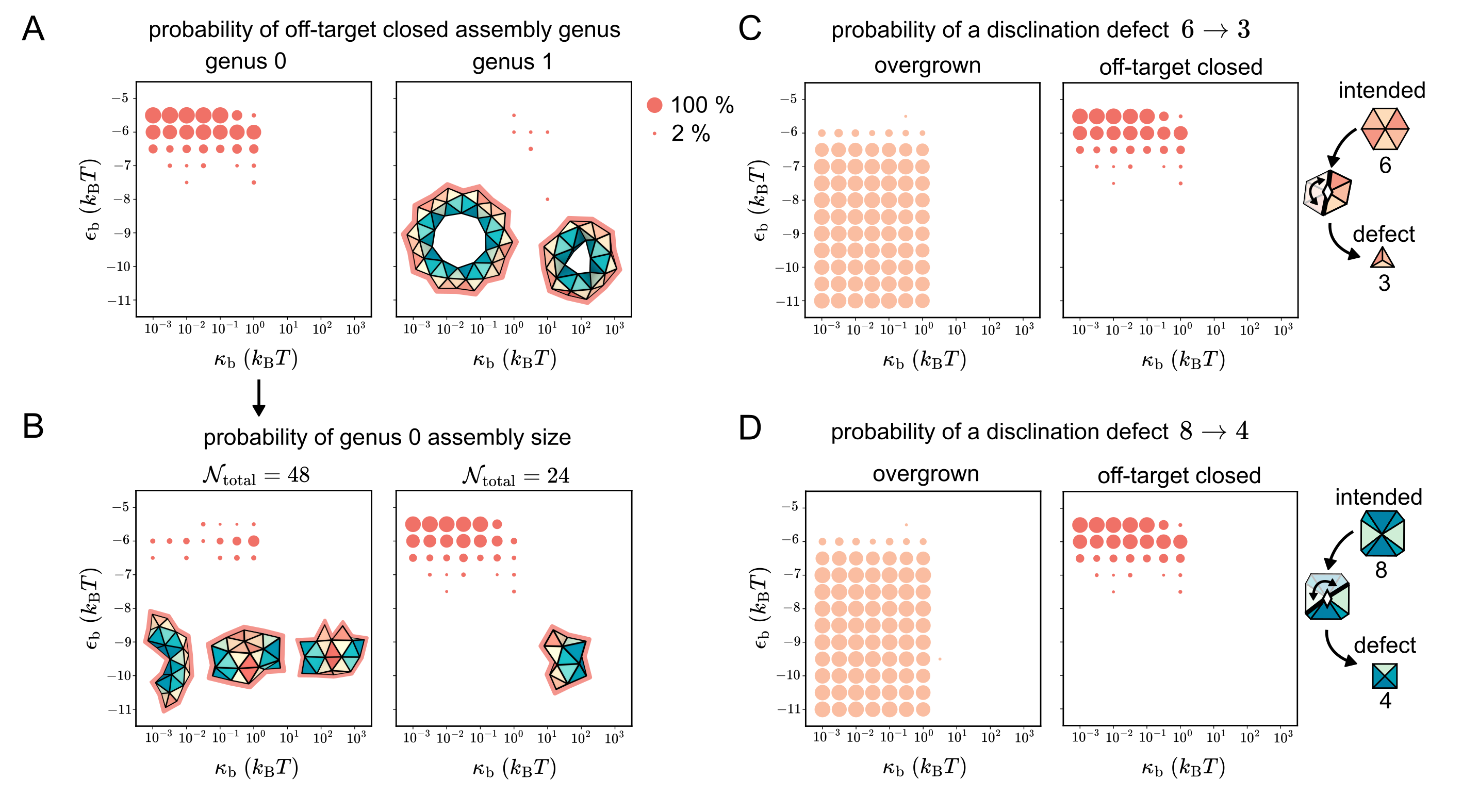}
 \caption{\textbf{Analysis of toroid failure modes.} (A) Classifying off-target closed assembly genus in the toroid simulation results reveals that genus 1 off-target assemblies are rare in comparison to genus 0. (B) Measuring the frequency of different sized assemblies with genus 0 reveal that smaller assemblies are more likely to close in the low bending modulus regime ($\kappa_{\rm b} \lesssim 1 \; k_{\rm b}T$). (C and D) Both the (C) 3-fold and (D) 4-fold disclination defects mediate closed and overgrown assemblies for $\kappa_{\rm b} \lesssim 1 \; k_{\rm b}T$. 
}
\label{SIfig:disclinations}

\end{figure*}

To characterize the frequency of different off-target assemblies, and the mechanisms guiding them, we analyze the simulations of the (4,2,2,1) toroid and look at three features of assembly outcomes. First, we measure the probability of both genus 0 (no hole) and genus 1 (one hole) off-target closed assemblies (Fig.~\ref{SIfig:disclinations}A). For low bending modulus, only genus 0 off-target assemblies form, while genus 1 off-target assemblies only form for $\kappa_{\rm b}>1 \ k_{\rm b}T$. Genus 0 assemblies occur due to the disclination defects discussed in the main text, and can occur without bond stretching. However, genus 1 assemblies require bond stretching to form. The fact that we see more 5-fold toroids in experiments suggests that we overestimate the stretching modulus in simulations.

Next, we look at the probability of different-sized assemblies for the genus 0 off-target closed structures (Fig.~\ref{SIfig:disclinations}B). We find that genus 0 assemblies tend to favor the smaller 24 triangle structures over the 48 triangle ones. Although this may be surprising at first given that there are more examples of unique structures with 48 triangles, and hence more unique configurations in which the triangles could be arranged, it makes sense from the perspective that smaller assemblies are entropically preferred~\cite{Hagan2021}.

Lastly, we look at the probability of an assembly having a disclination defect: either a $6 \to 3$ in which an intended 6-fold coordinated vertex crossing a symmetry axis closes into a 3-fold defect, or an $8 \to 4$ in which an intended 8-fold coordinated vertex closes into a 4-fold defect (Fig.~\ref{SIfig:disclinations}C. Nearly all off-target assemblies have both defect types in the region where $\kappa_{\rm b} \lesssim 1 \; k_{\rm b}T$. For $\kappa_{\rm b} \gtrsim 1 \ k_{\rm b}T$, off-target assemblies can form without disclination defects since a larger bending modulus strongly prohibits changing the geometry of a vertex. Instead, these assemblies overgrow the target structure and contain sheet-like regions that have open boundaries.

\clearpage

\section{Toroids and helical tubules beyond $(T,L,D,R)$ scheme} \label{sec:otherclass}

Here we show four examples of how the $(T,L,D,R)$ scheme could be extended to generate other classes of toroids and helical tubules. We note that there may be other ways to extend the $(T,L,D,R)$ scheme or toroid construction that does not follow this scheme at all~\cite{stewart1980adventures}.

\begin{figure*}[htb]
 \centering
 \includegraphics[width=0.99\linewidth]{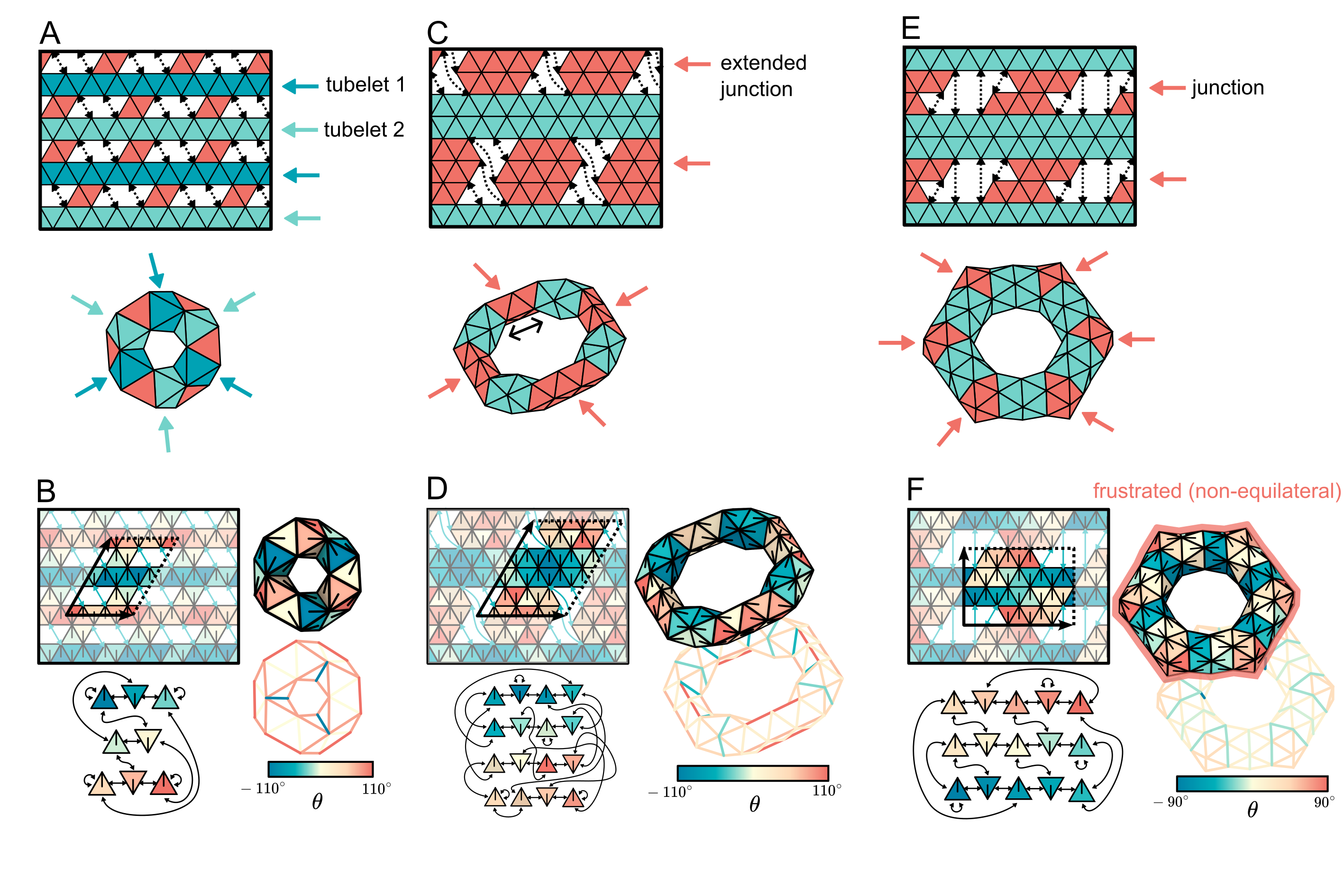}
 \caption{\textbf{Toroid extensions of $(T,L,D,R)$.} (A) Alternating tubelet segments can close to form a toroid. (B) Derived interactions and binding angles for the alternating tubelet scheme. (C) The junction region of a toroid can be extended. (D) Derived interactions and binding angles for the extended junction scheme. (E) The junction can be a completely different shape, forming a toroid with 6-fold symmetry about its axis. (F) Derived interactions and binding angles for the frustrated 6-fold scheme.
}
 \label{SIfig:otherToroids}
\end{figure*}

\subsection{Altering tubulets}
First, instead of connecting tubeletes of a single type together, we show an example in which two instances of a $(3,0)$ tubelet are connected together by junctions to form an ``alternating'' sequence of tubelets that close into a toroid (Fig.~\ref{SIfig:otherToroids}A,B). This is different from the TLDR construction because here, the two tubelet types are not equivalent to one another under the symmetry of the toroid. This could be further applied to other tubelet geometries to form other toroids and helical tubules composed of alternating tubelets.  

\subsection{Elongating tubulet or junction region}
Looking at the 3D geometry of a complete toroid, an entirely new class of toroids emerges by elongating the length of the junction region. We note that elongating the \textit{tubulet} region is still within the $(T,L,D,R)$ framework: we can increase $L$ to the next even number, and by tuning $R$ we will obtain a toroid with an elongated tubulet region. On the other hand, elongating junctions by inserting an additional row leads to a valid toroid, but this lies beyond the $(T,L,D,R)$ framework because the junction is no longer a diamond in the 2D unfolded holey tiling. For example, in Fig.~\ref{SIfig:otherToroids}C,D, we take the $(4,2,2,1)$ toroid and extend the junction region by one row. In this case, the toroid is still 4-fold symmetric. Further, we note that the 8-fold vertex is separated into two 7-fold vertices. Extending junction regions could be applied to any toroid or helical tubule in the $(T,L,D,R)$ framework to obtain new geometries with different sizes without changing the symmetry of the original structure. 

\subsection{Introducing non-parallelogram junctions}
Third, we consider another design in which junctions have a completely different shape. As an example we show $(5,0)$ tubelets connected together to form a toroid with 6-fold symmetry about its axis (Fig.~\ref{SIfig:otherToroids}E,F). In this case, a pair of 5-fold vertices lie along a meridian of the toroid instead of the equator, and are balanced by a pair of 7-fold vertices. This geometry is inspired by a design introduced and studied for toroidal carbon nanotubes (TCNT)~\cite{tamura2005positive, chuang2009generalized}. We note that this toroid is slightly frustrated because in order to close into a toroid, some of the triangle edges must stretch, making them non-equilateral. This further highlights the unpredictability of commensurability for toroids, especially when exploring different hole shapes. 

\subsection{Zigzag tubulet} \label{subsec:zigzag}
When considering the initial construction of helical tubules and toroids, we showed an example of when the wrapping vector for the cylinder is aligned with a lattice direction of the triangular tiling. We consider the other extreme, where one of the lattice directions of the sheet aligns with the tubule axis. In this limit, one of the lattice directions of the triangular lattice will lie perpendicular to the wrapping direction. Again, we define a rhombus with equal-length sides that we can cut out of a triangulated sheet. This time we choose the wrapping vector such that it lies in the (1,1) direction, using the ($h,k$) nomenclature defined before. This creates tubules that have a jagged edge, which we call a ``zigzag'' tubule.

\begin{figure*}[htb]
 \centering
 \includegraphics[width=0.99\linewidth]{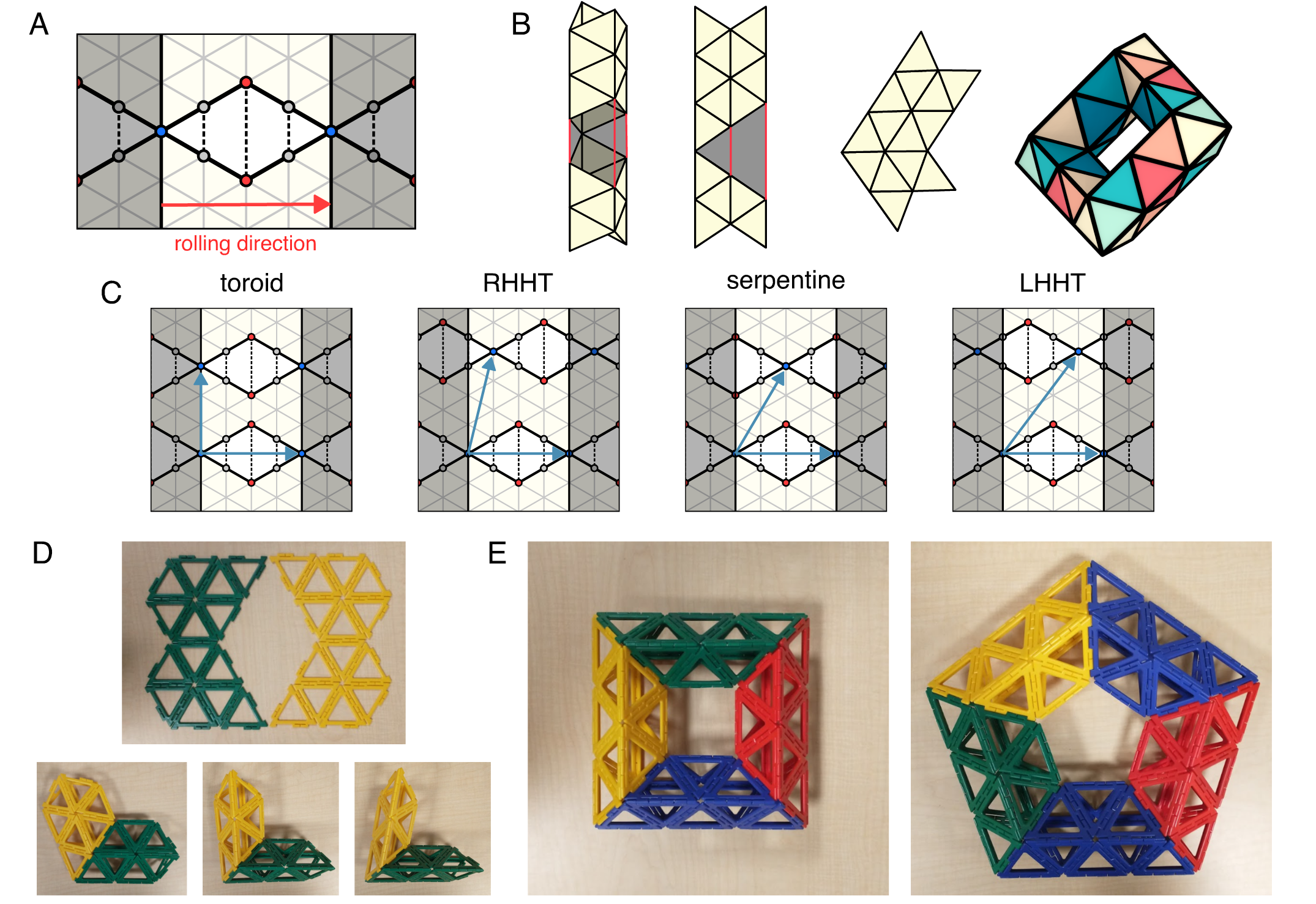}
 \caption{ \textbf{``Zigzag'' tubule kirigami construction.} (A) Placement of rhomboidal hole on a triangular lattice with a rolling direction of ($T,T$). (B) Illustration of the closure of the junction shown in (A). This junction can be repeated to close into a 4-fold toroid. (C) Examples of how one can shift the holes with respect to each other along the cylinder axis. For the example of a (2,2) cylinder shown, there are four options that yield a toroid, a serpentine tubule, and both right- and left-handed helical tubules. (D) Construction of a junction using rigid triangles that can snap together at the edge. The upper image shows the unfolded version of the junction in (A). The lower row shows that the junction is flexible and allows for variability in the junction angle. (E) Images of a 4-fold and 5-fold toroid made from the same junction construction.
}
 \label{SIfig:zigzagscheme}
\end{figure*}

An interesting feature of this ``zigzag'' construction is that it allows more than one $N$-fold toroid from the same ($T,L,D,R$) parameters and stitching pattern. We hypothesize that this is because a (2,2) tubule has a soft mode that allows for pure shear in the plane orthogonal to the tubule axis. In contrast, a (4,0) tubule does not have such a soft mode, it is a fully rigid structure. This soft mode of the (2,2) carries over, even after the addition of the junction, making the angle at which you add the next junction-tubelet pair flexible (Fig.~\ref{SIfig:zigzagscheme}D). We have found that this construction allows for 4- and 5-fold toroids (Fig.~\ref{SIfig:zigzagscheme}E). The limit at which a 6-fold toroid occurs is the case where all of the triangles flatten into a single plane, and is thus disallowed.

We can generalize the placement of the hole on the 2D tiling to work with arbitrary cylindrical tubules. As shown in Fig.~2 of the main text, any pair of vectors ($h_1, k_1$) and ($h_2, k_2$) form a primitive unit cell that describes the periodicity of the holes placed on the triangular sheet. Then it was chosen that ($h_1, k_1)=(T,0$) in the following section. What we showed above was the case where ($h_1,k_1)=(T,T)$. In general, our kirgami strategy should work for arbitrary choices of these vectors, as long as the holes do not overlap with each other. We hypothesize that the choices of ($T,0$) and ($T,T$) wrapping vectors are the only ones that can allow for the closure of toroids since they are the only choices that produce tubules that are achiral.

\subsection{Stitching rules}

\begin{figure*}[htb]
 \centering
 \includegraphics[width=0.99\linewidth]{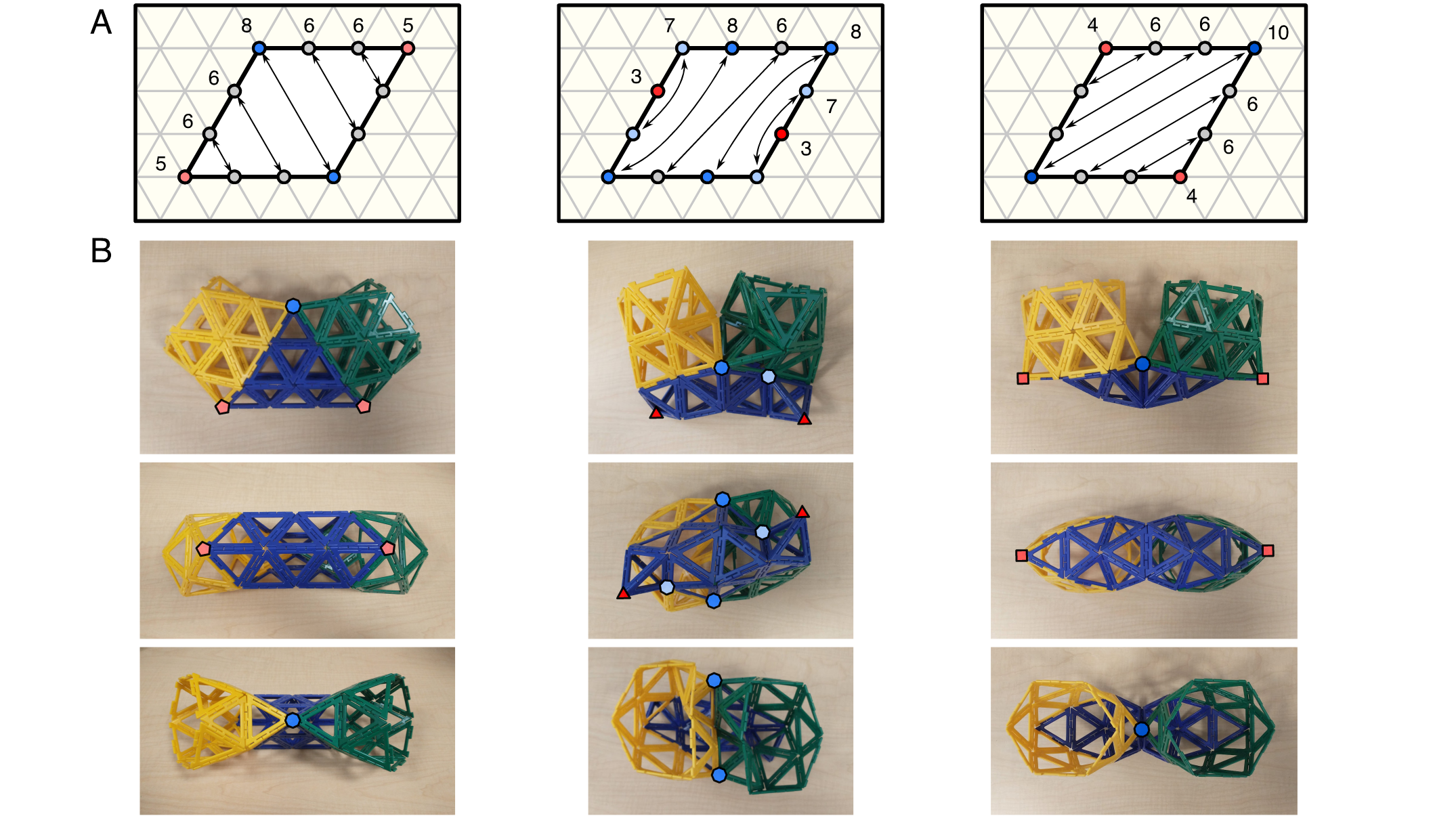}
 \caption{ \textbf{Examples of different hole stitching}. (A) Three variants of stitching for a $D=3$ hole. Lines denote the stitched vertices, and the adjacent number counts the number of triangles at the stitched vertex. (B) Constructions of junctions following the hole scheme in (A) with $T=6$ tubelets. Colored markers are placed on non-six-fold vertices in the junction.
} 
 \label{SIfig:holeclosure}
\end{figure*}

Another interesting avenue to consider is changing the hole's stitching scheme. In the main text, we limited our example to a single type of hole closure, but there are other valid ways to do the stitching. In Fig.~\ref{SIfig:holeclosure}A, we show three examples of stitching, including the one from the main text. In Fig.~\ref{SIfig:holeclosure}B, we show constructions of junctions based on the hole stitching. All three of these stitches can be folded up and cause the angle of the junction to change. The middle stitch pattern produces a chiral junction, which we had not seen previously. We think this is related to the stitch pattern not having mirror symmetry across the hole. It will be interesting to explore how different stitches change the geometry of these structures and what states (toroids, helical tubules, serpentine tubules) are allowed from them.

\clearpage

\section{Simulation methods}\label{sec:simmethod}

In this section we provide details about the model and kinetic Monte Carlo simulations that we used to generate the results in the main text. Our model is extended from previous works that studied the self-assembly of microcompartments by  Rotskoff and Geissler (which we refer to as the RG model henceforth) \cite{Rotskoff2018} and tubules with self-limited lengths due to geometric frustration by Tyukodi et al. \cite{Tyukodi2022Jun}. While those works considered only a single subunit species, in Duque et al. \cite{duque2024limits} (which modeled the assembly of size-controlled triply-periodic polyhedral), in Tyukodi et al. \cite{tyukodi2024magic} (which modeled the assembly of icosahedra)  and this work, we extend the model to an arbitrary number of species and inequivalent edge-types. We consider flexible triangular subunits that can bind to each other along edges with a set of preferred dihedral angles that set the preferred curvatures of the assembling sheet. We perform kinetic Monte Carlo (KMC) simulations  at fixed $\{\mu_i\} |{V T}$, with $\mu_i$ the chemical potential of subunit species $i$ in the bath.  Each Monte Carlo simulation involves a single cluster undergoing assembly and disassembly, with subunits taken from or returned to the bath respectively, as well as structural relaxation moves. Our description here closely follows that given in the supplementary information of Tyukodi et al. \cite{Tyukodi2022Jun, tyukodi2024magic} and Duque et al. \cite{duque2024limits}.

\subsubsection{Energies}
Each of the three edges of the triangular subunits may be of different types, $t(p)=1,2,3, ...$, for edge index $p$ and only complementary edge types may bind.

The total energy of the system is given by 
\begin{equation}
 E = \sum_{p}^{3 \ns} \Estretch^p + \frac{1}{2} \sum_{\langle pq \rangle} \left(\Ebend^{pq} + \Ebond^{pq}\right)
\end{equation}
where the first sum goes over all edges, with $\ns$ the number of subunits in the cluster.  The second sum only goes over bound edges (i.e. non-boundary, adjacent edges, so there are $2 \nb$ terms in the sum, with $\nb$ as the number of bonds). The $1/2$ factor corrects for double counting. 

The stretching energy is defined as:
\begin{eqnarray}
 \Estretch^p&=& \Es \frac{(l^p - l_0^{t(p)})^2}{2}
\end{eqnarray}
where $\Es$ is the stretching modulus, $l^t$ is the instantaneous length, and $l_0^{t}$ is the stress-free (rest) length of a type $t$ edge. We set the stretching modulus equal for all edges.

The bending energy is quadratic in deviations from the preferred dihedral angle:
\begin{equation}
 \Ebend^{pq} = \Kb \frac{\left(\theta^{pq} - \theta_0^{t(p)t(q)}\right)^2}{2}
\end{equation}
with $p$ and $q$ adjacent edges and $t(p), t(q)$ the edge types. $\Kb$ is the bending modulus and is set equal for all edge types. $\theta_0^{t(p)t(q)}$ is the preferred dihedral angle between edges with types $t(p)$ and $t(q)$.

Binding energies corresponding to all matching edge pairs are set equal, to $\Ebond^{pq} \equiv \Eb$.

In addition to the above terms, each subunit has at its center of mass a spherical excluder of radius $0.2 l_0$ to prevent subunit overlaps. Finally, to prevent extreme distortions of subunits, maximum edge-length fluctuations are limited to $l_0/2 < l < 3 l_0/2$.

\subsubsection{Coarse-graining}

As the model is motivated by the triangular DNA origami subunits presented in the main text, in which subunits bind through `patches' along subunit edges, we define attractive bonds along subunit edges (rather than at vertices as in the RG model). In particular, attractive bonds occur at each shared pair of subunit edges with the same type. Because the interactions in the experimental system are driven by base-pairing of short (5 base-pair or $\sim1.7$ nm) strands, they are extremely short-ranged in comparison to the subunit size (the subunit edge-lengths are approximately 50 nm). Therefore, in our simulations we avoid resolving the short length scale fluctuations in separation distance between bound edges and their associated vertices by coarse-graining as follows. 

\begin{figure}
\begin{center}
\includegraphics[width=0.7\columnwidth]{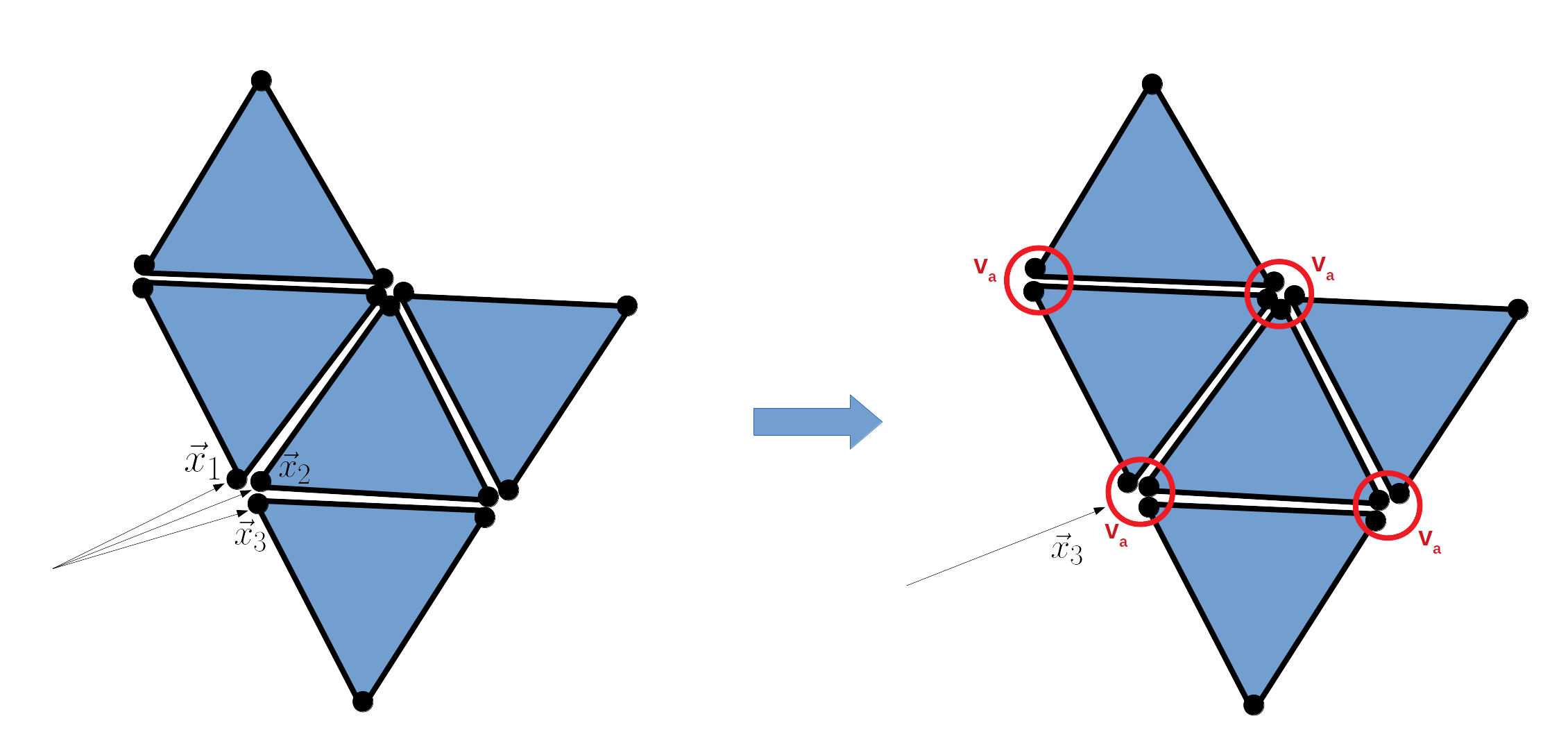}
\caption{Coarse-graining of an example cluster configuration. In this configuration, the number of subunits is $\ns=5$, the number of initial (before coarse-graining) vertices is $3 \ns = 15$, and the number of vertices after coarse-graining is $\nv=7$. The red circles indicate bound vertex groups, and the number of vertex degrees of freedom that have been eliminated by coarse-graining in this configuration is $\nvb = 1 + 3 + 2 + 2 = 8 = 3 \ns - \nv$. Motivated by DNA origami subunits, the attractive interactions (i.e. `bonds') in this model occur along edge-pairs of the same type shared by two subunits. In this configuration there are $\nb=4$ bonds. 
} 
\label{fig:coarsegraining}
\end{center} 
\end{figure}

A microstate $i$ is defined as the position of all the $3 \ns$ vertices of $\ns$ subunits: $i \to (\vec{x}_1, \vec{x}_2, ... \vec{x}_{3 \ns})$ The grand canonical probability \emph{density} of finding the system around state $i$ is
\begin{equation}
 f (i) = \frac{P (\vec{x}_1, \vec{x}_1 + d\vec{x}_1; ...; \vec{x}_1, \vec{x}_{n_s} + d\vec{x}_{n_s})}{d\vec{x}_1d\vec{x}_2...d\vec{x}_{3 \ns}} = \frac{1}{Z_\Omega} \frac{e^{\beta \sum_k \nsk \mu_k}}{\lambda^{9 \ns}} e^{-\beta E_i}
\end{equation}
where $\mu_k$ is the chemical potential of species $k$, $\ns$ is the total number of subunits and $\lambda^3$ is the standard state volume. This probability density has the dimensions of $1/\mathrm{volume}^{3 \ns}$ corresponding to all the $3 \ns$ vertices of the subunits. Due to bonds, however, some pairs of vertices are confined within a \emph{binding volume} $\vva$. We consider a square-well potential so that the binding energy is constant within this volume. Analogous to Ref.~\cite{Rotskoff2018}, we can then coarse-grain to avoid resolving intra-bond fluctuations. We assume that fluctuations of bound edges are sufficiently small that each pair of vertices at either end of a bound edge pair are constrained within the binding volume $\vva$. Note that we constrain vertices rather than edges so that the coarse-grained microstate can be represented in terms of positions of vertices rather than edges, which is easier to implement computationally.
In the coarse-grained system, a coarse microstate is specified by the coordinates corresponding to the independent vertex degrees of freedom (with 1 degree of freedom for each bound vertex group and unbound vertex): $\Gamma \to (\vec{x}_1, \vec{x}_2, ... \vec{x}_{\nv})$, where $\nv$ is the number of independent bound vertex groups and free vertices. The probability of such a coarse-grained state is given by the net weight of all the corresponding fine-grained microstates:

\begin{equation}
 \rho (\Gamma) = \int_{\{\vva\}} f (i) \mathrm{d^{\nvb}} \vec{x}
\end{equation}
where $\nvb$ is the number of vertex-bonds and is given by $\nvb = 3 \ns - \nv$. For simplicity, we take the limit in which $\sqrt[3]{\vva}$ is small in comparison to the length scale over which the elastic energy varies, so that the energy is constant within the bound volume $\vva$. Then $f(i)$ is a constant, and the probability density is given by
\begin{equation}
  \rho (\Gamma) =\frac{1}{Z_\Omega} \vva^{\nvb} \frac{e^{\beta \sum_k \nsk \mu_k}}{\lambda^{9 \ns}} e^{-\beta E_\Gamma}
\end{equation}
where $E_\Gamma$ is the total energy of state $\Gamma$ (including stretching, bending and binding energies). The coarse graining process is illustrated in Fig. \ref{fig:coarsegraining}.

\textit{Differences from the RG model \cite{Rotskoff2018}.}
First, in our model attractive bonds are defined along shared edges, while attractive bonds are counted at grouped vertices in the RG model. Thus, in our model the number of bonds $\nb$ for a given configuration is equal to the number of shared edges, whereas in the RG model it would be equal to the number vertex bonds, $\nvb$. However, the Hamiltonian for the elastic energy of the triangulated sheet and the corresponding Monte Carlo algorithm are much more straightforward to define in terms of a set of vertex positions rather than edge positions and orientations. Therefore, in our simulations we track vertex positions, by following the coarse-graining procedure described above. The distinction between attractions at edge- or vertex-pairs does not lead to qualitative differences between our model in the RG model, but  it does imply that the binding affinity values $\Eb$ and the binding volume $\vva$ must be treated as independent parameters in our model, since in general the number of bound vertices is not identical to the number of bound edges. Note that both definitions (bonds along edges or bonds at vertex pairs) involve approximations to the rotational and translational entropy penalties associated with subunit binding, since they assume a constant entropy penalty for each vertex degree of freedom that is lost within a bound vertex group, independent of the local environment (i.e., the number of bonds that a given subunit has) \cite{Hagan2011,Hagan2006a}. 

Second, because we are modeling independent triangular subunits binding along edges, each unbound edge in the graph (those edges at the boundary of the structure) corresponds to a single physical subunit edge, while bound edges (those in the structure interior) each correspond to two physical subunit edges. Therefore, in our model we set the stretching modulus for a bound edge to be twice that of a free edge. In the RG model all edges have the same stretching modulus.

\subsubsection{Implementation and data structure}

\begin{figure}
\begin{center}
\includegraphics[width=0.3\columnwidth]{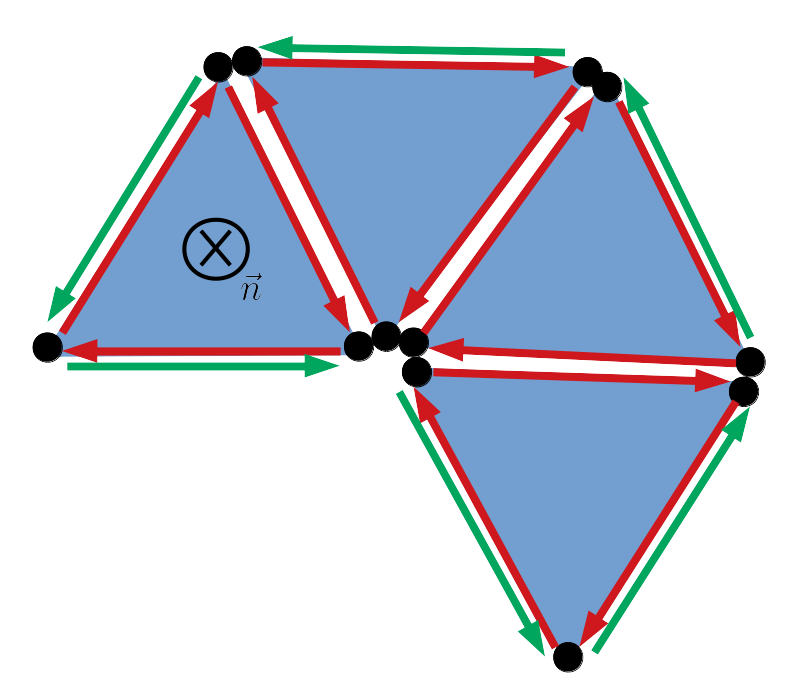}
\caption{The halfedge data structure used by OpenMesh. Each edge is represented by two directed edges. Boundary edges are no exception and thus are represented by a non-boundary halfedge and a boundary halfedge (in green). This latter is irrelevant for our model. Directed edges allow for the unambiguous definition of face normals, for efficient iterations of the element's neighborhood as well as boundary iterations.}
 
\label{fig:diredges}
\end{center} 
\end{figure}
The simulation is implemented on top of the OpenMesh library \cite{OpenMesh}. Subunits are implemented as triangular mesh elements. OpenMesh uses the halfedge data structure which is suitable to implement triangles with directed normals (Fig \ref{fig:diredges}). The directed halfedges allow for a clockwise iteration through the boundary of a triangle, which makes the two faces of the triangles distinguishable. Only halfedges with opposite orientations can bind together, making it impossible to form a Mobius strip, for example. The data structure and the resulting iterators in OpenMesh allow for an easy and efficient iteration over the neighborhood of mesh elements (vertices, edges and faces). The implementation of mesh element rearrangements is less straightforward, but we implemented it via the insertion and removal of virtual triangles. In addition, OpenMesh allows for the storage of various properties on mesh elements, allowing storage of edge types and face types stored on the elements. To improve readability in the upcoming sections, we will not represent halfedges separately.

\section*{The Monte Carlo moves}
In this section we detail the Monte Carlo moves of the simulation. Our algorithm has 13 moves: vertex displacement, simple subunit insertion/deletion, wedge insertion/deletion, hole insertion/deletion, wedge fusion/fission, crack fusion/fission, and edge fusion/fission. 

\textit{Detailed balance.}
For the transition between state $\Gamma$ and $\Gamma'$ detailed balance corresponds to \cite{Rotskoff2018, Frenkel1996}:
\begin{equation}
 P(\Gamma) \times \alpha({\Gamma \to \Gamma'}) \times \pacc(\Gamma \to \Gamma') =  P(\Gamma') \times \alpha({\Gamma' \to \Gamma}) \times \pacc(\Gamma' \to \Gamma)
\label{eq:DB}
\end{equation}
where $\alpha(\Gamma \to \Gamma')$ is the probability of generating a $\Gamma \to \Gamma'$ move attempt (trial), $\pacc (\Gamma \to \Gamma')$ is the probability of accepting the move, and $P(\Gamma) = \rho (\Gamma) \mathrm{d^{\nv(\Gamma)}} \vec{x}$ is the equilibrium probability of finding a system in a voxel of volume $\mathrm{d^{\nv(\Gamma)}} \vec{x}$.

Next, we use  Eq.~\eqref{eq:DB} to define the acceptance criteria for each MC move. The acceptance criteria are derived in detail for the wedge fusion/fission move; the steps to follow are the same for all other moves. 

\subsection{Vertex displacement}
\begin{figure}
\begin{center}
\includegraphics[width=0.5\columnwidth]{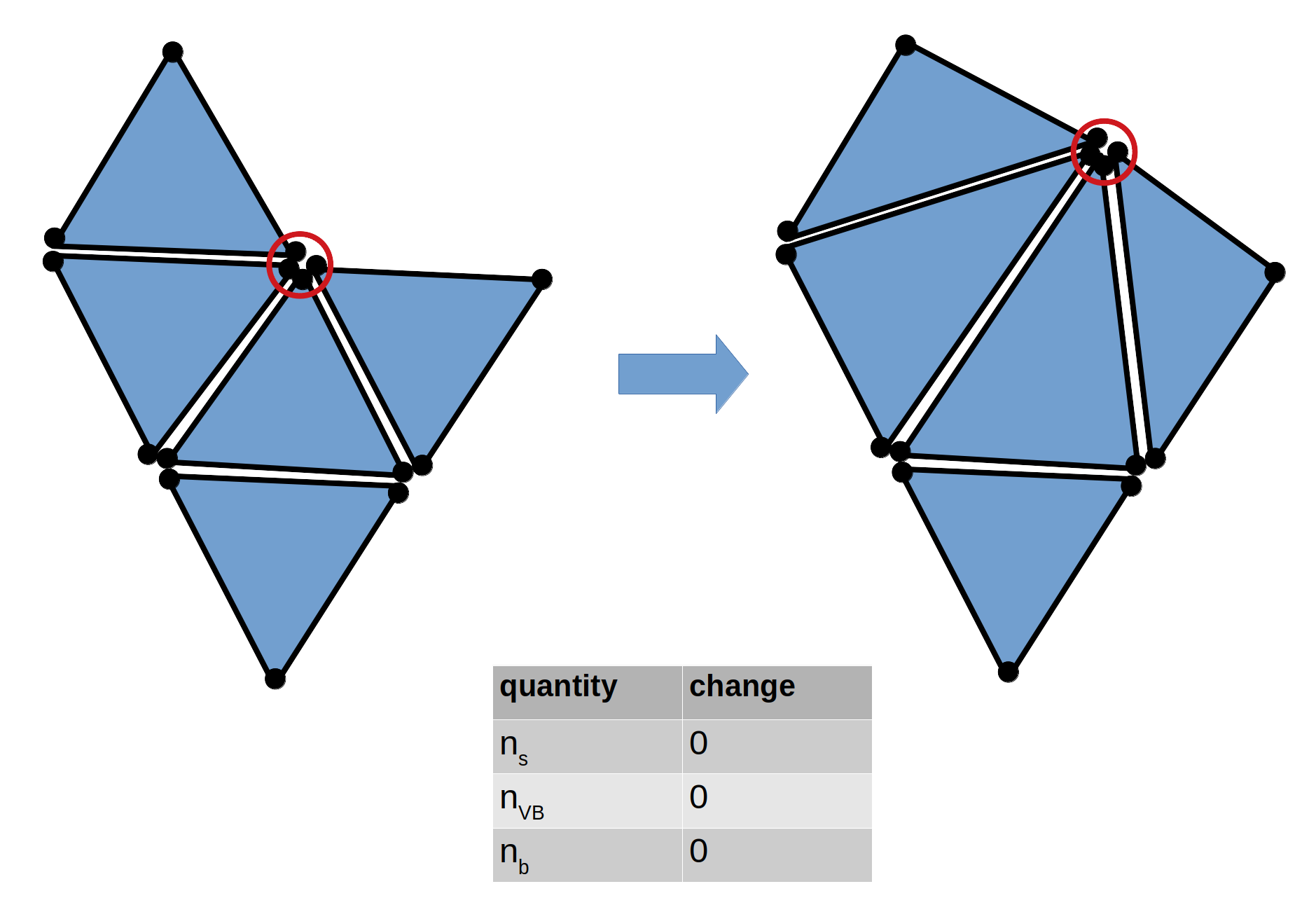}
\caption{Vertex move. A vertex is randomly displaced and the move is accepted according to the usual Metropolis probability.
} 
\label{fig:vertexmove}
\end{center} 
\end{figure}

In this move, a vertex is randomly selected, a random uniform displacement is drawn, and the vertex is displaced to its new position according to:
\begin{eqnarray}
 x &\to& x + \mathcal{U}(-\dm, \dm) \\
 y &\to& y + \mathcal{U}(-\dm, \dm) \\
 z &\to& z + \mathcal{U}(-\dm, \dm) 
\end{eqnarray}
with $\dm$ the maximum displacement.
The move is accepted with a probability $\pacc = \exp(-\Delta E/ \kt)$ where $\Delta E$ is the (bending plus stretching) energy change due to the displacement. The parameter $\dm$ can be adjusted during a burn-in period to optimize convergence to equilibrium.  Generally optimal values are on the order of the typical length scale of thermal fluctuations dictated by the elastic energy, leading to acceptance probabilities on the order of 50\%. In our simulations typical values are between $\dm=[0.01 l_0, 0.1 l_0]$. The vertex displacement move is illustrated in Fig \ref{fig:vertexmove}: the number of subunits $\ns$, number of vertices $\nv$, number of vertex bonds $\nvb$ and number of bonds $\nb$ remains unchanged during this move. 

\subsection{Simple insertion / removal}
\subsubsection{Simple insertion}
In this move, an edge is randomly selected from the set of all boundary edges, where a new subunit will be attached. The number of such boundary edges is $\ned$. Subunits can be inserted in $\nr$ different rotations, where $\nr$ is the number of distinct rotational states for a subunit which has one edge aligned with the edge of a neighboring subunit. For our triangular subunits with three distinct edge types, $\nr=3$. In our algorithm, during insertion of a subunit its rotational state is chosen randomly from the set of three possibilities. If the aligned edge is not complementary to the  type of the boundary edge, then the move is rejected. 

The positions of two of the new subunit's vertices (those at either end of the edge being bound) are set equal to the positions of the corresponding vertices of the boundary edge to which it is binding. The third vertex position is randomly chosen from within a volume $\vvadd$ centered at the equilibrium position of the new vertex.

Thus, the attempt probability for a simple insertion is given by:
\begin{equation}
 \alpha (i \to j) = \ned \ki \tau \nr \times \frac{1}{\ned \nr  (\vvadd/\mathrm{d}\vec{x})}
\label{eq:simpleInsertion}.
\end{equation}
Then, applying Eq.~\eqref{eq:DB} and the attempt probability for the reverse move (simple deletion, presented next, Eq.~\eqref{eq:simpleDeletion}), the acceptance probabilities for a simple insertion is
\begin{equation}
\pacc (i \to j) =\min \left[1, \frac{ \vva^2 \vvadd } { \lambda^9} \exp[ -(\Delta E_{i \to j} - \mu) / \kt]  \right]
\label{eq:pAccSimpleInsertion}.
\end{equation}
$\Delta E_{i \to j}$ is the energy change due to the move and includes the stretching energy of the newly inserted subunit, its bending energy along the shared edge, and the binding energy due to the creation of an extra bond.
During this move, one new (edge) bond and two new vertex bonds are created; i.e. $\nb\to\nb+1$ and $\nvb\to\nvb+2$. Moreover, the number of vertices in the structure increases by one, $\nv\to\nv+1$. 

\subsubsection{Simple removal}
The reverse move to simple insertion is simple removal. Subunits that can be deleted with this move are those with two boundary edges. The number of simply removable subunits is $\nsr$. One of these is selected randomly, so the attempt probability is 
\begin{equation}
 \alpha (j \to i) = \nsr \ki \tau \times \frac{1}{\nsr}
\label{eq:simpleDeletion}
\end{equation}
and, using Eq.~\eqref{eq:DB} and Eq.~\eqref{eq:simpleInsertion}, the acceptance probability is
\begin{equation}
 \pacc (j \to i) =\min \left[ 1, \frac{\lambda^9}{\vva^2 \vvadd} \exp[ -(\Delta E_{j \to i} + \mu) / \kt] \right]
\end{equation}
During this move, the structure loses one (edge) bond and two vertex bonds; $\nb\to\nb-1$ and $\nvb\to\nvb-2$. The number of vertices in the structure decreases by one, $\nv\to\nv-1$. 

If there are multiple species with chemical potentials $\mu_k$, detailed balance must be satisfied for each species, individually. Moreover, each species can have different insertion rates $\ki^k$.

To keep $\alpha<1$, we ensure that the insertion rate $\ki$ constrained by
\begin{eqnarray}
 \ned \ki \tau \nr <1 \\
 \nsr \ki \tau <1
\end{eqnarray}

In equilibrium, one can use adaptive rates, i.e. reduce $k_i$ on the run if the above condition is not satisfied. In that case, sampling is not taken for the ensuing several time steps. Alternatively, the rates may be set to a low enough value from the beginning and only tested on the run to ensure that the $\alpha<1$ condition is satisfied. This latter technique is appropriate for dynamical runs as it keeps the rates constant throughout the simulation.

Moreover, we must ensure that $\vvadd$ is large enough so that the vertex does not leave the $\vvadd$ volume during structural relaxation moves; otherwise the insertion/deletion moves would not be reversible and the detailed balance would be violated. For a better convergence, one could choose a gaussian distribution $\mathcal{N} (\vec{r})$ for the position of the new vertex instead of a uniform distribution $1/\vvadd$. In this case, this distribution has to be accounted for in the acceptance probabilities $\pacc (i \to j)$ and $\pacc(j \to i)$.

\begin{figure}
\begin{center}
\includegraphics[width=0.5\columnwidth]{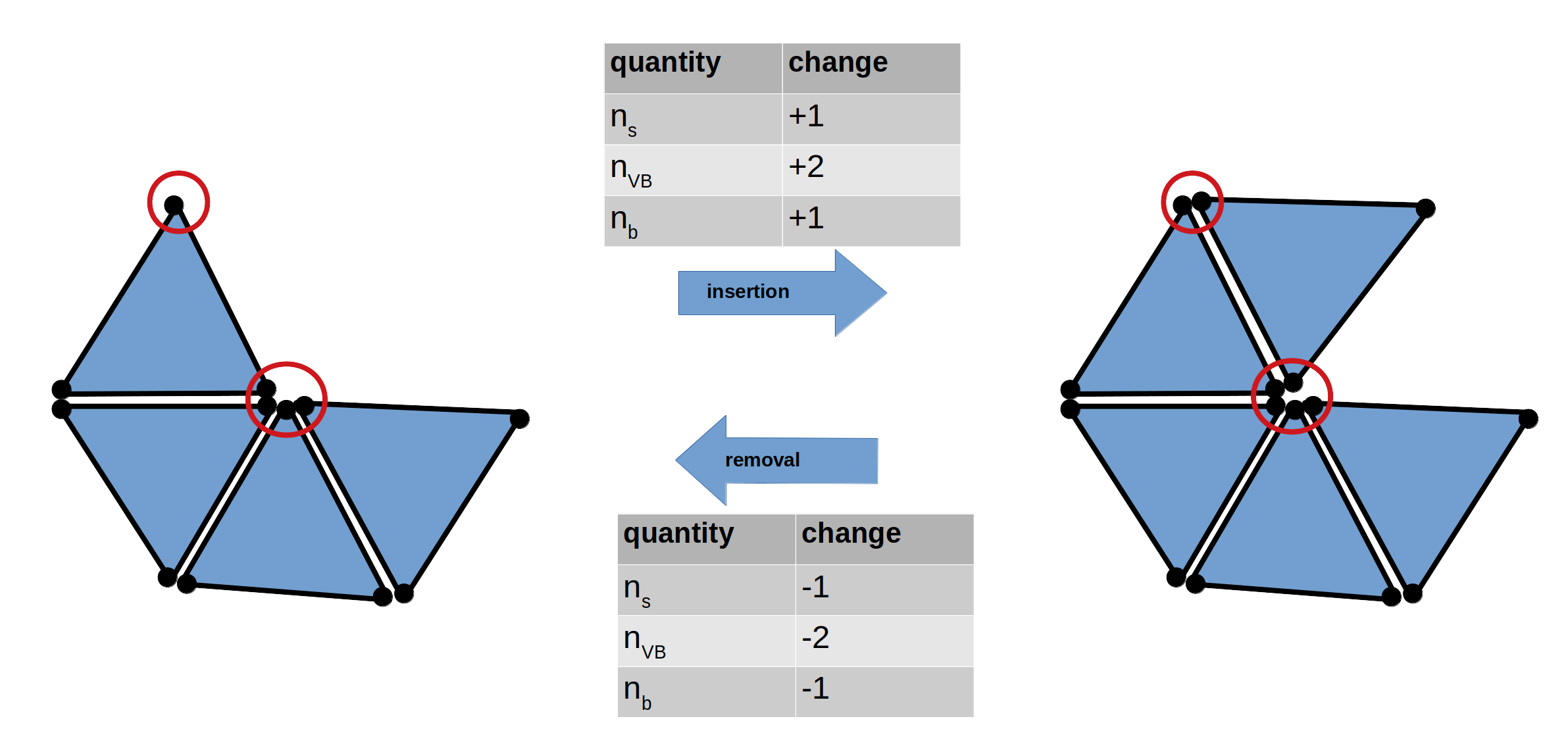}
\caption{Simple insertion and removal.
} 
\label{fig:simpleinsertion}
\end{center} 
\end{figure}

\subsection{Wedge insertion/removal}
\subsubsection{Wedge insertion}
\begin{figure}
\begin{center}
\includegraphics[width=0.5\columnwidth]{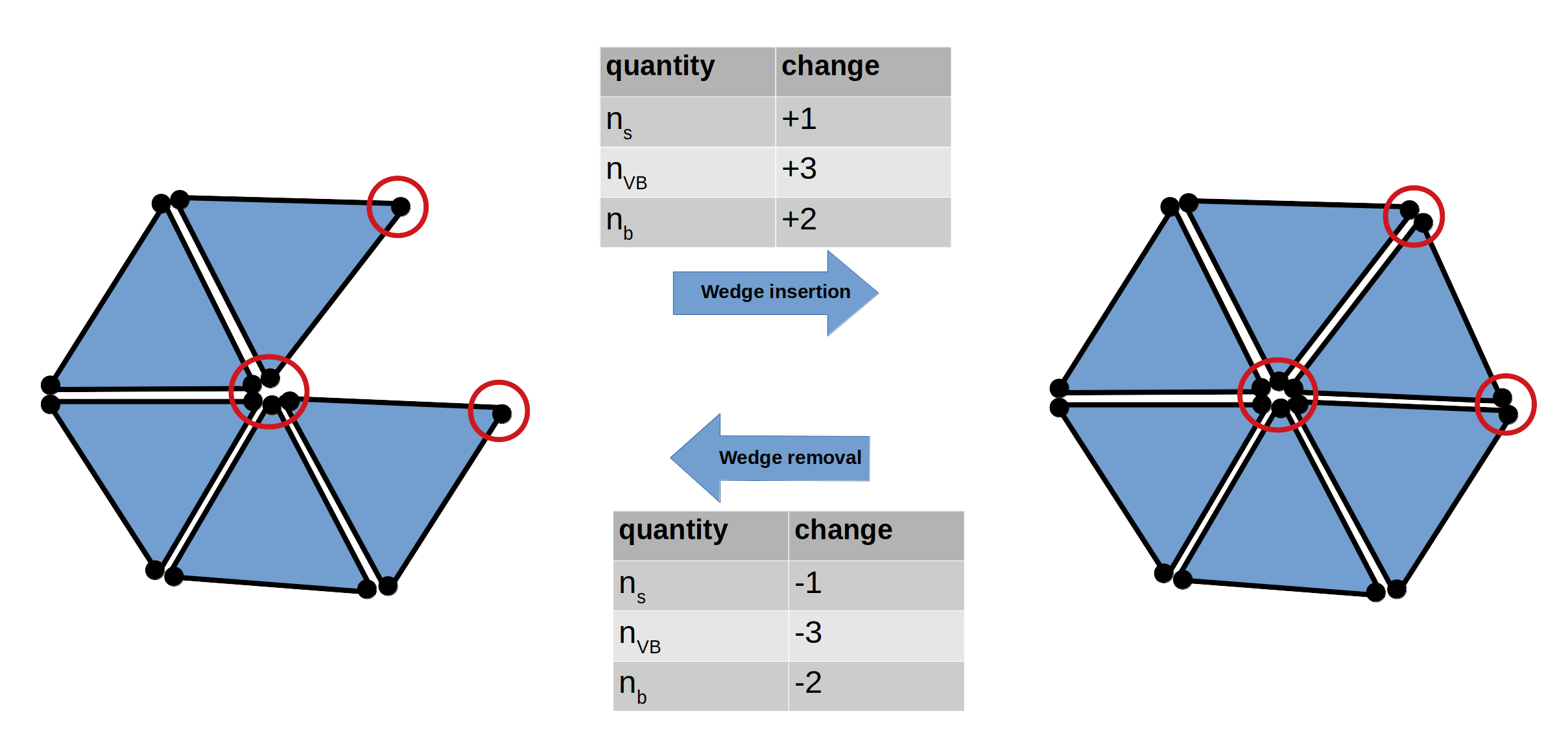}
\caption{Wedge insertion and removal.
} 
\label{fig:wedgeinsertion}
\end{center} 
\end{figure}

Wedges are positions in the structure where a triangle can be inserted via attaching to two edges (Fig.~\ref{fig:wedgeinsertion}). In a wedge move, we pick randomly from the set of available wedge positions in the structure, and pick a random orientation for the new subunit. Denoting the number of wedge positions in a given structure as $\nw$, the attempt probability for a wedge insertion move is
\begin{equation}
 \alpha(i \to j) = \nw \ki \tau \nr \times \frac{1}{\nr \nw}
\label{eq:wedgeInsertion}
\end{equation}
In contrast to the simple insertion move, there is no need for random vertex displacement in a wedge insertion move because all three vertices of the new subunit are fixed by the three vertices of the wedge position. Combining Eq.~\eqref{eq:wedgeInsertion} and the attempt probability for wedge removal (Eq.~\eqref{eq:wedgeRemoval}), 
The acceptance probability for a wedge insertion is
\begin{equation}
 \pacc(i \to j) = \min \left[ 1, \frac{\vva^3}{\lambda^9}\exp[ -(\Delta E_{i \to j} - \mu) / \kt] \right].
\end{equation}
During a wedge insertion, two edge bonds and three vertex bonds are created; i.e., $\nb\to\nb+2$ and $\nvb\to\nvb+3$, but the number of vertices is unchanged, $\nv\to\nv$. $\Delta E_{i \to j}$ includes the binding energy of the two newly formed bonds, the two bending energies along the two newly bound edges and the stretching energy of the newly inserted subunit.

\subsubsection{Wedge removal}
The reverse move of wedge insertion is wedge removal. In a wedge removal, we randomly choose one of the removable wedges from a given structure.  With the number of removable wedges as $\nwr$, the attempt probability is
\begin{equation}
 \alpha (j \to i) = \nwr \ki \tau \times \frac{1}{\nwr}
\label{eq:wedgeRemoval}.
\end{equation}
Using Eq.~\eqref{eq:wedgeInsertion}, the acceptance probability for a wedge removal is then
\begin{equation}
 \pacc(j \to i ) = \min \left[ 1, \frac{\lambda^9}{\vva^3}\exp[ -(\Delta E_{j \to i} + \mu) / \kt] \right] .
\end{equation}

We have the following constraints on rates $\ki$ for wedge insertion/removal:
\begin{eqnarray}
  \nw \ki \tau \nr <1 \\
 \nwr \ki \tau <1 
\end{eqnarray}

As for simple insertion and removal, in the case of multiple species, detailed balance is satisfied for each species separately for wedge insertion/removal.

\subsection{Hole insertion/removal}
\subsubsection{Hole insertion}
Holes are positions in the structure where a triangle can be inserted via attaching to three edges (and thus closing a triangular hole, Fig.~\ref{fig:wedgeinsertion}). In a hole insertion move, we pick randomly from the set of available triangular hole positions in the structure, and pick a random rotational orientation for the new subunit. Denoting the number of hole positions in a given structure as $\nh$, the attempt probability for a hole insertion move is
\begin{equation}
 \alpha(i \to j) = \nh \ki \tau \nr \times \frac{1}{\nr \nh}
\label{eq:holeInsertion}
\end{equation}
Similarly to the wedge insertion move, there is no need for random vertex displacement in a hole insertion move because all three vertices of the new subunit are fixed by the three vertices of the hole position. Combining Eq.~\eqref{eq:holeInsertion} and the attempt probability for hole removal (Eq.~\eqref{eq:holeRemoval}), 
the acceptance probability for a hole insertion is
\begin{equation}
 \pacc(i \to j) = \min \left[ 1, \frac{\vva^3}{\lambda^9}\exp[ -(\Delta E_{i \to j} - \mu) / \kt] \right].
\end{equation}
During a hole insertion, three edge bonds and three vertex bonds are created; i.e., $\nb\to\nb+3$ and $\nvb\to\nvb+3$, but the number of vertices is unchanged, $\nv\to\nv$. $\Delta E_{i \to j}$ includes the binding energy of the three newly formed bonds, the three bending energies along the three newly bound edges and the stretching energy of the newly inserted subunit.

\subsubsection{Hole removal}
The reverse move of hole insertion is the removal of a subunit bound at each three edges. In such a removal, we randomly choose one of the subunits with no boundary edges from a given structure.  With the number of removable subunits as $\nhr$, the attempt probability is
\begin{equation}
 \alpha (j \to i) = \nhr \ki \tau \times \frac{1}{\nhr}
\label{eq:holeRemoval}.
\end{equation}
Using Eq.~\eqref{eq:holeInsertion}, the acceptance probability for a hole removal is then
\begin{equation}
 \pacc(j \to i ) = \min \left[ 1, \frac{\lambda^9}{\vva^3}\exp[ -(\Delta E_{j \to i} + \mu) / \kt] \right] .
\end{equation}

We have the following constraints on rates $\ki$ for wedge insertion/removal:
\begin{eqnarray}
  \nh \ki \tau \nr <1 \\
 \nhr \ki \tau <1 
\end{eqnarray}

As for simple insertion and removal, in the case of multiple species, detailed balance is satisfied for each species separately for hole insertion/removal.

\begin{figure}
\begin{center}
\includegraphics[width=0.5\columnwidth]{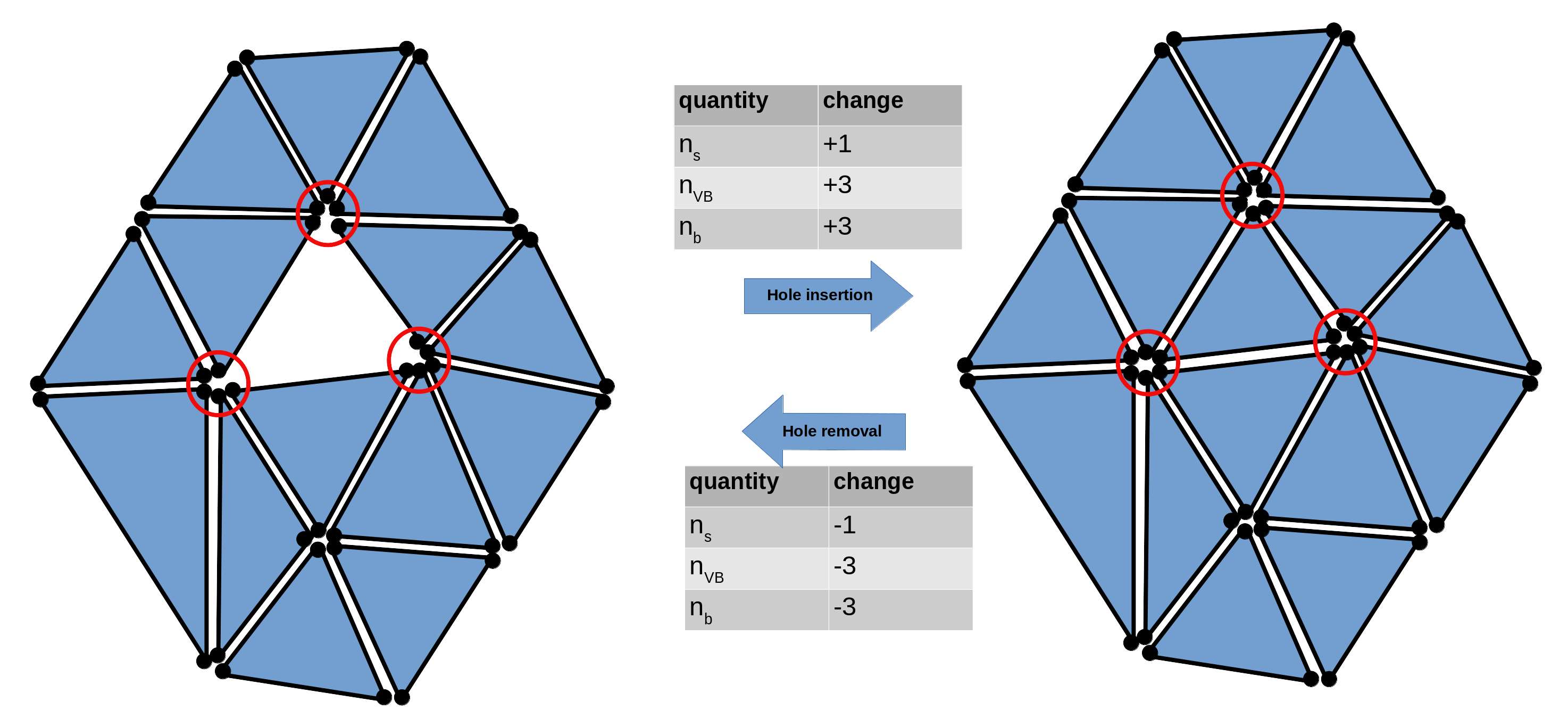}
\caption{Hole insertion and removal.
} 
\label{fig:holeinsertion}
\end{center} 
\end{figure}

\subsection{Wedge fusion / fission}
\subsubsection{Wedge fusion}
\begin{figure}
\begin{center}
\includegraphics[width=0.5\columnwidth]{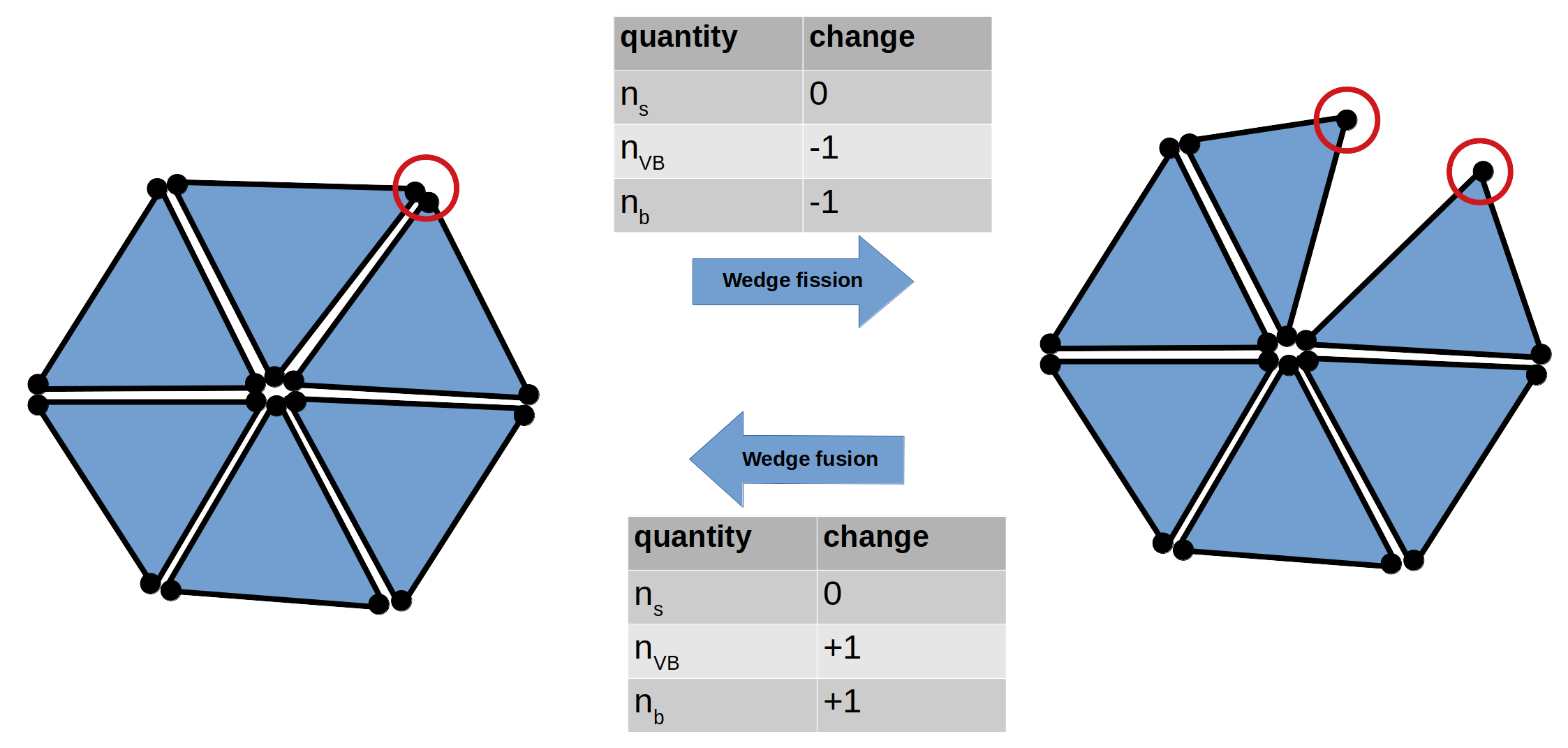}
\caption{Wedge fusion and fission.
} 
\label{fig:wedgefission}
\end{center} 
\end{figure}

In this move, a \emph{fusable wedge} is closed, without inserting a new subunit (Fig \ref{fig:wedgefission}).  That is, the two vertices on either side of the wedge opening are merged into a single vertex. Fusable wedges are vertex pairs that i) form a wedge (as in the case of wedge insertion) and ii) are within a separation distance of $\lfuse$. 

Denoting the number of fusible wedge positions as $\nw$, in each MC step, a wedge fusion is attempted with probability $\nw \kwf \tau$, where $\kwf$ is an adjustable parameter controlling the relative probability of attempting wedge fusion. Then, a wedge position is selected randomly from the set of all $\nw$ fusible wedges. The attempt probability is thus
\begin{equation}
 \alpha (i \to j) = \nw \kwf\tau \times \frac{1}{\nw}
\label{eq:wedgeFusion}.
\end{equation}
Using Eqs.~\eqref{eq:wedgeFusion} and \eqref{eq:wedgeFission}, the acceptance probability for fusion moves is  
\begin{equation}
 \pacc(i \to j) = \min \left[1, \frac{\vva}{ \vfuse} \exp(-\Delta E_{i \to j} / \kt) \right]
\label{eq:pAccwedgeFusion}
\end{equation}
where $\vfuse = (4 \pi / 3) (\lfuse/2)^3$ is the volume of a sphere with diameter $\lfuse$, and $\Delta E_{i \to j}$ is the energy change due to the fusion, including changes in bending, stretching, and binding energies. A fusion move increases the number of edge bonds and vertex bonds by one, $\nb\to\nb+1$ and $\nvb\to\nvb+1$; the factor of $\vva$ appears in Eq.~\eqref{eq:pAccwedgeFusion} to account for the latter.

\subsubsection{Wedge fission}
Wedge fission, in which a wedge is opened, is the reverse of the wedge fusion move. Fissionable edges are those edges that can be split along their boundary vertex to obtain a wedge. Denoting the number of such edges as $\nf$, the probability of attempting a wedge fission move during an MC step is $\nf \kwf\tau$. If a fission move is attempted, then an edge is selected randomly from the $\nf$ fissionable edges. The position of one of the new vertices is selected randomly within the sphere of volume $\vfuse$ centered at the original position of the merged vertices, and the other new vertex is placed in the opposite direction from the original position, at the same distance. Thus, 
the attempt generation probability is 
\begin{equation}
 \alpha (j \to i) = \nf \kwf\tau \times \frac{1}{\nf (\vfuse/\mathrm{d}\vec{x})} 
\label{eq:wedgeFission}
\end{equation}
and the acceptance probability is
\begin{equation}
 \pacc(j \to i) = \min \left[1, \frac{\vfuse}{\vva} \exp(-\Delta E_{j \to i} / \kt) \right]
\label{eq:pAccwedgeFission}
\end{equation}

We verify that detailed balance holds between wedge fusion and fission as follows. There are two cases to consider: 
\begin{enumerate}
 \item $ (\vfuse / \vva) \exp(-\Delta E_{j \to i} / \kt) <1 \Leftrightarrow (\vva/ \vfuse) \exp(-\Delta E_{i \to j} / \kt) >1$
 
 In this case, $\pacc(i \to j)=1$ and $\pacc(j \to i) = (\vfuse/\vva)  \exp(-\Delta E_{j \to i} / \kt)$. Then
 \begin{eqnarray}
 P_i \times \alpha(i \to j) \times \pacc(i \to j) &=& \frac{1}{Z_\Omega} \vva^{n_{\text{VB},i}} \exp[ -(E_i - \mu n_{s,i} )/\kt  ] \frac{1}{\lambda^{9 n_{s,i}}}\times  \mathrm{d^{n_{v,i}}}  \vec{x} \times k_\text{wf}\tau
\end{eqnarray}
 and
\begin{eqnarray}
 P_j \times \alpha(j \to i) \times \pacc(j \to i) &=& \frac{1}{Z_\Omega} \vva^{n_{\text{VB},j}} \exp[ -(E_j - \mu n_{\text{s},j} )/\kt  ] \frac{1}{\lambda^{9 n_{\text{s},j}}}\times  \mathrm{d^{n_{v,j}}}  \vec{x} \\ 
 &\times& k_\text{wf}\tau \mathrm{d} \vec{x} / \vfuse \times (\vfuse/\vva) \exp(-\Delta E_{j \to i} / \kt) \nonumber \\
\end{eqnarray}
Using: $\Delta E_{j \to i} = E_i - E_j$, $n_{s,i}=n_{s,j}$ (because the move leaves the subunit number unchanged), $n_{\text{V}B,i}=n_{\text{VB},j}-1$ (one vertex bond is broken upon fission) and $n_{v,i} = n_{v,j}+1$ (an extra vertex is being born upon fission), we see that the two are equal and detailed balance holds. 
 
 \item $ (\vfuse / \vva) \exp(-\Delta E_{j \to i} / \kt) >1 \Leftrightarrow (\vva/ \vfuse) \exp(-\Delta E_{i \to j} / \kt) <1$
 
In this case, $\pacc(i \to j)=  (\vva / \vfuse) \exp(-\Delta E_{i \to j} / \kt)$ and $\pacc(j \to i) = 1$. Then
\begin{eqnarray}
 P_i \times \alpha(i \to j) \times \pacc(i \to j) &=& \frac{1}{Z_\Omega} v_{a}^{n_{\text{VB},i}} \exp[ -(E_i - \mu n_{s,i} )/\kt  ] \frac{1}{\lambda^{9 n_{s,i}}}\times  \mathrm{d^{n_{v,i}}}  \vec{x} \times k_\text{wf}\tau \\
 &\times& (\vva / \vfuse) \exp(-\Delta E_{i \to j} / \kt)
\end{eqnarray}
and 
\begin{eqnarray}
 P_j \times \alpha(j \to i) \times \pacc(j \to i) &=& \frac{1}{Z_\Omega} v_{a}^{n_{\text{VB},j}} \exp[ -(E_j - \mu n_{s,j} )/\kt  ] \frac{1}{\lambda^{9 n_{s,j}}}\times  \mathrm{d^{n_{v,j}}}  \vec{x} \\ 
 &\times& k_\text{wf}\tau \mathrm{d} \vec{x} / \vfuse  \nonumber \\
\end{eqnarray}  
\end{enumerate}
Using again $\Delta E_{j \to i} = E_i - E_j$, $n_{s,i}=n_{s,j}$, $n_{VB,i}=n_{\text{VB},j}-1$ and $n_{v,i} = n_{v,j}+1$, detailed balance holds.

Note that detailed balance is satisfied regardless of the values of $\kwf \tau$ or $\vfuse$, but as with all of the move frequencies these parameters can be optimized during burn-in to accelerate convergence to the equilibrium distribution $P(i)$. 
In our simulations, we find that the optimal value of $\vfuse$ is on the order of the optimal value of $\dm$ for analogous reasons: if $\vfuse$ is too small there will be very few vertex pairs identified as fusable, so $\nw$ will be low. If $\vfuse$ is too large, there will be many fusion candidates but most fusion attempts will be rejected due to the large elastic energy change necessary for the merging deformation.

Most importantly, we note the constraint on the parameters $\kwf \tau$ to ensure that generation probabilities do not become larger than unity. Because each attempt is generated as a three step process, using three probabilities, one has to ensure that all those probabilities are less than 1. Specifically,
\begin{eqnarray}
 \nw \kwf \tau <1 \\
 \nf \kwf \tau <1 .
\end{eqnarray}

\subsection{Crack fusion / fission}
\subsubsection{Crack fusion}
\begin{figure}
\begin{center}
\includegraphics[width=0.5\columnwidth]{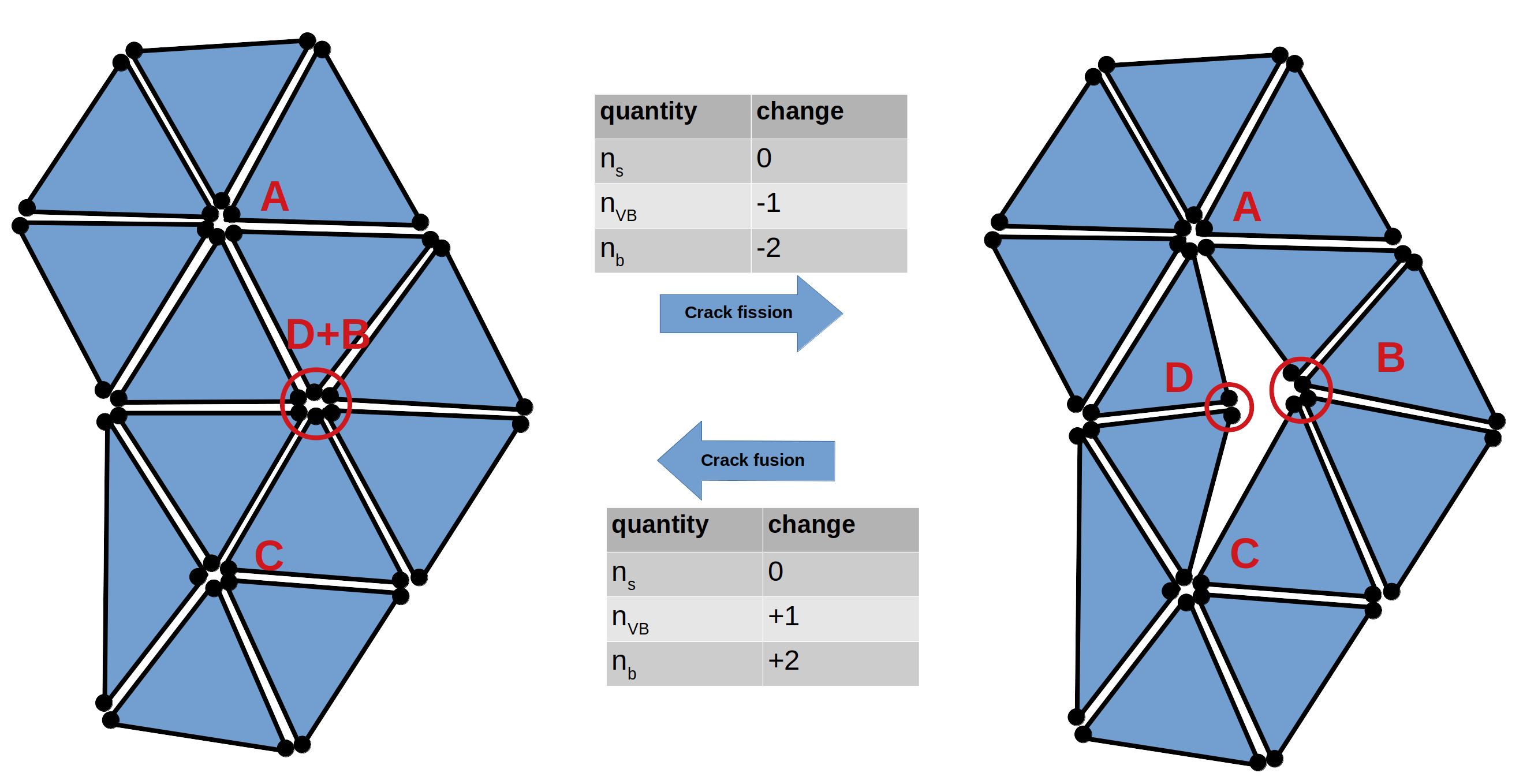}
\caption{Crack fusion and fission.
} 
\label{fig:crackfission}
\end{center} 
\end{figure}

Crack fusion closes a crack within the structure; i.e., two adjacent pairs of edges are merged (Fig. \ref{fig:crackfission}). Cracks are identified as 4-edge-length holes inside the structure. If the vertices of the hole are labeled A, B, C, D then the polygon ABCD forms a closed loop (see Fig. \ref{fig:crackfission}). The crack can be closed by either merging vertices A and C (and correspondingly edges CD to DA and AB to BC) or by merging vertices B and D (and correspondingly edges AD to AB and CD to CB). Each 4-edge-length loop thus defines two potential fusable cracks. However, an additional condition for a crack to be fusable is that its merging vertices must be within a distance $\lfuse$ (A and C or D and B in this example). In this work, we have set the crack fusion volume to be the same as that for wedge fusion to reduce the number of parameters, but it is not necessary that they be the same.
and the acceptance probability is
\begin{equation}
 \pacc (i \to j) = \min \left[1, \frac{\vva}{\vfuse} \exp(-\Delta E_{i \to j} / \kt) \right]
\end{equation}
There are two edge bonds and one vertex bond formed during a crack fusion.

\subsubsection{Crack fission}
The reverse move for crack fusion is crack fission. With the number of potential cracks as $\ncf$:
\begin{equation}
 \alpha (j \to i) = \ncf \kcf\tau \times \frac{1}{\ncf (\vfuse/\mathrm{d}\vec{x})} 
\end{equation}
\begin{equation}
 \pacc (j \to i) = \min \left[1,  \frac{\vfuse}{ \vva }\exp(-\Delta E_{j \to i} / \kt) \right]
\end{equation}

As for the case of wedge fusion/fission, the crack fusion attempt frequency parameter $\kcf$ is constrained by the conditions maintaining probabilities smaller than unity:
\begin{eqnarray}
 \nc \kcf \tau <1 \\
 \ncf \kcf \tau <1 \\
\end{eqnarray}

\subsection{Edge fusion / fission}
\subsubsection{Edge fusion}
\begin{figure}
\begin{center}
\includegraphics[width=0.5\columnwidth]{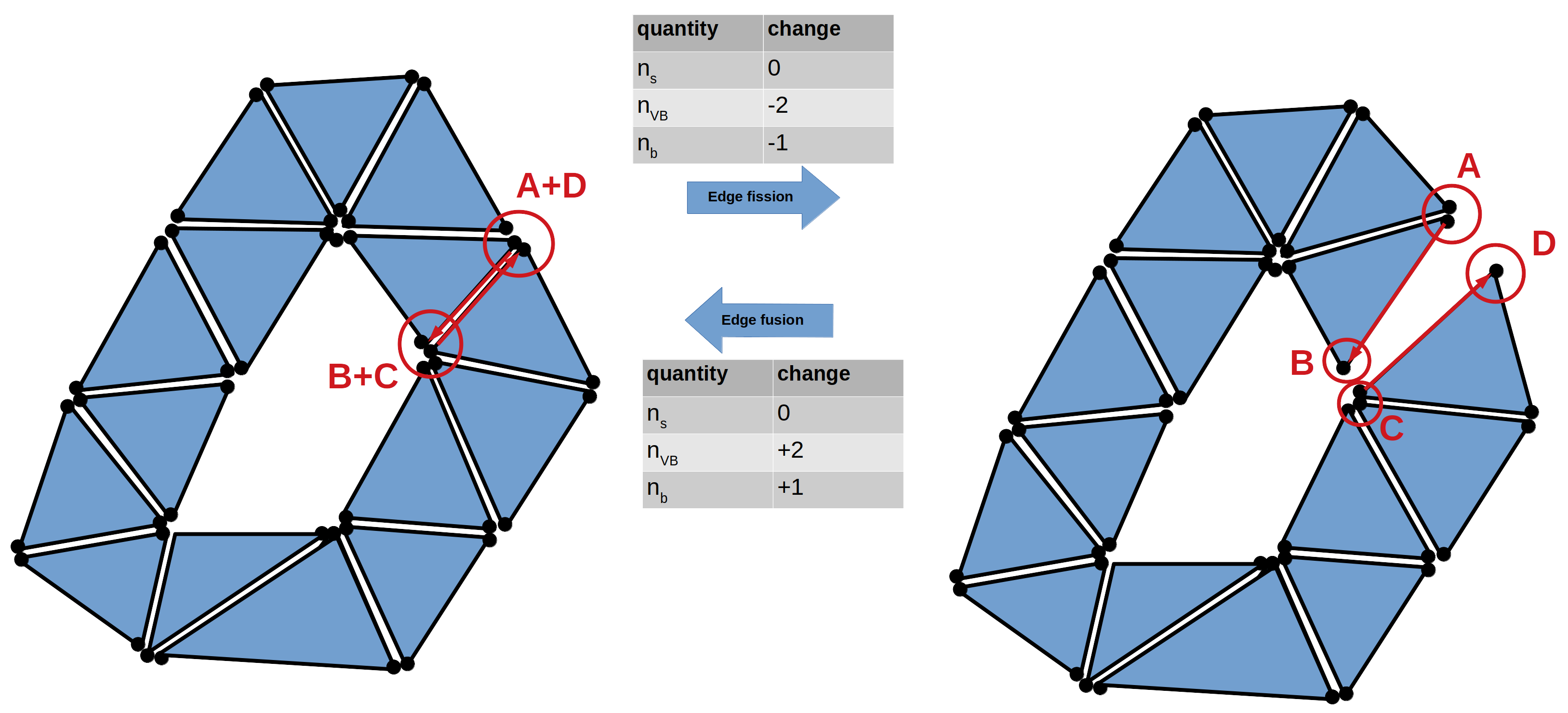}
\caption{Edge fusion and fission.
} 
\label{fig:edgefission}
\end{center} 
\end{figure}

During this move two non-neighbor edges are fused (Fig. \ref{fig:edgefission}). Fusable edges are non-neighboring edge pairs whose corresponding vertices are within a separation distance $\lfuse$. Since edges are directed, they can only fuse such that, after fusion, they point in the opposite direction. Assuming the edges to be fused are $A \to B$ and $C \to D$ (see Fig. \ref{fig:edgefission}), vertex $A$ will merge into vertex $D$ and vertex $B$ will merge into vertex $C$. Edges are counted as fusable if $A$ is within a distance $\lfuse$ to $D$ and $B$ is also within a distance $\lfuse$ to $C$.  The attempt probability is analogous to that for wedge and crack fusion/fission,
\begin{equation}
 \alpha (i \to j) = \ned \kef\tau \times \frac{1}{\ned}
\end{equation}
with $\ned$ the number of fusable edges and $\kef$ the edge fusion frequency parameter. The acceptance probability is
\begin{equation}
 \pacc (i \to j) = \min \left[1,  \left(\frac{\vva}{\vfuse}\right)^2 \exp(-\Delta E_{i \to j} / \kt) \right]
\end{equation}
During edge fusion, one edge bond and two vertex bonds are created.

\subsubsection{Edge fission}
Edge fission is the reverse move of edge fusion. $\nef$ is the number of breakable edges, that is, those edges that have both vertices on the boundary and which would not result in breaking the structure apart.
\begin{equation}
 \alpha (j \to i) = \nef \kef\tau \times \frac{1}{\nef (\vfuse/\mathrm{d}\vec{x})^2} 
\end{equation}
The factor $1/(\vfuse)^2$ arises because we must select a random position for each pair of vertices, independently. The acceptance probability is then
\begin{equation}
 \pacc (j \to i) = \min \left[1, { \left(\frac{\vfuse}{\vva} \right)^2} \exp(-\Delta E_{j \to i} / \kt) \right].
\end{equation}

To maintain probabilities within unity, the edge fusion frequency parameter $\kef$ is constrained by
\begin{eqnarray}
 \ned \kef \tau <1 \\
 \nef \kef \tau <1.
\end{eqnarray}

Values used for the various rates are summarized in Table \ref{tab:parameters}.
\begin{table}[!h]
\begin{center}
\begin{tabular}{ c c c }
insertion rate & $\ki$ & $10^{-4}$ \\ 
wedge fusion rate & $\kwf$ & $10^{-3}$ \\  
edge fusion rate & $\kef$ & $10^{-2}$ \\   
crack fusion rate & $\kcf$ & $5\times10^{-4}$ \\
\end{tabular}
\end{center}
\caption{Table with simulation metaparameter values used in the main text. $\vvadd$, $\lfuse$, $\dm$ are set such that they are commensurate with typical thermal fluctuations at the given elastic parameters.}
\label{tab:parameters}
\end{table}

\subsection*{Setting the subunit concentrations}
We note that the subunit concentrations are set to be the same because all species appear in equal stoichiometry for all of the target structures considered here. We choose a fixed chemical potential of $\mu=-4.5 k_BT$ for all species.

\clearpage

\begin{figure*}[htbp]
 \centering
 \includegraphics[width=0.99\linewidth]{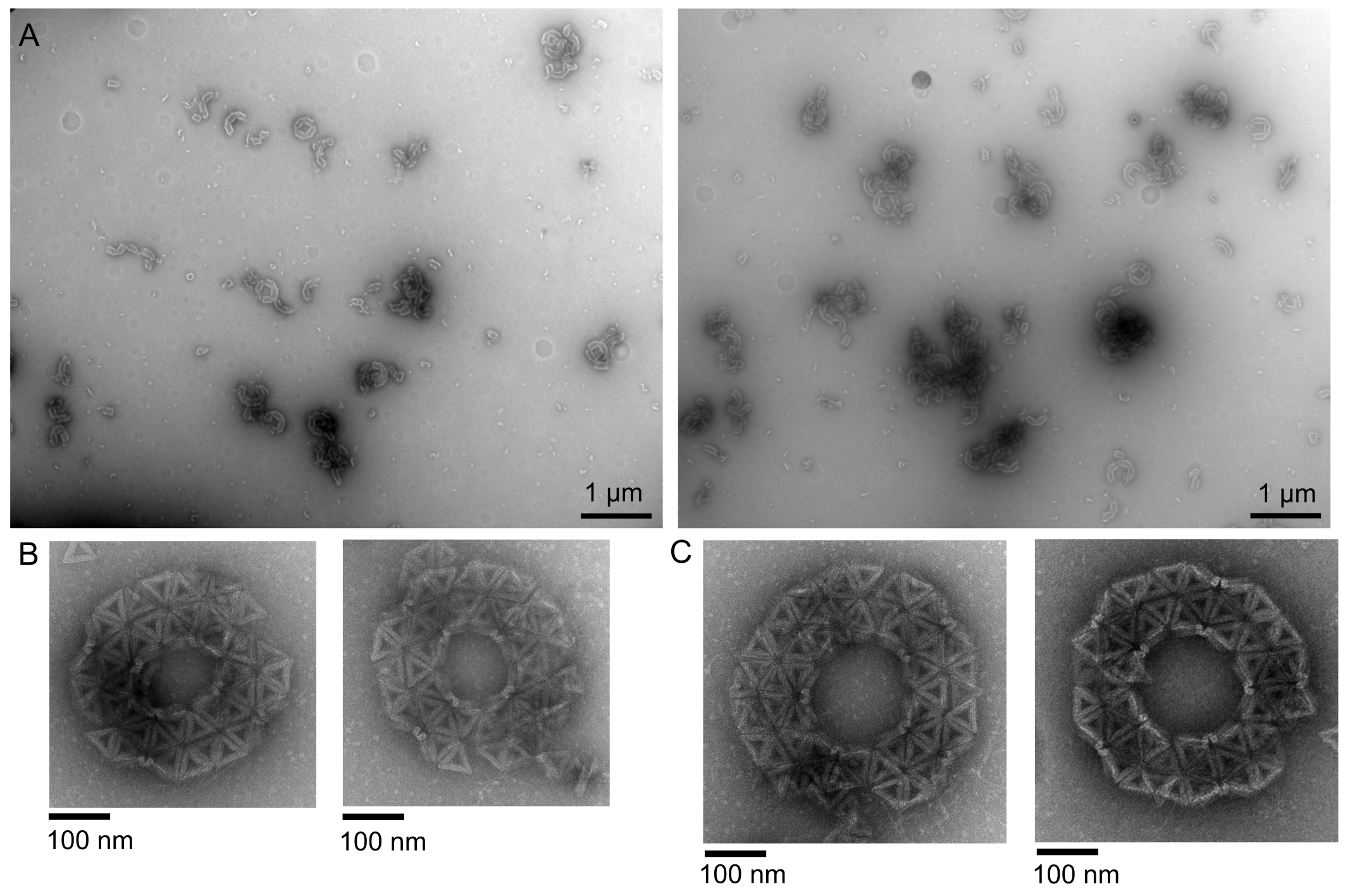}
 \caption{\textbf{Toroid TEM.} (A) Wide-field images of toroids under TEM. (B) Close-up TEM images of 4-fold toroids. (C) Close-up TEM images of 5-fold toroids. 
}
\end{figure*}

\begin{figure*}[htbp]
 \centering
 \includegraphics[width=0.99\linewidth]{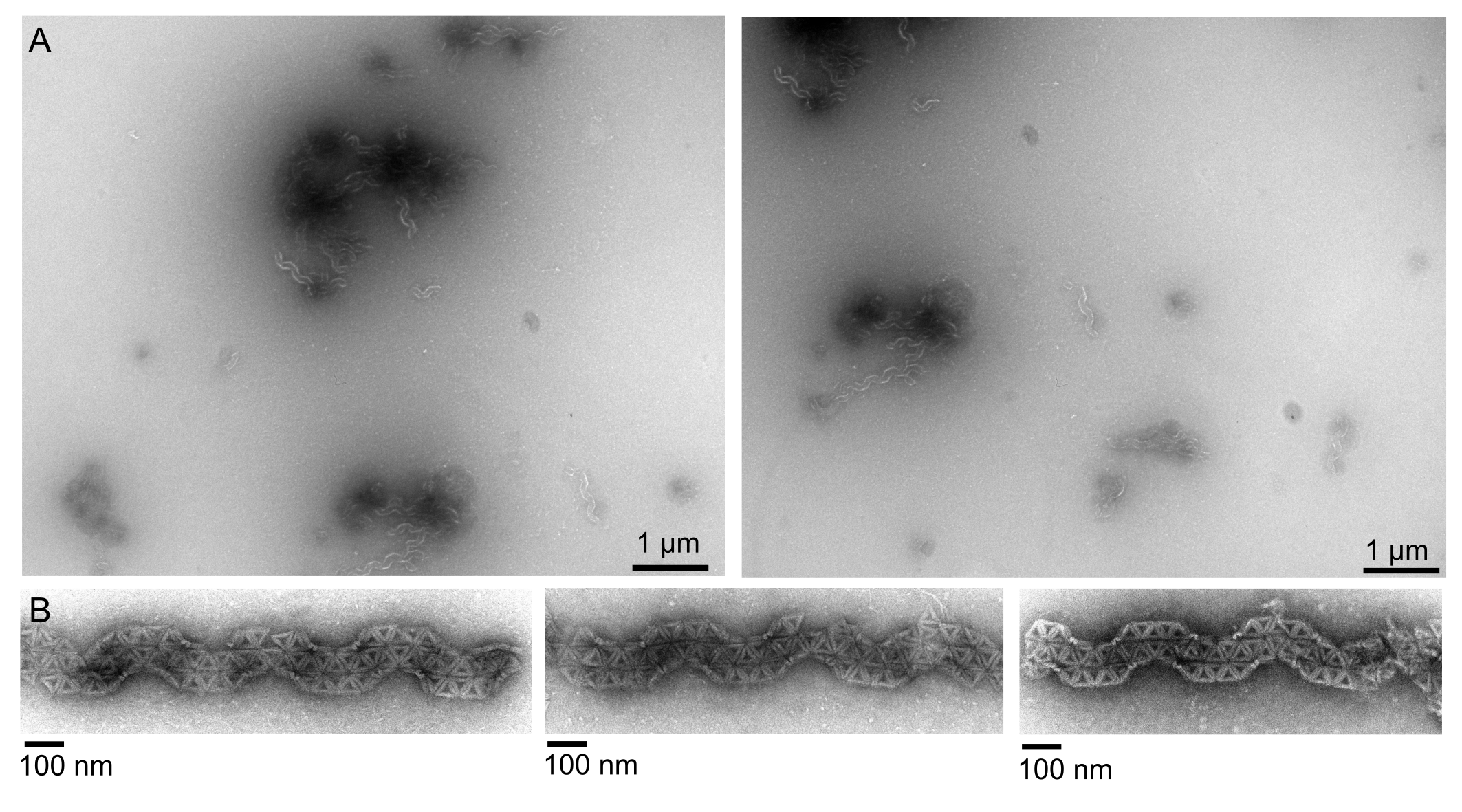}
 \caption{\textbf{Achiral serpentine tubule TEM.} (A) Wide-field images of serpentine tubules under TEM. (B) Close-up TEM images. All images are contrast adjusted.
}
\end{figure*}

\begin{figure*}[htbp]
 \centering
 \includegraphics[width=0.99\linewidth]{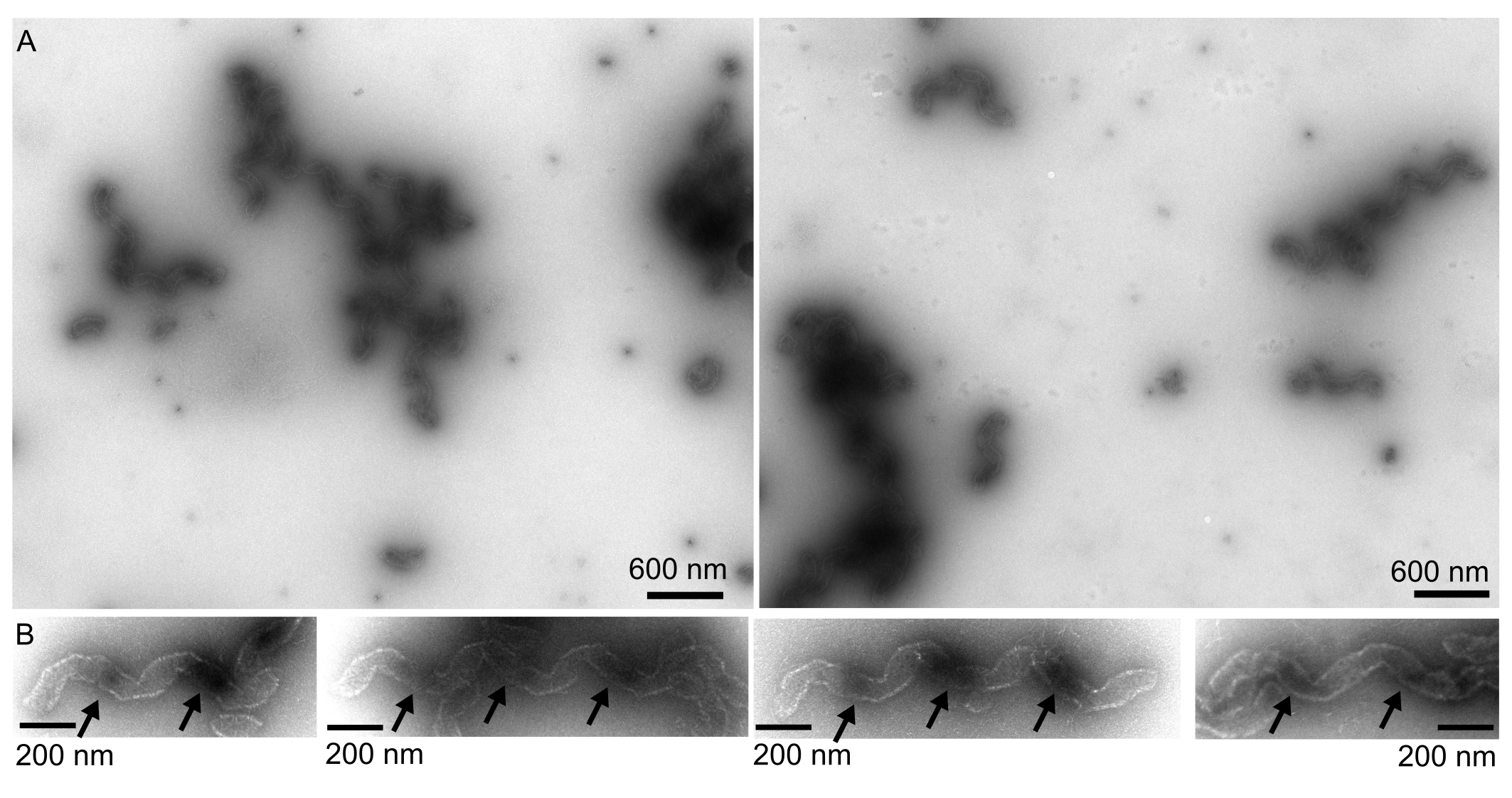}
 \caption{\textbf{Left-handed helical tubule TEM.} (A) Wide-field images of left-handed helical tubules under TEM. (B) Close-up TEM images. The arrows indicate repeating dark regions, which correspond to lower regions of the assembly. All images are contrast adjusted. 
}
\end{figure*}

\begin{figure*}[htbp]
 \centering
 \includegraphics[width=0.99\linewidth]{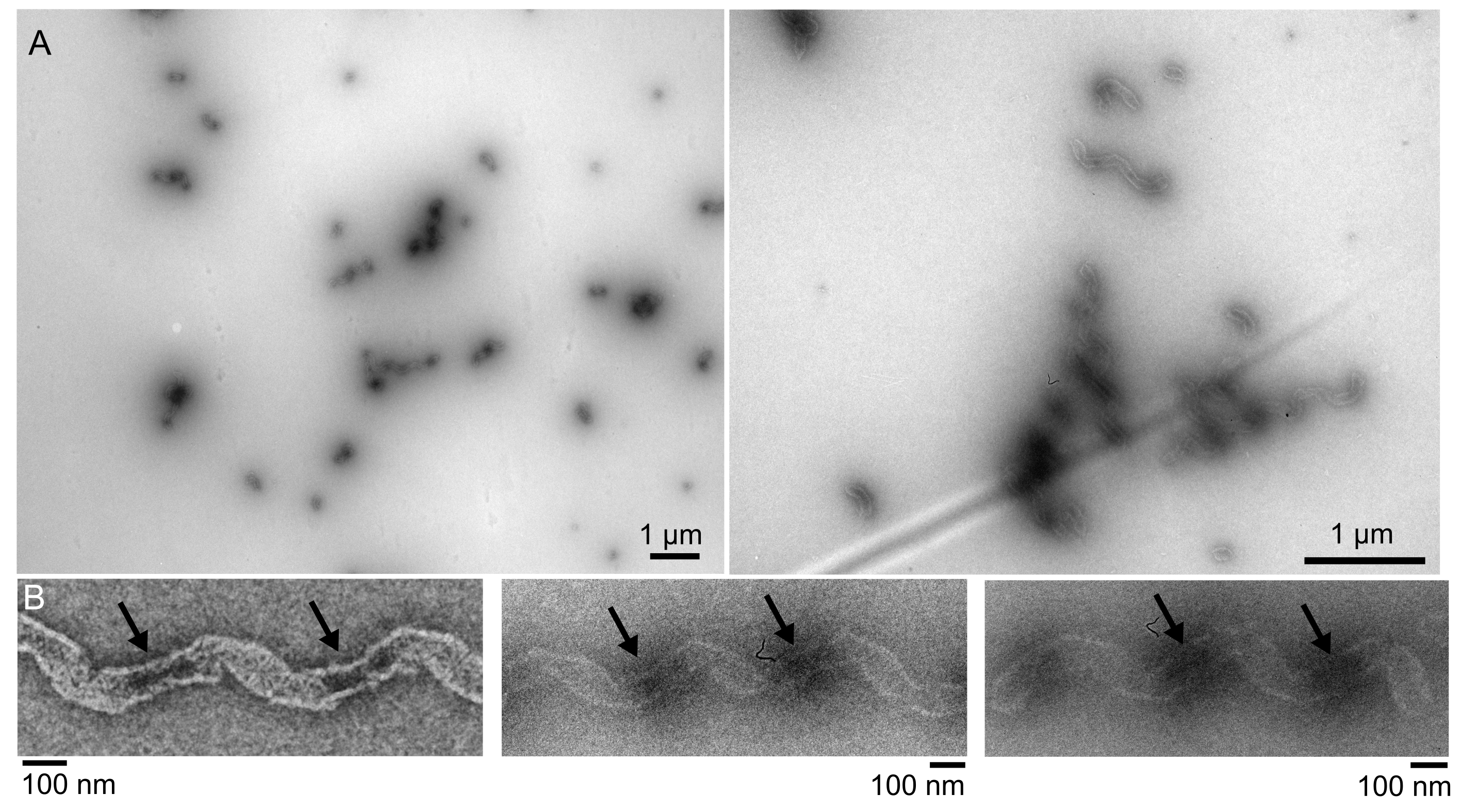}
 \caption{\textbf{Right-handed helical tubule TEM.} (A) Wide-field images of right-handed helical tubules under TEM. (B) Close-up TEM images. The arrows indicate repeating dark regions, which correspond to lower regions of the assembly. All images are contrast adjusted.
}
\end{figure*}

\clearpage

\begin{figure}
\begin{center}
\includegraphics[width=0.99\columnwidth]{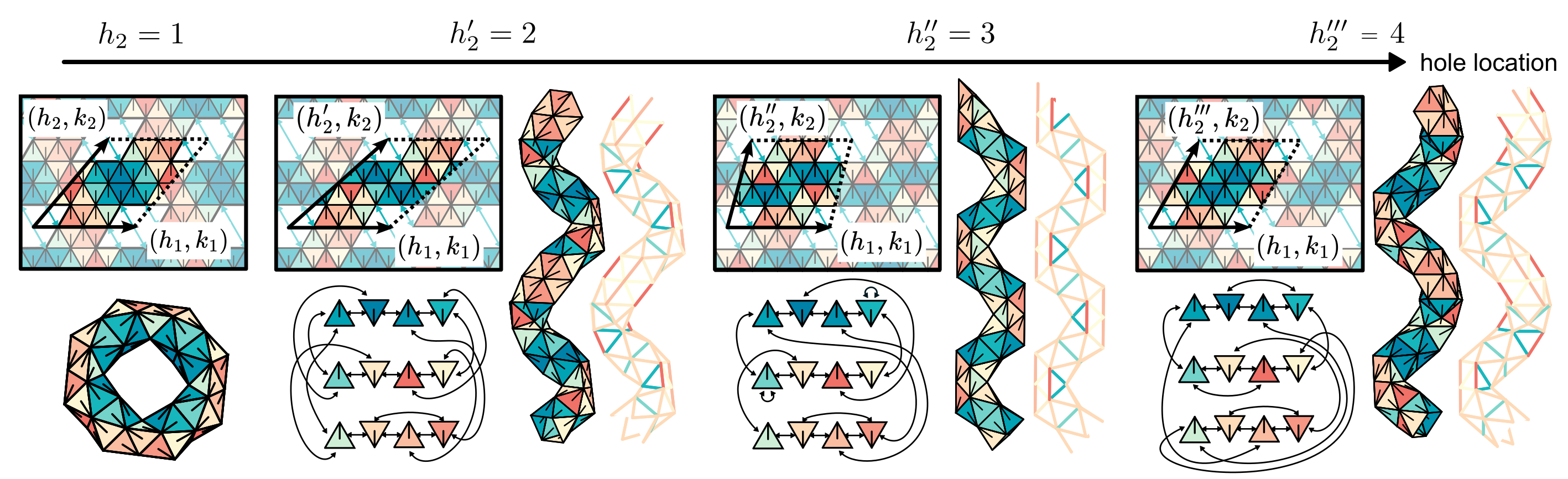}
\caption{\textbf{Edge-edge interactions for the assembled helical tubules.} For the helical tubules shown in the subsection of the main text `Programmable self-assembly of curved tubules using DNA origami triangles', we apply the same symmetry-guided inverse design of components to derive their interactions and binding angles using the corresponding holey tilings and 3D embedding, respectively. All of these holey tilings have the same shape hole, and number of components (12). The only thing that changes is the second periodic vector $(h_2, k_2)$ varies, changing the spacing of the holes. }
\end{center}
\end{figure}
\clearpage

\begin{longtable}{p{0.09\linewidth} p{0.09\linewidth} p{0.09\linewidth} p{0.09\linewidth}p{0.09\linewidth} p{0.09\linewidth} p{0.09\linewidth} p{0.09\linewidth} p{0.09\linewidth}}
\caption{\textbf{DNA sequence for multispecies assemblies.} A list of the set of eight interaction sequences that make up a side interaction of a monomer. As shown in Fig.~3C, Position 1 binds to Position 4 and Position 2 binds to Position 3. Sequences only interact with other sequences on the same helix. A lowercase `s' indicates that the sequence is self-complementary. An asterisk denotes a sequence that is complementary to the original.} \label{tab:sidesequence} \\
\input{Tab-HandleSequences}
\end{longtable}

\begin{longtable}{p{0.15\linewidth} p{0.10\linewidth} p{0.10\linewidth} p{0.18\linewidth}p{0.25\linewidth} }
\caption{\textbf{Toroid interactions.}} \label{tab:triInteractions} \\
\input{Tab-ToroidInteractions}
\end{longtable}

\begin{longtable}{p{0.15\linewidth} p{0.10\linewidth} p{0.10\linewidth} p{0.18\linewidth}p{0.25\linewidth} }
\caption{\textbf{Right-handed helical tubule interactions.}} \label{tab:RHHTInts} \\
\input{Tab-RHHT}
\end{longtable}

\begin{longtable}{p{0.15\linewidth} p{0.10\linewidth} p{0.10\linewidth} p{0.18\linewidth}p{0.25\linewidth} }
\caption{\textbf{Serpentine tubule interactions.}} \label{tab:SerpentInts} \\
\input{Tab-SerpentInt}
\end{longtable}

\begin{longtable}{p{0.15\linewidth} p{0.10\linewidth} p{0.10\linewidth} p{0.18\linewidth}p{0.25\linewidth} }
\caption{\textbf{Left-handed helical tubule interactions.}} \label{tab:LHHTInts} \\
\input{Tab-LHHT}
\end{longtable}

\bibliography{main.bib}
\clearpage

%% file: Tab-geometry_HT.tex
Structure ID	&	Number of species	&	Handedness	&	Helical radius	&	Tube radius	&  Pitch length & $\phi$ \\	\hline
\endfirsthead

Structure ID	&	Number of species	&	Handedness	&	Helical radius	&	Tube radius	&  Pitch length & $\phi$ \\	\hline
\endhead

(3,1,1,1)	&	4	&	left	&	0.6	&	0.55	&	2.83	&	0.12	\\
(3,1,1,2)	&	4	&	right	&	0.6	&	0.55	&	2.83	&	0.64	\\
(3,1,1,3)	&	4	&	achiral	&	0.35	&		&	2	&	0.96	\\
(3,2,1,2)	&	7	&	right	&	0.42	&	0.59	&	4.34	&	1.02	\\
(3,2,1,3)	&	7	&	left	&	0.42	&	0.59	&	4.32	&	1.02	\\
(3,3,1,1)	&	10	&	right	&	1.32	&	0.59	&	6.63	&	0.67	\\
(3,3,1,2)	&	10	&	achiral	&	0.64	&		&	4.59	&	1.06	\\
(3,3,1,3)	&	10	&	left	&	1.32	&	0.59	&	6.62	&	0.67	\\
(4,1,2,1)	&	8	&	left	&	0.95	&	0.66	&	2.96	&	0.46	\\
(4,1,2,2)	&	8	&	right	&	0.96	&	0.66	&	2.95	&	0.45	\\
(4,1,2,3)	&	8	&	right	&	0.34	&	0.67	&	3.71	&	1.05	\\
(4,1,2,4)	&	8	&	left	&	0.33	&	0.67	&	3.72	&	1.06	\\
(4,2,2,2)	&	12	&	right	&	0.87	&	0.69	&	5.42	&	0.78	\\
(4,2,2,3)	&	12	&	achiral	&	0.57	&		&	4.29	&	1.08	\\
(4,2,2,4)	&	12	&	left	&	0.89	&	0.69	&	5.36	&	0.76	\\
(4,3,2,1)	&	16	&	right	&	1.94	&	0.7	&	5.51	&	0.42	\\
(4,3,2,2)	&	16	&	right	&	0.75	&	0.73	&	6.44	&	0.94	\\
(4,3,2,3)	&	16	&	left	&	0.74	&	0.73	&	6.45	&	0.95	\\
(4,3,2,4)	&	16	&	left	&	1.79	&	0.7	&	5.14	&	0.43	\\

%% file: Tab-geometry_toroids.tex
Structure ID	&	Number of species	&	Symmetry	&	Toroid radius	&	Tube radius	& Commensurate \\	\hline
\endfirsthead
Structure ID	&	Number of species	&	Symmetry	& Toroid radius	&	Tube radius	& Commensurate \\	\hline
\endhead
(3,2,1,1)	&	7	&	5	&	1.55	&	0.58	&	No	\\
(3,4,1,3)	&	13	&	5	&	2.74	&	0.61	&	No	\\
(4,2,2,1)	&	12	&	4	&	1.7	&	0.69	&	Yes	\\
(4,4,2,4)	&	20	&	4	&	2.64	&	0.71	&	Yes	\\
(6,2,2,3)	&	16	&	5	&	1.79	&	0.91	&	No	\\
(7,2,3,3)	&	23	&	5	&	2.43	&	0.97	&	No	\\
(8,2,4,3)	&	32	&	4	&	2.27	&	1.15	&	Yes	\\
(9,2,4,4)	&	34	&	5	&	2.72	&	1.34	&	No	\\
(9,4,4,3)	&	52	&	5	&	4.02	&	1.39	&	No	\\

%% file: Tab-HandleSequences.tex
& Helix3 & Helix3 & Helix3 & Helix3 & Helix1 or 5 & Helix1 or 5 & Helix1 or 5 & Helix1 or 5 \\ 
Name & Pos1 & Pos2 & Pos3 & Pos4 & Pos1 & Pos2 & Pos3 & Pos4 \\ 
\hline
\endfirsthead
\hline
     & Helix3 & Helix3 & Helix3 & Helix3 & Helix1 or 5 & Helix1 or 5 & Helix1 or 5 & Helix1 or 5 \\ 
Name & Pos1 & Pos2 & Pos3 & Pos4 & Pos1 & Pos2 & Pos3 & Pos4 \\ 
\hline 
\endhead

\endfoot

\endlastfoot

sA & ACTAG & AGTTA & TAACT & CTAGT & TTCAA & CCATT & AATGG & TTGAA \\ 
sB & TTAAC & TCGAC & GTCGA & GTTAA & TCAGA & GTCTA & TAGAC & TCTGA \\ 
A & CAATA & TGATT & CTAGG & CACAT & ATGAC & TACAG & AACCT & GAGAC \\ 
A* & ATGTG & CCTAG & AATCA & TATTG & GTCTC & AGGTT & CTGTA & GTCAT \\ 
B & GCATC & TATTC & AGATT & TTCTC & AGATA & TTCCT & TTCCA & GATAT \\ 
B* & GAGAA & AATCT & GAATA & GATGC & ATATC & TGGAA & AGGAA & TATCT \\ 
C & AGTTC & CGATT & ATTCT & ATTCA & GGATA & TCATC & GGTAT & GGTAA \\ 
C* & TGAAT & AGAAT & AATCG & GAACT & TTACC & ATACC & GATGA & TATCC \\ 
D & CACAT & CAATA & ACGAA & TGATT & GAGAC & ATGAC & GACAG & TACAG \\ 
D* & AATCA & TTCGT & TATTG & ATGTG & CTGTA & CTGTC & GTCAT & GTCTC \\ 
E & TTCTC & GCATC & CTGTG & TATTC & GATAT & AGATA & ATGCA & TTCCT \\ 
E* & GAATA & CACAG & GATGC & GAGAA & AGGAA & TGCAT & TATCT & ATATC \\ 
F & ATTCA & AGTTC & CTTGA & CGATT & GGTAA & GGATA & ACTGA & TCATC \\ 
F* & AATCG & TCAAG & GAACT & TGAAT & GATGA & TCAGT & TATCC & TTACC \\ 
G & ACGAA & CTAGG & ACCTG & CAATA & GACAG & AACCT & ACTAA & ATGAC \\ 
G* & TATTG & CAGGT & CCTAG & TTCGT & GTCAT & TTAGT & AGGTT & CTGTC \\ 
H & CTGTG & AGATT & TCGTA & GCATC & ATGCA & TTCCA & AACAT & AGATA \\ 
H* & GATGC & TACGA & AATCT & CACAG & TATCT & ATGTT & TGGAA & TGCAT \\ 
I & CTTGA & ATTCT & GTAGA & AGTTC & ACTGA & GGTAT & AGAGA & GGATA \\ 
I* & GAACT & TCTAC & AGAAT & TCAAG & TATCC & TCTCT & ATACC & TCAGT \\ 
J & CAATA & GACCT & GGTAT & ATTCA & ATGAC & TGATT & CCTAT & GGTAA \\ 
J* & TGAAT & ATACC & AGGTC & TATTG & TTACC & ATAGG & AATCA & GTCAT \\ 
K & GCATC & AGTCA & AACCT & CACAT & AGATA & TATTC & CGATG & GAGAC \\ 
K* & ATGTG & AGGTT & TGACT & GATGC & GTCTC & CATCG & GAATA & TATCT \\ 
L & AGTTC & CGTCC & TTCCA & TTCTC & GGATA & CGATT & CTTGT & GATAT \\ 
L* & GAGAA & TGGAA & GGACG & GAACT & ATATC & ACAAG & AATCG & TATCC \\ 
M & ATTCA & CAATA & GACAG & GACCT & GGTAA & ATGAC & CTTGA & TGATT \\ 
M* & AGGTC & CTGTC & TATTG & TGAAT & AATCA & TCAAG & GTCAT & TTACC \\ 
N & CACAT & GCATC & TAACA & AGTCA & GAGAC & AGATA & ACGAA & TATTC \\ 
N* & TGACT & TGTTA & GATGC & ATGTG & GAATA & TTCGT & TATCT & GTCTC \\ 
O & TTCTC & AGTTC & AGTAT & CGTCC & GATAT & GGATA & CTGTG & CGATT \\ 
O* & GGACG & ATACT & GAACT & GAGAA & AATCG & CACAG & TATCC & ATATC \\ 
P & GACAG & GGTAT & AACAT & CAATA & CTTGA & CCTAT & AACTA & ATGAC \\ 
P* & TATTG & ATGTT & ATACC & CTGTC & GTCAT & TAGTT & ATAGG & TCAAG \\ 
Q & TAACA & AACCT & AGAGA & GCATC & ACGAA & CGATG & TCTTC & AGATA \\ 
Q* & GATGC & TCTCT & AGGTT & TGTTA & TATCT & GAAGA & CATCG & TTCGT \\ 
R & AGTAT & TTCCA & ACTAA & AGTTC & CTGTG & CTTGT & GTATG & GGATA \\ 
R* & GAACT & TTAGT & TGGAA & ATACT & TATCC & CATAC & ACAAG & CACAG 

%% file: Tab-ToroidInteractions.tex
Particle Label	&	Side ID	&	Sequence	&	Design Angle [deg.]	&	Angle Domain Length [bp]	\\	\hline
\endfirsthead
Particle Label	&	Side ID	&	Sequence	&	Design Angle [deg.]	&	Angle Domain Length [bp]	\\	\hline
\endhead
A	&	1	&	A	&	52.4	&	+7	\\	
	&	2	&	B	&	52.4	&	+7	\\	
	&	3	&	C	&	-49.1	&	-7	\\	
B	&	1	&	A*	&	52.4	&	+7	\\	
	&	2	&	D	&	52.4	&	+7	\\	
	&	3	&	sA	&	-27.7	&	-4	\\	
C	&	1	&	E	&	52.4	&	+7	\\	
	&	2	&	D*	&	52.4	&	+7	\\	
	&	3	&	F*	&	-49.1	&	-7	\\	
D	&	1	&	EI	&	52.4	&	+7	\\	
	&	2	&	G	&	52.4	&	+7	\\	
	&	3	&	H	&	-27.7	&	-4	\\	
E	&	1	&	I	&	52.4	&	+7	\\	
	&	2	&	G*	&	52.4	&	+7	\\	
	&	3	&	J*	&	21.4	&	+3	\\	
F	&	1	&	I*	&	52.4	&	+7	\\	
	&	2	&	K	&	52.4	&	+7	\\	
	&	3	&	sB	&	-27.7	&	-4	\\	
G	&	1	&	L	&	52.4	&	+7	\\	
	&	2	&	K*	&	52.4	&	+7	\\	
	&	3	&	M	&	21.4	&	+3	\\	
H	&	1	&	L*	&	52.4	&	+7	\\	
	&	2	&	B*	&	52.4	&	+7	\\	
	&	3	&	H*	&	-27.7	&	-4	\\	
I	&	1	&	N	&	109.5	&	+15	\\	
	&	2	&	F*	&	-49.1	&	-7	\\	
	&	3	&	C*	&	-49.1	&	-7	\\	
J	&	1	&	N*	&	109.5	&	+15	\\	
	&	2	&	O	&	0	&	0	\\	
	&	3	&	P	&	0	&	0	\\	
K	&	1	&	Q	&	70.5	&	+10	\\	
	&	2	&	O*	&	0	&	0	\\	
	&	3	&	M*	&	21.4	&	+3	\\	
L	&	1	&	Q*	&	70.5	&	+10	\\	
	&	2	&	J	&	21.4	&	+3	\\	
	&	3	&	P*	&	0	&	0	\\	

%% file: Tab-RHHT.tex
Particle Label	&	Side ID	&	Sequence	&	Design Angle [deg.]	&	Angle Domain Length [bp]	\\	\hline
\endfirsthead
Particle Label	&	Side ID	&	Sequence	&	Design Angle [deg.]	&	Angle Domain Length [bp]	\\	\hline
\endhead
A	&	1	&	A	&	52.4	&	+7	\\	
	&	2	&	B	&	52.4	&	+7	\\	
	&	3	&	M	&	21.4	&	+3	\\	
B	&	1	&	A*	&	52.4	&	+7	\\	
	&	2	&	D	&	52.4	&	+7	\\	
	&	3	&	H	&	-27.7	&	-4	\\	
C	&	1	&	E	&	52.4	&	+7	\\	
	&	2	&	D*	&	52.4	&	+7	\\	
	&	3	&	C	&	-49.1	&	-7	\\	
D	&	1	&	E*	&	52.4	&	+7	\\	
	&	2	&	G	&	52.4	&	+7	\\	
	&	3	&	H*	&	-27.7	&	-4	\\	
E	&	1	&	I	&	52.4	&	+7	\\	
	&	2	&	G*	&	52.4	&	+7	\\	
	&	3	&	F*	&	-49.1	&	-7	\\	
F	&	1	&	I*	&	52.4	&	+7	\\	
	&	2	&	K	&	52.4	&	+7	\\	
	&	3	&	R	&	-27.7	&	-4	\\	
G	&	1	&	L	&	52.4	&	+7	\\	
	&	2	&	K*	&	52.4	&	+7	\\	
	&	3	&	J*	&	21.4	&	+3	\\	
H	&	1	&	L*	&	52.4	&	+7	\\	
	&	2	&	B*	&	52.4	&	+7	\\	
	&	3	&	R*	&	-27.7	&	-4	\\	
I	&	1	&	N	&	109.5	&	+15	\\	
	&	2	&	F*	&	-49.1	&	-7	\\	
	&	3	&	C*	&	-49.1	&	-7	\\	
J	&	1	&	N*	&	109.5	&	+15	\\	
	&	2	&	O	&	0	&	0	\\	
	&	3	&	P	&	0	&	0	\\	
K	&	1	&	Q	&	70.5	&	+10	\\	
	&	2	&	O*	&	0	&	0	\\	
	&	3	&	M*	&	21.4	&	+3	\\	
L	&	1	&	Q*	&	70.5	&	+10	\\	
	&	2	&	J*	&	21.4	&	+3	\\	
	&	3	&	P*	&	0	&	0	\\	

%% file: Tab-SerpentInt.tex
Particle Label	&	Side ID	&	Sequence	&	Design Angle [deg.]	&	Angle Domain Length [bp]	\\	\hline
\endfirsthead
Particle Label	&	Side ID	&	Sequence	&	Design Angle [deg.]	&	Angle Domain Length [bp]	\\	\hline
\endhead
A	&	1	&	A	&	52.4	&	+7	\\	
	&	2	&	B	&	52.4	&	+7	\\	
	&	3	&	J*	&	21.4	&	+3	\\	
B	&	1	&	A*	&	52.4	&	+7	\\	
	&	2	&	D	&	52.4	&	+7	\\	
	&	3	&	H	&	-27.7	&	-4	\\	
C	&	1	&	E	&	52.4	&	+7	\\	
	&	2	&	D*	&	52.4	&	+7	\\	
	&	3	&	M	&	21.4	&	+3	\\	
D	&	1	&	E*	&	52.4	&	+7	\\	
	&	2	&	G	&	52.4	&	+7	\\	
	&	3	&	sA	&	-27.7	&	-4	\\	
E	&	1	&	I	&	52.4	&	+7	\\	
	&	2	&	G*	&	52.4	&	+7	\\	
	&	3	&	C	&	-49.1	&	-7	\\	
F	&	1	&	I*	&	52.4	&	+7	\\	
	&	2	&	K	&	52.4	&	+7	\\	
	&	3	&	H*	&	-27.7	&	-4	\\	
G	&	1	&	L	&	52.4	&	+7	\\	
	&	2	&	K*	&	52.4	&	+7	\\	
	&	3	&	F*	&	-49.1	&	-7	\\	
H	&	1	&	L*	&	52.4	&	+7	\\	
	&	2	&	B*	&	52.4	&	+7	\\	
	&	3	&	sB	&	-27.7	&	-4	\\	
I	&	1	&	N	&	109.5	&	+15	\\	
	&	2	&	F*	&	-49.1	&	-7	\\	
	&	3	&	C*	&	-49.1	&	-7	\\	
J	&	1	&	N*	&	109.5	&	+15	\\	
	&	2	&	O	&	0	&	0	\\	
	&	3	&	P	&	0	&	0	\\	
K	&	1	&	Q	&	70.5	&	+10	\\	
	&	2	&	O*	&	0	&	0	\\	
	&	3	&	M*	&	21.4	&	+3	\\	
L	&	1	&	Q*	&	70.5	&	+10	\\	
	&	2	&	J*	&	21.4	&	+3	\\	
	&	3	&	P*	&	0	&	0	\\	

%% file: Tab-LHHT.tex
Particle Label	&	Side ID	&	Sequence	&	Design Angle [deg.]	&	Angle Domain Length [bp]	\\	\hline
\endfirsthead
Particle Label	&	Side ID	&	Sequence	&	Design Angle [deg.]	&	Angle Domain Length [bp]	\\	\hline
\endhead

A	&	1	&	A	&	52.4	&	+7	\\	
	&	2	&	B	&	52.4	&	+7	\\	
	&	3	&	F*	&	-49.1	&	-7	\\	
B	&	1	&	A*	&	52.4	&	+7	\\	
	&	2	&	D	&	52.4	&	+7	\\	
	&	3	&	R	&	-27.7	&	-4	\\	
C	&	1	&	E	&	52.4	&	+7	\\	
	&	2	&	D*	&	52.4	&	+7	\\	
	&	3	&	J*	&	21.4	&	+3	\\	
D	&	1	&	E*	&	52.4	&	+7	\\	
	&	2	&	G	&	52.4	&	+7	\\	
	&	3	&	H	&	-27.7	&	-4	\\	
E	&	1	&	I	&	52.4	&	+7	\\	
	&	2	&	G*	&	52.4	&	+7	\\	
	&	3	&	M	&	21.4	&	+3	\\	
F	&	1	&	I*	&	52.4	&	+7	\\	
	&	2	&	K	&	52.4	&	+7	\\	
	&	3	&	H*	&	-27.7	&	-4	\\	
G	&	1	&	L	&	52.4	&	+7	\\	
	&	2	&	K*	&	52.4	&	+7	\\	
	&	3	&	C	&	-49.1	&	-7	\\	
H	&	1	&	L*	&	52.4	&	+7	\\	
	&	2	&	B*	&	52.4	&	+7	\\	
	&	3	&	R*	&	-27.7	&	-4	\\	
I	&	1	&	N	&	109.5	&	+15	\\	
	&	2	&	F*	&	-49.1	&	-7	\\	
	&	3	&	C*	&	-49.1	&	-7	\\	
J	&	1	&	N*	&	109.5	&	+15	\\	
	&	2	&	O	&	0	&	0	\\	
	&	3	&	P	&	0	&	0	\\	
K	&	1	&	Q	&	70.5	&	+10	\\	
	&	2	&	O*	&	0	&	0	\\	
	&	3	&	M*	&	21.4	&	+3	\\	
L	&	1	&	Q*	&	70.5	&	+10	\\	
	&	2	&	J*	&	21.4	&	+3	\\	
	&	3	&	P*	&	0	&	0	\\	